\documentclass[floats,twocolumn,nofootinbib,superscriptaddress,fleqn,usenatbib]{mnras}

\usepackage{amssymb}
\usepackage{amsmath}

\usepackage{graphicx}
\usepackage{graphics}
\usepackage{color}
\usepackage{fancyhdr} 
\usepackage{graphicx}
\usepackage{float}
\usepackage[dvipsnames]{xcolor}
\usepackage{comment}

\usepackage{subfig}

\usepackage[utf8]{inputenc}

\usepackage[justification=centerlast, format=plain, labelfont=bf]{caption}

\def\VEV#1{\left\langle #1 \right\rangle}

    \newcommand{\be}{\begin{equation}}
  \newcommand{\ee}{\end{equation}}
    \newcommand{\ba}{\begin{align}}
  \newcommand{\ea}{\end{align}}
   
\newcommand{\Tcmb}{T_{\rm CMB}}

\newcommand{\Msun}{M_{\odot}}
\newcommand{\Mpcinv}{ {\rm Mpc}^{-1} }
\newcommand{\fesc}{f_{\rm esc}}
\newcommand{\xHI}{x_{\rm HI}}

\newcommand{\vcb}{ {v_{\rm cb}}  }
\newcommand{\vrms}{ {v_{\rm rms}}  }

\title{The Impact of the First Galaxies on Cosmic Dawn and Reionization}

\author[J.B.~Mu\~noz et al.]{
Julian B.~Mu\~noz,$^{1}$\thanks{E-mail: julianmunoz@cfa.harvard.edu}
Yuxiang Qin,$^{2,3}$
Andrei Mesinger,$^{4}$
Steven G. Murray,$^{5}$
\newauthor{
Bradley Greig,$^{2,3}$
and Charlotte Mason$^{1,6,7}$
}
\\
$^{1}$Harvard-Smithsonian Center for Astrophysics, 60 Garden St., Cambridge, MA 02138, USA \\
$^{2}$School of Physics, University of Melbourne, Parkville, VIC 3010, Australia\\
$^{3}$ARC Centre of Excellence for All Sky Astrophysics in 3 Dimensions (ASTRO 3D)\\
$^{4}$Scuola Normale Superiore, 56126 Pisa, PI, Italy\\
$^{5}$School of Earth and Space Exploration, Arizona State University, Tempe, AZ\\
$^{6}$Cosmic Dawn Center (DAWN)\\
$^{7}$Niels Bohr Institute, University of Copenhagen, Jagtvej 128, 2200 København N, Denmark
}

\date{Accepted XXX. Received YYY; in original form ZZZ}

\pubyear{2021}

\begin{document}
\label{firstpage}
\pagerange{\pageref{firstpage}--\pageref{lastpage}}
\maketitle

\begin{abstract}
The formation of the first galaxies during cosmic dawn and reionization (at redshifts $z=5-30$), triggered the last major phase transition of our universe, as hydrogen evolved from cold and neutral to hot and ionized. 
The 21-cm line of neutral hydrogen will soon allow us to map these cosmic milestones and study the galaxies that drove them.
To aid in interpreting these observations, we upgrade the publicly available code {\tt 21cmFAST}.  We introduce a new, flexible parametrization of the additive feedback from: an inhomogeneous, $H_2$-dissociating (Lyman-Werner; LW) background; and dark matter -- baryon relative velocities; which recovers results from recent, small-scale hydrodynamical simulations with both effects.
We perform a large, ``best-guess" simulation as the 2021 installment
of the Evolution of 21-cm Structure (EOS) project.  This improves the previous release with a galaxy model that reproduces the observed UV luminosity functions (UVLFs), and by including a population of molecular-cooling galaxies. 
The resulting 21-cm global signal and power spectrum are significantly weaker, primarily due to a more rapid evolution of the star-formation rate density required to match the UVLFs.
Nevertheless, we forecast high signal-to-noise detections for both HERA and the SKA.
We demonstrate how the stellar-to-halo mass relation of the unseen, first galaxies can be inferred from the 21-cm evolution.
Finally, we show that the spatial modulation of X-ray heating due to  relative velocities
provides a unique acoustic signature that is detectable at $z \approx 10-15$ in our fiducial model.
Ours are the first public simulations with joint inhomogeneous LW and relative-velocity feedback across the entire cosmic dawn and reionization, and we make them available at \href{https://www.dropbox.com/sh/dqh9r6wb0s68jfo/AACc9ZCqsN0SQ_JJN7GRVuqDa?dl=0}{this link}.
\end{abstract}


\maketitle

\section{Introduction}

The epoch of reionization (EoR) and the cosmic dawn (CD) represent fundamental milestones in the history of our universe, and are rapidly becoming the next frontier in astrophysics.
These two eras witnessed the last major phase change of our universe, as the intergalactic medium (IGM) evolved from being cold and neutral following recombination, to being hot and ionized due to the radiation emitted by the first galaxies.
At present we only hold some pieces of this cosmic puzzle~(e.g., \citealt{loeb2013first, Mesinger16}).
Nevertheless, the next few years will see the advent of different observations targeting these cosmic eras~\citep{Bowman:2018yin,Beardsley:2016njr,DeBoer:2016tnn,Mellema:2012ht,Greig:2020suk,Mertens:2020llj}, which will open a window to the astrophysics of the early universe.

While the stellar content of low-$z$ galaxies is relatively well understood (see, e.g., \citealt{Wechsler:2018pic} for a review), much less is known about the first galaxies that started the CD.
Given the hierarchical paradigm of structure formation, we expect the first (Pop III) stars to form in small molecular-cooling galaxies (MCGs) at $z\sim $ 20--30~\citep{Tegmark:1996yt,Abel:2001pr,Bromm:2003vv,Haiman:2006si,Trenti:2010hs}, residing in haloes with virial temperatures $T_{\rm vir}\lesssim 10^4$ K (which corresponds to total halo masses $M_h\lesssim 10^8\,\Msun$ during the EoR and CD).
Feedback eventually quenches star formation in these MCGs, and hierarchical evolution ushers in the era of heavier, atomically cooled galaxies (ACGs; $M_h\gtrsim 10^8\,\Msun$).  
With most ACGs forming out of pre-enriched MCG building blocks, it is likely that second-generation (Pop II) stars drove the bulk of cosmic reionization at $z\sim$ 5--10 (e.g., \citealt{Aghanim:2018eyx, Greig:2018rts, Mason:2019ixe, Choudhury:2020vzu, Qin:2021gkn}).

Current data, from ultraviolet (UV) luminosity functions (UVLFs; \citealt{Bouwens:2013hxa,Bouwens:2014fua,Livermore:2016mbs, Atek:2015axa,Atek:2018nsc,Ishigaki_2018,Oesch_2018,Bouwens2021}), the Lyman-$\alpha$ forest~\citep{Becker:2014oga,Bosman:2018xxh,  Qin:2021gkn}, and the evolution of the volume-averaged hydrogen neutral fraction ($\bar{x}_{\rm HI}$; \citealt{Aghanim:2018eyx,Mason:2019oeg}), provide some constraints~\citep{Park:2018ljd,Naidu:2019gvi} on the halo-galaxy connection for ACGs at $z\lesssim 10$.
However, very little is known about MCGs or the $z > 10$ universe.
The brightest galaxies in this regime will be reached by the James Webb Space Telescope (JWST) and the Nancy Grace Roman Space Telescope ({\it Roman}).  Unfortunately, the bulk of ACGs and MCGs are too faint to be seen with these telescopes, and must be studied indirectly with 21-cm observatories.  These include both global-signal efforts like the Experiment to Detect the Global EoR Signature (EDGES), the Shaped Antenna measurement of the background Radio Spectrum (SARAS), the Large-aperture Experiment to Detect the Dark Ages (LEDA), as well as interferometers like the Low-Frequency Array (LOFAR), the Murchison Widefield Array (MWA), the Hydrogen Epoch of Reionization Array (HERA) and the Square Kilometre Array (SKA). Therefore, it is of paramount importance to develop flexible and robust models of the astrophysics of the CD and the EoR to compare against these upcoming data.
In this work we build upon the public {\tt 21cmFAST} code to include MCGs with all of the relevant sources of feedback.

Stellar formation in MCGs is easily disrupted by different processes, given their shallow potential wells.
In addition to feedback from photo-heating and supernovae~\citep[][which also affects stellar formation in ACGs]{Draine:1996hna,Barkana:1999apa,Wise:2007nb,Sobacchi:2013ww}, MCGs suffer from two distinct sources of feedback.
The first is driven by Lyman-Werner (LW) radiation, composed of photons in the 11.2-13.6 eV band, which efficiently photo-dissociate molecular hydrogen (H$_2$), hampering gas cooling and subsequent star formation in MCGs~\citep{Machacek:2000us,Johnson:2006mt,OShea:2007ita,SafranekShrader:2012qn,Visbal:2014fta,Schauer2017,Skinner_Wise_2020_selfshielding}.
The second are driven by the dark matter (DM)-baryon streaming velocities ($\vcb$), which impede gas from efficiently accreting and cooling onto DM haloes, thus slowing the formation of the first stars~\citep{Tseliakhovich:2010bj,Greif:2011iv,Naoz:2012fr,OLeary:2012gem,Hirano:2017znw,Schauer:2018iig}.
Until recently it was not clear how these two feedback channels interacted.
A clearer picture has emerged from the small-scale hydrodynamical simulations of PopIII star formation in \citet{Schauer:2020gvx} and \citet{Kulkarni:2020ovu}.
In this picture, the two sources of feedback are additive, enhancing the minimum mass that an MCG ought to have to be able to form stars.
We synthesize the results from these simulations into a flexible fitting formula that jointly includes both LW and $\vcb$ feedback.
Along with a calibration for this formula, we assume MCGs host PopIII stars with a simple stellar-to-halo mass relation (SHMR), distinct from that of ACGs hosting PopII stars.

We include these results into the semi-numeric simulation code {\tt 21cmFAST}\footnote{Publicly available at {\url{https://github.com/21cmfast/21cmFAST}}.}~\citep{Mesinger:2007pd,Mesinger:2010ne,Murray:2020trn}.
This builds upon the implementation of the streaming velocities in {\tt 21cmvFAST} by \citet[][though in that work it was assumed MCGs and ACGs shared the same SHMR and that LW feedback was isotropic]{Munoz:2019rhi,Munoz:2019fkt}, and the first implementation of MCGs with inhomogeneous LW feedback in {\tt 21cmFAST} by \citet[][though there relative velocities were not considered and the old LW prescription from \citealt{Machacek:2000us} was used]{Qin:2020xyh,Qin:2020pdx}.
This code is now able to self-consistently evolve the anisotropic LW background, the streaming velocities, as well as the X-ray, ionizing, and non-ionizing UV backgrounds, necessary to predict the 21-cm signal.

We run a large simulation suite  (1.5 Gpc comoving on a side, with $1000^3$ cells), as the 2021 installment of the Evolution Of 21-cm Structure (EOS) project, updating those in~\citet{Mesinger:2016ddl}.
This represents our state of knowledge about cosmic dawn and reionization.
We dub this EOS simulation {\it AllGalaxies}, and we make its most relevant lightcones public\footnote{EOS lightcones can be downloaded at \href{https://www.dropbox.com/sh/dqh9r6wb0s68jfo/AACc9ZCqsN0SQ_JJN7GRVuqDa?dl=0}{this link}
\label{footnote:linkEOS}
}.
This simulation goes beyond the previous {\it FaintGalaxies} model, both by including PopIII-hosting MCGs as well as including a SHMR that fits current UVLFs. 
As a consequence, we predict a slower evolution of the 21-cm signal, and 21-cm fluctuations that are nearly an order of magnitude smaller (see also \citealt{Mirocha2016_UVLF_GS, Park:2018ljd}).
Our {\it AllGalaxies} (EOS2021) simulation represents the current state-of-knowledge of the evolution of cosmic radiation fields during the first billion years.

Furthermore, we explore how parameters governing star formation and feedback in MCGs impact the 21-cm signal and other observables.
We find that MCGs likely dominate the star formation rate density (SFRD) at $z\gtrsim 12$, implying they are important for the timing of the CD but not the EoR (see e.g.,~\citealt{Wu:2021kch,Qin:2020xyh}).
We also search for the velocity-induced acoustic oscillations (VAOs) that appear due to the acoustic nature of the streaming velocities~\citep{Munoz:2019rhi,Dalal:2010yt,Visbal:2012aw,McQuinn:2012rt,Fialkov:2012su}, and for the first time identify VAOs in a full 21-cm lightcone, as opposed to a co-eval (i.e., fixed-$z$) box.
We predict significant VAOs for redshifts as low as $z=10-15$, which act as a standard ruler to the cosmic dawn~\citep{Munoz:2019fkt}.

This paper is structured as follows.
In Sec.~\ref{sec:FirstGalaxies} we introduce our model for the first galaxies, both atomic- and molecular-cooling, which we use
in Sec.~\ref{sec:CDEoR} to predict how the epochs of cosmic dawn and reionization unfold.
Secs.~\ref{sec:astroparams} and \ref{sec:VAOs} explore how the first galaxies shape the 21-cm signal, where in the former we vary the MCG parameters in our simulations, and in the latter we study the VAOs.
Finally, we conclude in Sec.~\ref{sec:Conclusions}.
In this work we fix the cosmological parameters to the best fit from {\it Planck} 2018 data (TT,TE,EE+lowE+lensing+BAO in \citealt{Aghanim:2018eyx}), and all distances are comoving unless  specified otherwise.

\section{Modeling the First Galaxies}
\label{sec:FirstGalaxies}

We begin by describing our model for the first galaxies.
As our semi-numerical {\tt 21cmFAST} simulations do not keep track of the metallicity and accretion history of each galaxy, we use population-averaged quantities to relate stellar properties to host halo masses.  Although these relations can allow for mixed stellar populations, 
we will make the simplifying assumption (consistent with simulations and merger-tree models; e.g., \citealt{Xu16,Mebane18}) that on average PopIII stars form in (first-generation, unpolluted) MCGs, whereas PopII stars form in (second-generation, polluted) ACGs.
ACGs are hosted by haloes with virial temperatures of $T_{\rm vir} \gtrsim 10^4$ K \cite[][corresponding to halo masses $M_h \gtrsim M_{\rm atom}(z)\sim 10^8\,\Msun$ at the relevant redshifts]{Oh:2001ex}, whereas MCGs reside in minihaloes with $10^3$ K $\lesssim T_{\rm vir} \lesssim 10^4$ K \cite[][and thus
$M_{\rm mol}<M_h<M_{\rm atom}$, with $M_{\rm mol}(z)\sim 10^6\,\Msun$ during cosmic dawn]{Tegmark:2008au}.
As larger halos form in the universe, star formation transitions from being dominated by Pop III stars to being dominated by Pop II stars (e.g.~\citealt{Scheider02, BL04}). 
We now describe how these galaxy populations are modeled and the feedback processes that affect them.

\subsection{Pop II star formation in atomic-cooling galaxies}
\label{subsec:ACGs}

For the better-understood PopII stars forming in ACGs we follow the model from \citet{Park:2018ljd}, which we now briefly review.

The key ingredient that enters our calculations is the (spatially varying) star-formation rate density (SFRD), which for ACGs we calculate as
\be
{\rm SFRD}^{\rm (II)} = \int dM_h \dfrac{dn}{dM_h} \dot M_*^{\rm (II)} f_{\rm duty}^{\rm (II)},
\label{eq:SFRDII}
\ee
where $dn/dM_h$ is the conditional halo mass function (HMF, e.g., \citealt{Barkana:2003qk}), $f_{\rm duty}^{\rm (II)}$ is a duty cycle (i.e., halo occupation fraction) described below, and $\dot M_*$ is the star-formation rate (SFR).  The superscripts $^{\rm (II)}$ denote that a quantity refers to PopII-dominated ACGs.

For both ACGs and MCGs, we assume that the SFR is proportional (on average) to the stellar mass divided by a characteristic timescale:
\be
\dot M_*^{(i)} = \dfrac{M_*^{(i)}}{t_* H^{-1}(z)},
\ee
where $H$ is the Hubble expansion rate, and $t_*$ is a free parameter, which for simplicity we fix to $t_*=0.5$ (as it will be degenerate with the normalization of the stellar fraction).

The SHMR is modeled as a simple power law:
\be
M_*^{\rm (II)} = f_{*,10}^{\rm (II)} \dfrac{\Omega_b}{\Omega_m} \left(\dfrac{M_h}{10^{10} \Msun}\right)^{\alpha_*^{\rm (II)}} M_h,
\label{eq:MstarACG}
\ee
where $\Omega_{b/m}$ are the baryon and matter densities.
This SHMR is governed by two parameters, the normalization $f_{*,10}^{\rm (II)}$ (defined at a characteristic scale $M_h=10^{10}\,\Msun$), and a power-law index $\alpha_*^{\rm (II)}$, which controls its mass dependence\footnote{We note that we could rephrase the galaxy-halo connection in terms of $\tilde{f_*} = \dot M_*/\dot M_h$ rather than $f_*$, and assume a halo-accretion rate $\dot M_h$, as in \citet{Mason:2015cna,Tacchella:2018qny} and \citet{Mirocha:2020slz}.
For an exponential accretion rate $\dot M_h = M_h/t_*$ (which provides a good fit to simulations; \citealt{Schneider:2020xmf}) these two formulations are identical.
}.
The value of this index could be set by supernovae (SNe) feedback, which is thought to regulate star formation inside these small galaxies (e.g.~\citealt{Wyithe2013MNRAS.428.2741W, Dayal14, Yung2019_UVLF}).

This model, without any additional redshift dependence, provides an excellent fit to current EoR observations at $5 \lesssim z \lesssim 10$, including the Lyman-$\alpha$ forest opacity fluctuations ~\citep{Qin:2021gkn} and the faint-end UVLFs~\citep{Park:2018ljd,Rudakovskyi:2021jyf} (or the entire luminosity range if a high-mass turnover is added to characterize AGN-induced feedback in rare, bright galaxies; e.g., \citealt{Sabti:2020ser}).

The final part of Eq.~\eqref{eq:SFRDII} is the duty fraction,
\be
f_{\rm duty}^{\rm (II)} = \exp\left(-M_{\rm turn}^{\rm (II)}/M_h\right),
\ee
which accounts for inefficient star formation below a characteristic scale $M_{\rm turn}^{\rm (II)}$.  Halos below $M_{\rm turn}^{\rm (II)}$ are exponentially less likely to host an ACGs, due to inefficient atomic cooling and/or feedback, as discussed below.

\subsubsection*{Feedback on ACGs}

As in previous work (e.g.,~\citealt{Qin:2020xyh}), we consider two sources of feedback in ACGs: photo-heating and supernovae.

Radiation during cosmic reionization heats the gas around low-mass galaxies (both atomic- and molecular-cooling), delaying its collapse and subsequent star formation (e.g., \citealt{TW96, NM14}).
We use the 1D collapse-simulation results from \citet{Sobacchi:2014rua} to calculate the (local) critical halo mass below which star formation becomes inefficient due to photo-heating (see also~\citealt{Hui:1997dp,Okamoto:2008sn,Ocvirk:2018pqh,Katz:2019due}):
\begin{align}\label{eq:mcrit_ion}
\dfrac{M_{\rm crit}^{\rm ion}}{2.8{\times}10^9 \Msun}  &= \left(\dfrac{\Gamma_{\rm ion}}{10^{-12}{\rm s^{-1}}}\right)^{0.17} \!\!\!
\left(\frac{10}{1{+}z}\right)^{2.1}
\!\!
\left[1 {-} \left(\frac{1{+}z}{1{+}z^{\rm ion}}\right)^{2}\right]^{2.5},
\end{align}
where $\Gamma_{\rm ion}$ is the local ionizing background, and $z^{\rm ion}$ is the reionization redshift of the simulation cell.

Therefore, for ACGs the turn-over mass is
\begin{equation}
\label{eq:mturn}
   M_{\rm turn}^{\rm (II)} = {\rm max} \left( M_{\rm crit}^{\rm ion}, M_{\rm atom} \right),
\end{equation}
where $M_{\rm atom} = 3.3\times 10^{7}\,\Msun [(1+z)/21]^{-3/2}$ during the epoch of interest for our fiducial cosmology.

We additionally include feedback from supernovae (SNe) 
or radiative processes 
by assuming a smaller fraction of gas is available to form stars inside smaller-mass halos.
Such ``feedback-limited" models result in a (positive) power-law scaling of the SHMR, c.f. the power-law index $\alpha_*^{\rm (II)}$ introduced in Eq.~\eqref{eq:MstarACG} (e.g., \citealt{Wyithe2013MNRAS.428.2741W}).
We fix this parameter to $\alpha_*^{\rm (II)}=0.5$ in this work, as it provides an excellent fit to UV LFs at $z=6$ -- 10 \citep{Qin:2021gkn}.
Different authors infer somewhat different values of $\alpha_*^{\rm (II)}$, as they can depend on the assumed star-formation histories (e.g.,~\citealt{Behroozi:2019kql, Tacchella:2018qny}).  Although $\alpha_*^{\rm (II)}$ could also evolve with time, $z=2-10$ LFs are consistent with a constant value (e.g.,~\citealt{Mason:2015cna,Tacchella:2018qny,Mirocha:2020slz}).
Our implementation also allows for the possibility that SNe feedback results in an additional characteristic mass scale (effectively adding $M_{\rm SNe}$ to the RHS of eq. \ref{eq:mturn}; c.f. \citealt{Qin:2020xyh}).  However, we do not consider that possibility in our fiducial model (setting $M_{\rm SNe} < {\rm max} [M_{\rm crit}^{\rm ion}, M_{\rm atom}] $); this is supported by $z=0$ data that seems to show no cutoff for haloes down to $M_h\approx 10^9\,\Msun$~\citep{DES:2019ltu} roughly at the $M_{\rm atom}$ threshold (which however are likely not actively star forming at $z=0$).

\subsection{Pop III star formation in molecular-cooling galaxies}
\label{subsec:MCGs}

The star formation and feedback mechanisms in the first MCG galaxies are still not fully understood.
For simplicity, we will assume their SHMR has the same, generic power-law form as ACGs, with a few modifications (see \citealt{Qin:2020xyh} for further details of the implementation in {\tt 21cmFAST}).
We take the PopIII SFRD\footnote{We introduce an approximation to analytically calculate the SFRD within {\tt 21cmFAST}.
This dramatically speeds up the calculation of the SFRD tables, reducing that computational cost by nearly two orders of magnitude.
This option can be turned on by setting {\tt FAST\_FCOLL\_TABLES = True}.
We encourage the reader to visit Appendix~\ref{app:fasttabs} for details.}
to be
\be
{\rm SFRD}^{\rm (III)} = \int dM_h \dfrac{dn}{dM_h} \dot M_*^{\rm (III)} f_{\rm duty}^{\rm (III)}.
\label{eq:SFRDIII}
\ee
The changes with respect to Eq.~\eqref{eq:SFRDII} are in the SFR and the duty cycle.
For the former, we assume the same relation between $\dot M_*$ and $M_*$ as Eq.~\eqref{eq:MstarACG}, though we allow MCGs to have a unique SHMR:
\be
M_*^{\rm (III)} = f_{*,7}^{\rm (III)}\dfrac{\Omega_b}{\Omega_m} \left(\dfrac{M_h}{10^{7} \Msun}\right)^{\alpha_*^{\rm (III)}} M_h,
\label{eq:MstarMCG}
\ee
which has two new free parameters, $f_{*,7}^{\rm (III)}$, and $\alpha_*^{\rm (III)}$.

The other difference is the MCG duty cycle,
\be
f_{\rm duty}^{\rm (III)} = \exp(-M_h/M_{\rm turn}^{\rm (III)}) \exp(-M_{\rm atom}/M_h),
\ee
which follows our previous assumption that there is a smooth transition from MCGs (below $M_{\rm atom}$) forming PopIII stars to ACGs (above that mass) forming PopII stars.
Further, there is a lower-mass cutoff $M_{\rm turn}^{\rm (III)}$ for MCG stellar formation, set by different feedback processes as detailed below.  {\it Our comprehensive treatment of $M_{\rm turn}^{\rm (III)}$ represents the main modeling improvement of this work.}

\subsubsection*{Feedback on MCGs}

MCGs reside in haloes with shallow gravitational potentials, which makes them highly sensitive to different effects that disrupt their gas distribution, abundance, or H$_2$ content.
In addition to the previously mentioned feedback, MCGs are also sensitive to: (i) Lyman-Werner photons, and (ii) the DM-baryon relative velocities.
As we did for ACGs, we set the turnover mass to be
\begin{equation}
    M_{\rm turn}^{(\rm III)}  = {\rm max} \left (M_{\rm crit}^{\rm ion},M_{\rm mol}\right),
\end{equation}
where the first term accounts for photo-ionization feedback, and the second accounts for the additional impact of LW and $\vcb$.

Both the LW and $\vcb$ feedback effects impede stellar formation, and the simulations from \citet{Schauer:2020gvx} and \citet{Kulkarni:2020ovu} show that their joint impact is cumulative.
We model the molecular-cooling turn-over mass as a product of three factors:
\be
M_{\rm mol}  = M_0(z) \,  f_{\vcb}(\vcb) \, f_{\rm LW}(J_{21}),
\label{eq:Mturnfit}
\ee
where $J_{21}$ is the LW intensity in units of $10^{-21}$ erg s$^{-1}$ cm$^{-2}$ Hz$^{-1}$ sr$^{-1}$,
and $M_0(z) = \tilde M_0 \times (1+z)^{-3/2}$ is the (no-feedback) molecular-cooling threshold, with $\tilde M_0=3.3 \times 10^7 \Msun$ (corresponding to $T_{\rm vir}=10^3$ K; \citealt{Tegmark:1996yt}).
The two factors $f_{\rm LW}$ and $f_{\vcb}$ encode the two new sources of feedback, and our chosen functional form assumes that they add coherently (as assumed previously in  \citealt{Fialkov:2012su} and \citealt{Munoz:2019rhi}) and are redshift independent.
This was shown to be the case in \citet{Schauer:2020gvx} for values of $J_{21}\leq 0.1$, as well as in \citet{Kulkarni:2020ovu} for $J_{21}=1$. 
For larger values, $J_{21}\geq 10$, however, the simulations of \citet{Kulkarni:2020ovu} show a deviation from this form (towards less suppression). This large-$J_{21}$ regime only occurs at $z\lesssim 6$ in our simulations, at which point ACGs completely dominate the global SFRD, making MCGs (and thus their feedback mechanisms) largely irrelevant when computing radiation backgrounds.

\subsubsection{Lyman-Werner Feedback}

Photons in the LW band (11.2-13.6 eV) can photo-dissociate H$_2$ molecules, thus impeding the ability of gas to cool onto MCGs
~\citep{Tegmark:1996yt,Abel:2001pr,Bromm:2003vv,Haiman:2006si,Trenti:2010hs}.
\citet{Machacek:2000us} showed that
the minimum halo mass for star formation in MCGs has a power-law dependence on $J_{21}$.
This result has been the standard in studies of CD and the EoR (see e.g., \citealt{Fialkov:2012su,Visbal:2014fta,Mirocha:2017xxz,Munoz:2019rhi,Qin:2020pdx,Qin:2020xyh}).
However, recent simulations have shown that self-shielding can be important, which reduces the suppression produced by a given LW background.
We include these improved results in our analysis.

\begin{figure}
	\includegraphics[width=0.48\textwidth]{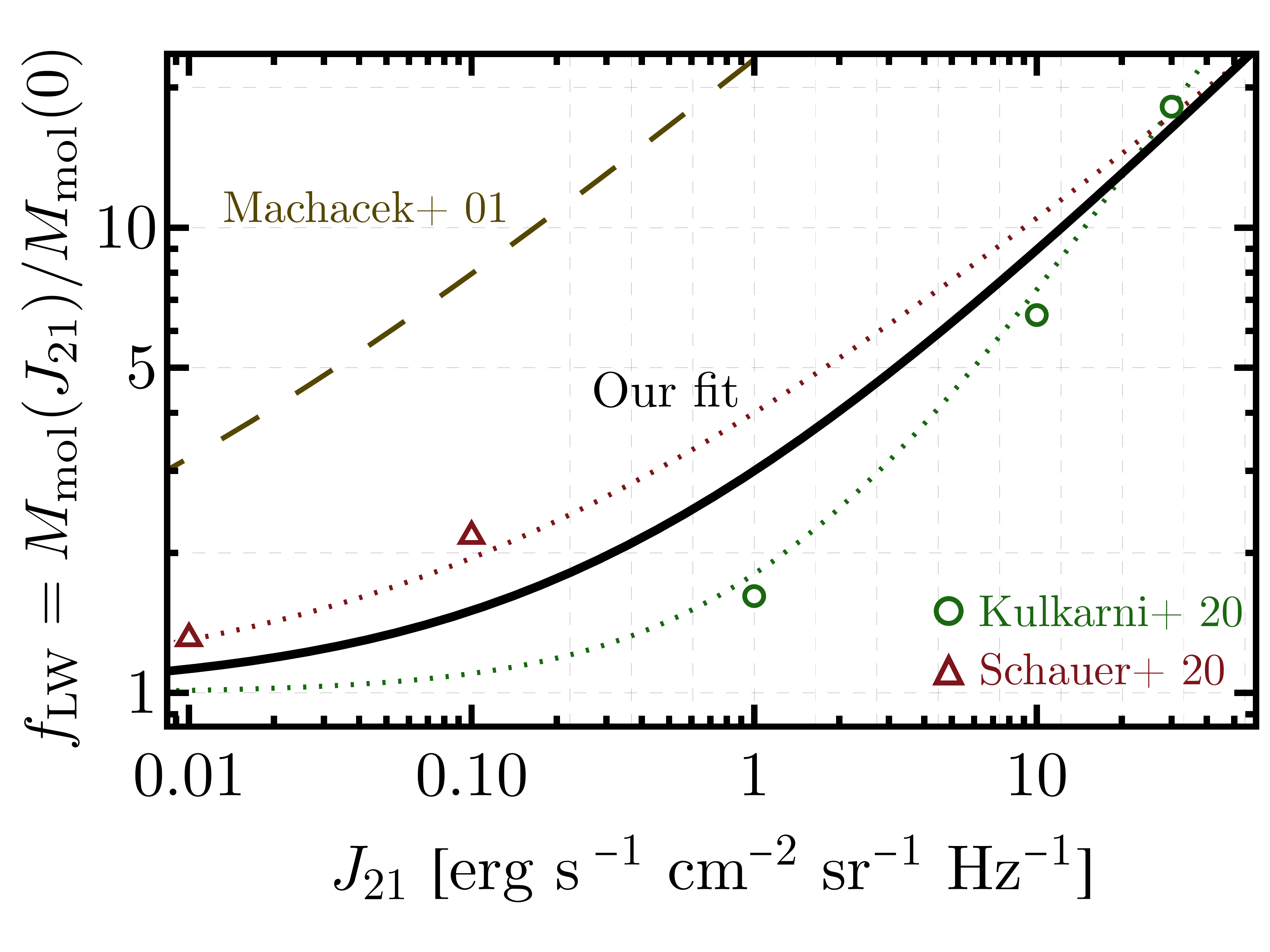}
	\caption{
	Feedback factor $f_{\rm LW}$ describing how the turn-over (minimum) mass for a molecular-cooling halo to form stars grows due to Lyman-Werner (LW) feedback, as a function of the LW intensity $J_{21}$ in units of $10^{-21}$ erg s$^{-1}$ cm$^{-2}$ sr$^{-1}$ Hz$^{-1}$.
	The data points are reduced from simulation data in~\citet[][green]{Kulkarni:2020ovu} and \citet[][red]{Schauer:2020gvx}.
	Our fit, which sits in the middle in solid black, follows Eq.~\eqref{eq:fLW} with $A_{\rm LW}=2$ and $\beta_{\rm LW}=0.5$.
	The parameters of Eq.~\eqref{eq:fLW} can be modified to closely follow each simulation, as the red and green lines are obtained with different parameter combinations.
	The dashed brown line is the fit from \citet{Machacek:2000us}, which did not include self-shielding.
	}	
	\label{fig:fLW_fit}
\end{figure}

In particular, we use the results from \citet{Kulkarni:2020ovu} and \citet[][see also \citealt{Skinner_Wise_2020_selfshielding} for similar conclusions]{Schauer:2020gvx}, which have independently tackled the issue of jointly simulating the effects of LW (including self-shielding) and relative-velocity feedback (which we will explore later) on the first stars. 
From Eq.~\eqref{eq:Mturnfit} we can separate the LW feedback factor
\begin{equation}
    f_{\rm LW} = \dfrac{M_{\rm mol}(J_{21})}{M_{\rm mol}(0)},
    \label{eq:fLWdef}
\end{equation}
and calculate it from simulation results, which factors out differences in the zero-point mass (i.e., on the mass $M_0$ in the absence of feedback).
We show the simulation results for $f_{\rm LW}$ in Fig.~\ref{fig:fLW_fit}, where it is clear that the \citet{Kulkarni:2020ovu} results are below the \citet{Schauer:2020gvx} ones, even for larger values of $J_{21}$.
However, a direct comparison of their results is difficult due to differences in the assumed LW intensities: \citet{Kulkarni:2020ovu} considered $J_{21}\geq 1$ (or zero), whereas \citet{Schauer:2020gvx} considered $J_{21}\leq 0.1$.

Rather than use a fit to $f_{\rm LW}$ from either group, we use a flexible parametrization inspired by \citet{Visbal:2014fta}:
\be
f_{\rm LW} = 1 + A_{\rm LW} (J_{21})^{\beta_{\rm LW}},
\label{eq:fLW}
\ee
with $ A_{\rm LW}$ and $\beta_{\rm LW}$ as free parameters that can be varied within {\tt 21cmFAST}.
The \citet{Kulkarni:2020ovu} simulation results can be well fit setting $\{A_{\rm LW}, \beta_{\rm LW}\}=\{0.8,0.9\}$, whereas the \citet{Schauer:2020gvx} ones require $\{A_{\rm LW}, \beta_{\rm LW}\}=\{3.0,0.5\}$, which indicates stronger feedback, though a weaker dependence with $J_{21}$.
Both of these fits are shown in Fig.~\ref{fig:fLW_fit}.
Instead of choosing between these two, we use the flexibility of Eq.~\eqref{eq:fLW} to propose a joint fit, which lands in between both simulation results, with $ A_{\rm LW}=2$ and $\beta_{\rm LW}=0.6$ (where the round numbers are purposefully chosen to avoid conveying more agreement than the simulations provide).
These will be our fiducial parameters for this work.
For reference, the old work of \citet{Machacek:2000us}, which did not account for self shielding, had $\{A_{\rm LW}, \beta_{\rm LW}\}=\{22,0.47\}$, producing a correction that was nearly a factor of 10 larger (also shown in Fig.~\ref{fig:fLW_fit}).
We emphasize that we assume that the LW feedback factor only depends on $J_{21}$, and not on $z$ or any halo property, as a simplifying assumption, which could however be revisited if required by further simulations.

\subsubsection{Relative-Velocity Feedback}

The DM-baryon relative velocities impede the formation of stars in the first galaxies in two main ways.  The first one, pointed out in \citet[][see also \citealt{Naoz:2011if}]{Tseliakhovich:2010bj}, is that regions of large velocity show suppressed matter fluctuations at small scales, as the baryons there contribute less to the growth of structure.
As a consequence, the abundance of small-mass halos is modulated by relative velocities: fewer collapsed halos reside in regions with large streaming motions.
This effect is difficult to include in our semi-numerical simulations, as it would require altering the power spectrum at each cell~\footnote{We note that the relative velocities are coherent on scales below 3 Mpc~\citep{Tseliakhovich:2010bj}, and thus can be taken to be constant within each of our cells.}.
Instead, we include this  effect on average by modifying the power spectrum in all cells.
We solve for the evolution of the baryon and DM overdensities following \citet[][based on \citealt{Tseliakhovich:2010bj}]{Munoz:2019rhi}, and find that the impact of $v_{\rm bc}$ on the power spectrum at a wavenumber $k$ is well captured by:
\be
\dfrac{P_m(k,z;\vcb)}{P_m(k,z;0)} = 1 - A_p  \exp \left[ - \dfrac{(\log[k/k_p])^2}{2 \sigma_p^2}, \right]
\label{eq:Pmatterfit}
\ee
where the three free parameters, $A_p$, $k_p$ and $\sigma_p$ depend on $\vcb$, and mildly on $z$.
Here we fix to the values at $z=20$ and at the root-mean square (rms) relative velocity ($\vcb=\vrms$), which are $A_p=0.24$, $k_p = 300\, \Mpcinv$, and $\sigma_p=0.9$.
This is conservative, in that it will yield no VAOs (as these quantities do not spatially fluctuate) and will suppress MCGs by roughly the average amount.
We encourage the interested reader to visit Appendix~\ref{app:Pmattervcb} for details of how this fit is obtained, as well as to find the fit as a function of $z$ and $\vcb$.
This matter-fluctuation suppression also affects ACGs, and therefore should be included even when MCGs are not.
However, we find this to be a subdominant effect to the one we will discuss next, and thus our average treatment will suffice. We leave for future work implementing the fluctuations on this suppression.

The second---and largest---effect of the velocities is to suppress star formation in MCGs \citep{Dalal:2010yt, Tseliakhovich:2010yw}.
Small halos in regions with large relative velocities have difficulties in accreeting gas, as well as cooling gas into stars (e.g., \citealt{Naoz:2012fr, Greif:2011iv,Schauer:2018iig}).

\begin{figure}
	\includegraphics[width=0.48\textwidth]{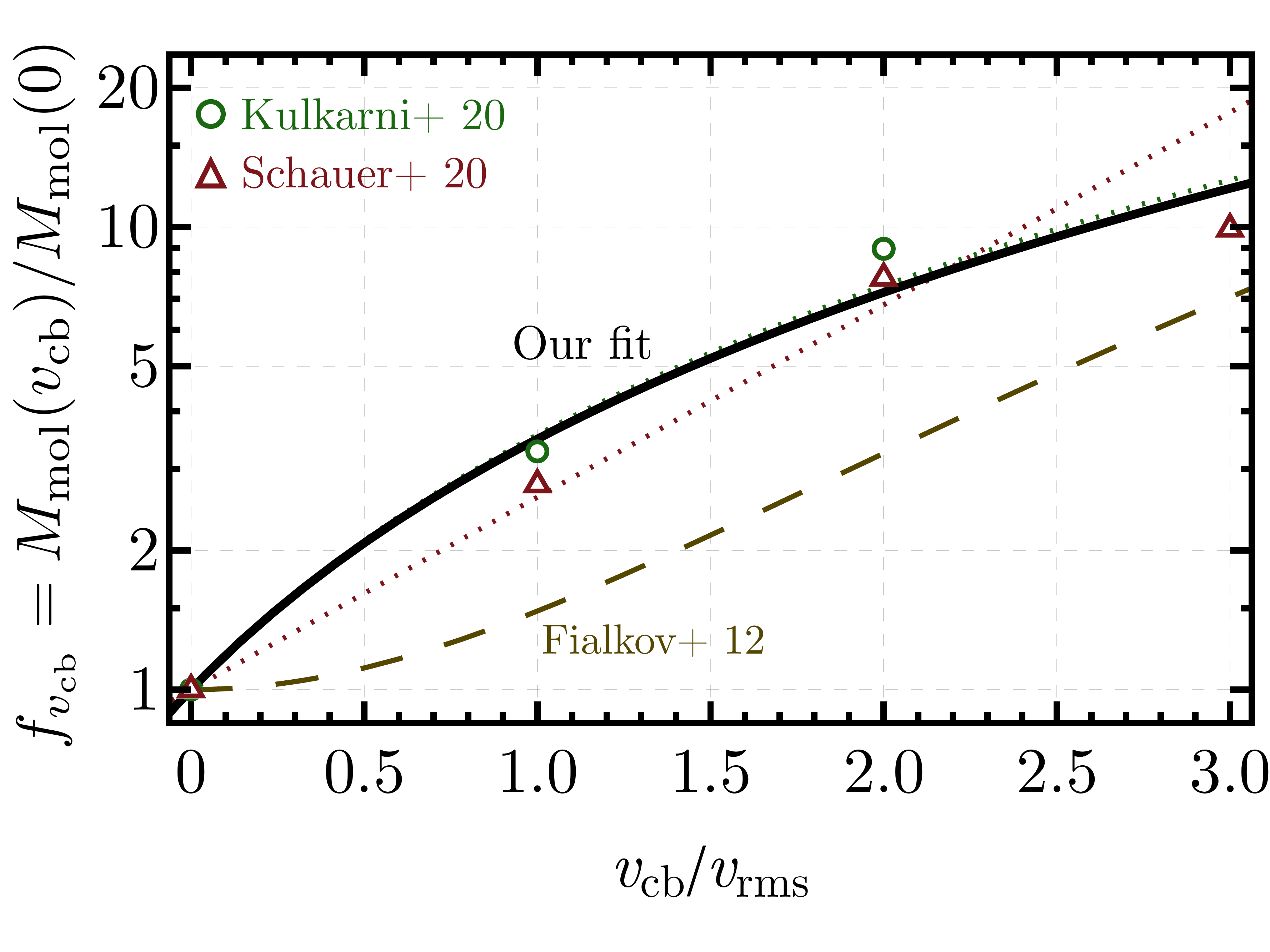}
	\caption{
	Same as Fig.~\ref{fig:fLW_fit} but for the DM-baryon relative velocity $v_{\rm cb}$ (here divided by its rms value $v_{\rm rms}$ so it is dimensionless).
	Our fit, in solid black, follows Eq.~\eqref{eq:fvcb}, and the red and green lines show the fits from each of the two references.
	The brown dashed line shows the formula from \citet{Fialkov:2011iw}, which underpredicts the star formation supression from relative velocities.
	}	
	\label{fig:fvcb_fit}
\end{figure}

The small-scale ($\sim$ Mpc) hydro simulations of \citet{Schauer:2020gvx} and \citet{Kulkarni:2020ovu} were the first to investigate the impact of streaming velocities together with LW feedback.
Both groups account for the relative velocities in a similar fashion, and solve for the evolution of gas inside MCGs to find if the conditions for star formation are met.
Both works find that the streaming motions impede star formation in the smallest galaxies, raising the minimum mass $M_{\rm mol}$ required for star formation (in the absence of photo-heating feedback).
Analogously to $f_{\rm LW}$, we define a feedback factor also for $\vcb$:
\begin{equation}
    f_{\vcb} = \dfrac{M_{\rm mol}(\vcb)}{M_{\rm mol}(0)}.
\end{equation}
We calculate this factor for the different simulation outputs of the two groups, and plot it in Fig.~\ref{fig:fvcb_fit}, which shows the excellent agreement between the two simulation suites.

We fit $f_{\vcb}$ as a $z$-independent function, following the functional form in \citet{Kulkarni:2020ovu}:
\begin{equation}
f_\vcb = \left(1 + A_\vcb \dfrac{\vcb}{\vrms}\right)^{\beta_\vcb},
\label{eq:fvcb}
\end{equation}
where $\vrms \approx 30$ km s$^{-1}$ is the rms velocity, and we use $A_\vcb=1$, and $\beta_\vcb=1.8$, very similar to the values in \citet{Kulkarni:2020ovu}.
We compare this fit with the previous formula from \citet{Fialkov:2011iw}, where it was assumed that the halo virial velocity at the supression scale follows:
\be
V_{\rm mol} = \left[ V_0^2 + \alpha_{\rm cb}^2 \vcb^2(z)  \right]^{1/2},
\ee
where $V_0=4$ km s$^{-1}$ is the virial velocity of molecular-cooling haloes in the absence of any feedback, and they find $\alpha_{\rm cb}=4$.
This translates into a scaling $f_\vcb = (
V_{\rm mol}(\vcb) /V_0)^{3}$, shown in Fig.~\ref{fig:fvcb_fit}, which underpredicts the impact of the relative velocities by roughly a factor of 2.

\begin{figure}
	\includegraphics[width=0.48\textwidth]{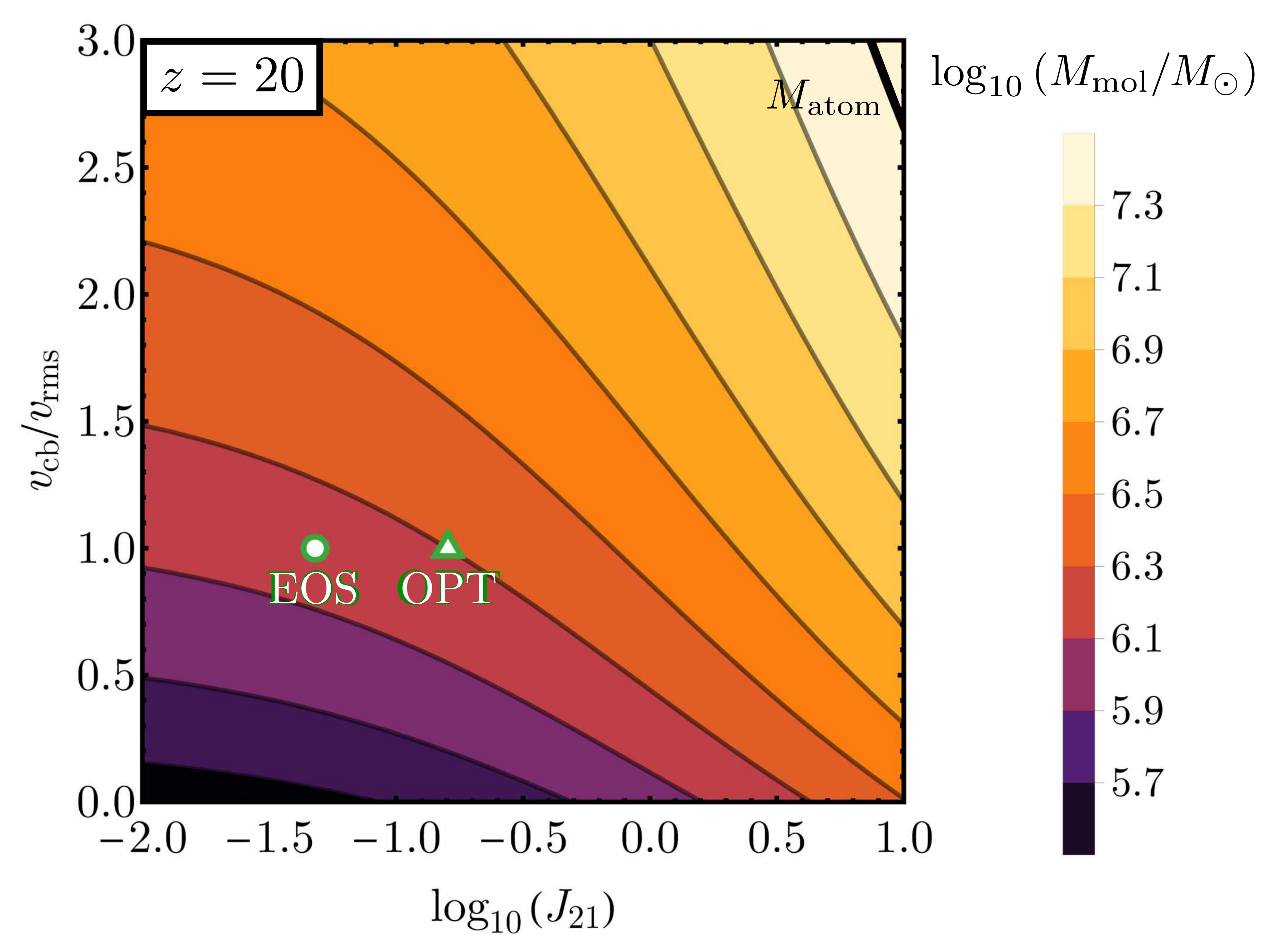}
	\caption{
	Value of the minimum-mass $M_{\rm mol}$ for a molecular-cooling halo to host PopIII stellar formation at $z=20$, as a function of the LW flux ($J_{21}$ in the customary units of $10^{-21}$ erg s$^{-1}$ Hz$^{-1}$ cm$^{-2}$ sr$^{-1}$) and the streaming velocity ($\vcb$ divided by its rms value $v_{\rm rms}$), following Eq.~\eqref{eq:Mturnfit}.
	We assume negligible photo-ionization feedback at this redshift.
	Marked green points represent the expected fluxes $J_{21}$ for our two fiducial simulation sets presented in Tab.~\ref{tab:Fids}.
	The black solid line shows $M_{\rm atom}$, above where there is no PopIII stellar formation (see Sec.~\ref{subsec:MCGs}).
	}	
	\label{fig:joint_feedback}
\end{figure}

With a prescription for both sources of feedback, we can evaluate $M_{\rm mol}$ in Eq.~\eqref{eq:Mturnfit}.
We show this turnover mass at $z=20$ in Fig.~\ref{fig:joint_feedback} for different values of $\vcb$ (in units of its rms value $v_{\rm rms}$, which is close to its median) and $J_{21}$ (we mark the values expected at $z=20$ for two of our fiducial parameter sets: EOS and OPT, see Tab.~\ref{tab:Fids}).
We predict $M_{\rm mol}\approx 10^6\, \Msun$ at $z=20$, though clearly this quantity depends strongly on $\vcb$.
This dependence will imprint $M_{\rm mol}$---and thus the SFRD---with the fluctuations of $\vcb$, giving rise to velocity-induced acoustic oscillations, which we study in Sec.~\ref{sec:VAOs}.

We find that $M_{\rm mol}$ depends less strongly on $J_{21}$.
In particular, for $J_{21}<10^{-1}$ the effect of LW feedback is rather weak, and $\vcb$ feedback dominates.
The situation is reversed for $J_{21}>10^{0.5}$, though we caution that the two feedback schemes may not add coherently in such a high LW flux regime ~\citep{Kulkarni:2020ovu}.
Luckily, this regime does not have a large impact on observables, as it produces $M_{\rm mol} \approx M_{\rm atom}$, so PopIII star formation in MCGs would be subdominant compared to PopII in ACGs.
Moreover, as we will see, values of $J_{21}>10^{0.5}$ only appear at very late times ($z\lesssim 10$) in our simulations, where PopII star formation far dominates.

\subsection{Comparison: PopII versus PopIII}

Armed with our two stellar populations (Pop II residing in ACGs, and PopIII in MCGs), and the feedback processes that can affect them (LW and relative velocities for MCGs, and stellar and photo-heating feedback for both), 
we can now compare their relative contribution to  different cosmic epochs.

We choose a set of fiducial parameters, which we dub EOS (later in Secs.~\ref{sec:astroparams} and~\ref{sec:VAOs} we will study parameter variations),
where the stellar fractions of ACGs and MCGs are set to
\begin{align}
   \log_{10} f_{*,10}^{\rm (II)} &= -1.25   \nonumber\\
   \log_{10} f_{*,7}^{\rm (III)} &= -2.5, \nonumber
\end{align}
with SHMR power-law indices given by
\begin{align}
   \alpha_{*}^{\rm (II)} &= 0.5 \nonumber\\
   \alpha_{*}^{\rm (III)} &= 0. \nonumber
\end{align}
The ACG parameters are very similar to the maximum a posteriori (MAP) from \citet{Qin:2021gkn}, which reproduces observations of the UVLFs, the cosmic-microwave background (CMB) optical depth, and the high-$z$ Lyman-$\alpha$ forest opacity fluctuations.
Our choice of $\log_{10} f_{*,7}^{\rm (III)} = -2.5$ is rather conservative (though slightly larger than found in the simulations of \citealt{Skinner_Wise_2020_selfshielding}), but it still allows PopIII stars to dominate the SFRD at early-enough times.
Given that the PopIII parameters are entirely unknown, we will focus on varying the latter in this work, though we note that the {\tt 21cmMC}~\citep{Greig:2015qca}~\footnote{\url{https://github.com/21cmfast/21CMMC}} and {\tt 21cmFish}~\citep{Mason_21cmFisher}~\footnote{\url{https://github.com/charlottenosam/21cmfish}} packages allow the user to vary all parameters simultaneously.

\begin{figure}
	\includegraphics[width=0.5\textwidth]{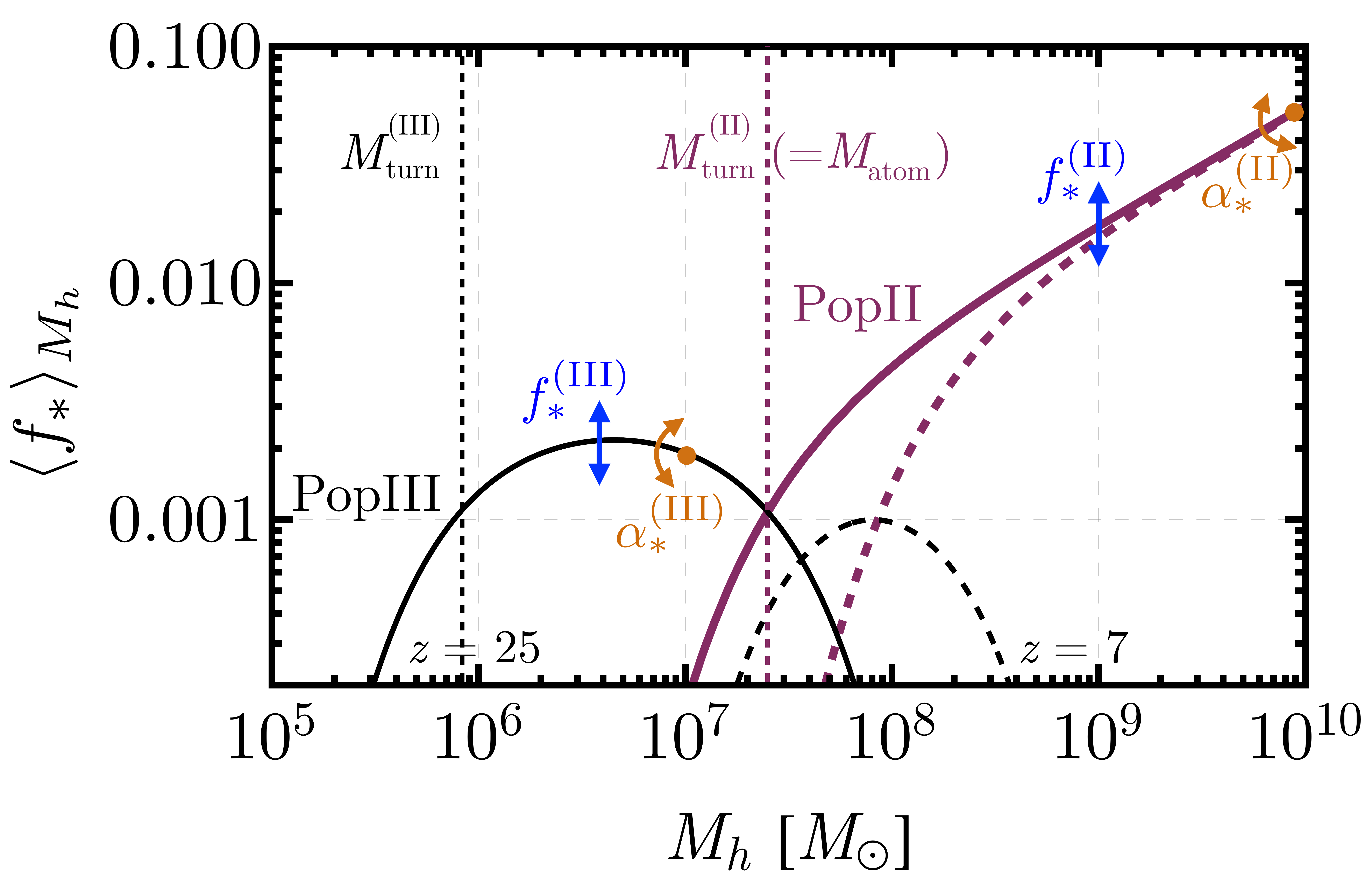}
	\caption{The fraction of galactic mass in stars, averaged over all halos of a given mass, $\VEV{f_*}_{M_h}$, for our fiducial (EOS2021) simulation. 
	In thick purple lines we show our model for PopII-forming ACGs and in thin black lines for PopIII-forming MCGs.
	The solid lines show this SHMR $\VEV{f_*}_{M_h}$ during cosmic dawn ($z=25$), whereas dashed are at the EoR ($z=7$).
	Vertical dotted lines represent the turn-over mass to form stars for MCGs (black, including feedback) and for ACGs (purple), both at $z=25$.
	Both SHMRs are parametrized via power-laws with indices $\alpha_*^i$, which vary the tilt as indicated by the orange arrows, and amplitudes $f_*^i$, which scale them up and down as per the blue arrows.
	This relation is expected to turn over at much higher masses ($M_h\gtrsim 10^{11}-10^{12}\,\Msun$), which we do not consider as such objects are too rare to appreciably contribute to cosmic radiation fields during these eras. 
	}	
	\label{fig:fstar}
\end{figure}

We begin by showing the SHMR ($f_*=M_*/M_h$) of PopII- and PopIII-hosting galaxies in Fig.~\ref{fig:fstar} at $z=25$ and 7.
The impact of the four free parameters is as follows: each index $\alpha_*^i$ changes the slope of the corresponding SHMR, whereas the $f_*^i$ factors re-scale them up and down.
Unlike at lower redshifts (see e.g.~\citealt{Behroozi:2012iw,Wechsler:2018pic,Tacchella:2018qny,Mason:2015cna,Yung2019_UVLF, Sabti:2021xvh,Sabti:2021unj,Sabti:2020ser, Rudakovskyi:2021jyf}), for the EoR/CD we are only interested in very low-mass haloes ($M_h\lesssim10^{11}\,\Msun$), as those dominate the photon budget compared to the brighter galaxies, which are rare at the redshifts of interest.
Therefore, we can ignore the turnover in the SHMR at $M_h\gtrsim 10^{11-12}\,\Msun$ commonly attributed to AGN feedback~\citep{Qin:2017lfx}, and focus on the fainter end of the ACGs that is well characterized by a single power-law.
Despite their relatively low $f_*^{\rm (III)}$, MCGs are abundant enough to dominate the SFRD at early times ($z\gtrsim 15$), as we quantify below.
In our model, MCGs only form stars for a narrow band of halo masses, which varies with $z$ depending on the dominant feedback process.
As a consequence of this narrow mass range, the main parameter that controls PopIII star formation in MCGs is $f_*^{\rm(III)}$, rather than the power-law index $\alpha_*^{\rm(III)}$.

At later times ($z=7$), feedback severely suppresses MCG star formation.  Furthermore, the evolution of the HMF means that their relative abundance (compared to ACGs) decreases.  As a result of these two effects, the contribution of MCGs to cosmic radiation fields becomes subdominant by $z\sim7$ in our fiducial model.

\begin{figure}
	\includegraphics[width=0.48\textwidth]{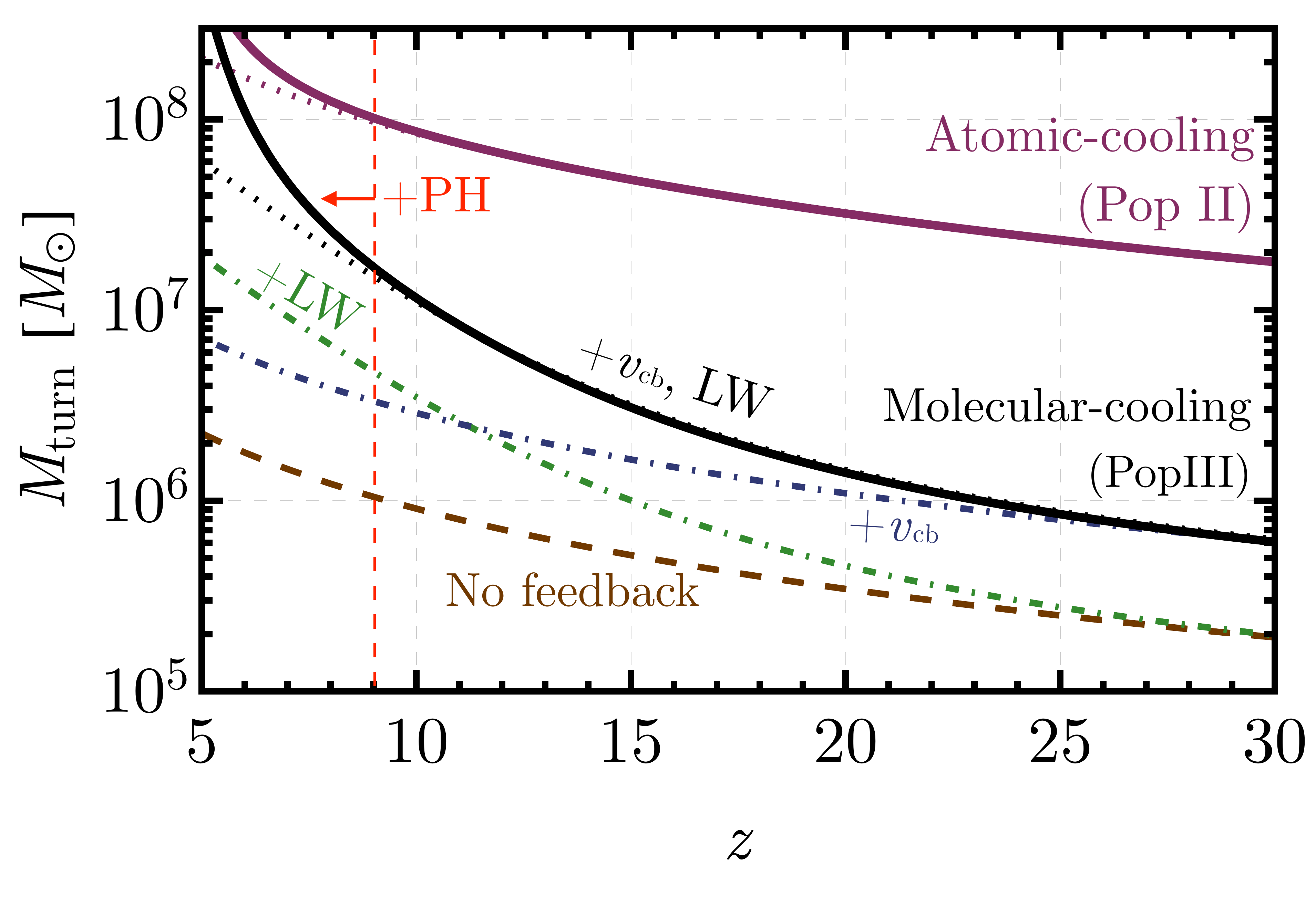}
	\caption{
	We show different turnover halo mass scales, which enter the SFRDs of PopII stars in Eq.~\eqref{eq:SFRDII} and PopIII stars in Eq.~\eqref{eq:SFRDIII}.
	Haloes above the purple line can efficiently cool gas through atomic-line transitions, and we assume they host \textcolor{purple}{\bf ACGs} forming PopII stars.
	The dotted purple line is the prediction from $T_{\rm vir}=10^4$ K, whereas the solid line is the simulation result that includes photoheating (\textcolor{red}{PH}) feedback (which is efficient during reionization, to the left of the red line).
	The dashed brown line represents the theoretical limit above which haloes hosting \textcolor{brown}{\bf MCGs} can form PopIII stars (through molecular cooling) in the absence of feedback.
	To illustrate the strength of different sources of feedback, we include them one at a time.
	First, in green we add only Lyman-Werner (\textcolor{ForestGreen}{LW}), and in blue only the streaming velocities ($\color{blue} \vcb$).
	The relative velocities have a bigger impact at high $z$, whereas LW feedback dominates at later times.
	We show their product as the black-dotted line, which represents their total feedback from Eq.~\eqref{eq:Mturnfit}.
	The solid black line is the result from the simulation, which again rises sharply during reionization due to photoheating feedback.
	}	
	\label{fig:Mturns}
\end{figure}

In order to understand how feedback evolves over time, we show the different characteristic mass scales in Fig.~\ref{fig:Mturns}.
Going from lowest to highest, we first show the mass for molecular cooling of gas in the absence of feedback, $M_0(z)$ in Eq.~\eqref{eq:Mturnfit}, which grows simply as $(1+z)^{-3/2}$.
We then include the impact of relative velocities (through $f_{\vcb}$), and LW feedback (through $f_{\rm LW}$, with the $J_{21}$ flux self-consistently and inhomogeneously computed in our simulation box) individually, as well as jointly.

The impact of $\vcb$ is notable at high redshifts, increasing the turnover mass by nearly an order of magnitude, to $M_{\rm turn}^{(\rm III)}\approx 10^6\,\Msun$ at $z=25$. 
At lower redshifts ($z\lesssim 12$ for our fiducial parameters), LW radiation dominates over $\vcb$ in setting the MCGs turnover mass.
At even later times (below $z\sim 8$), photoheating feedback from reionization steeply increases $M_{\rm turn}^{(\rm III)}$, so by the end of our simulations at $z=5$ there is essentially no star formation in MCGs. 
We remind the reader that in our model PopII stars form above the atomic-cooling threshold, also shown in Fig.~\ref{fig:Mturns}, and those are also affected by photoheating feedback.

\begin{figure}
	\includegraphics[width=0.48\textwidth]{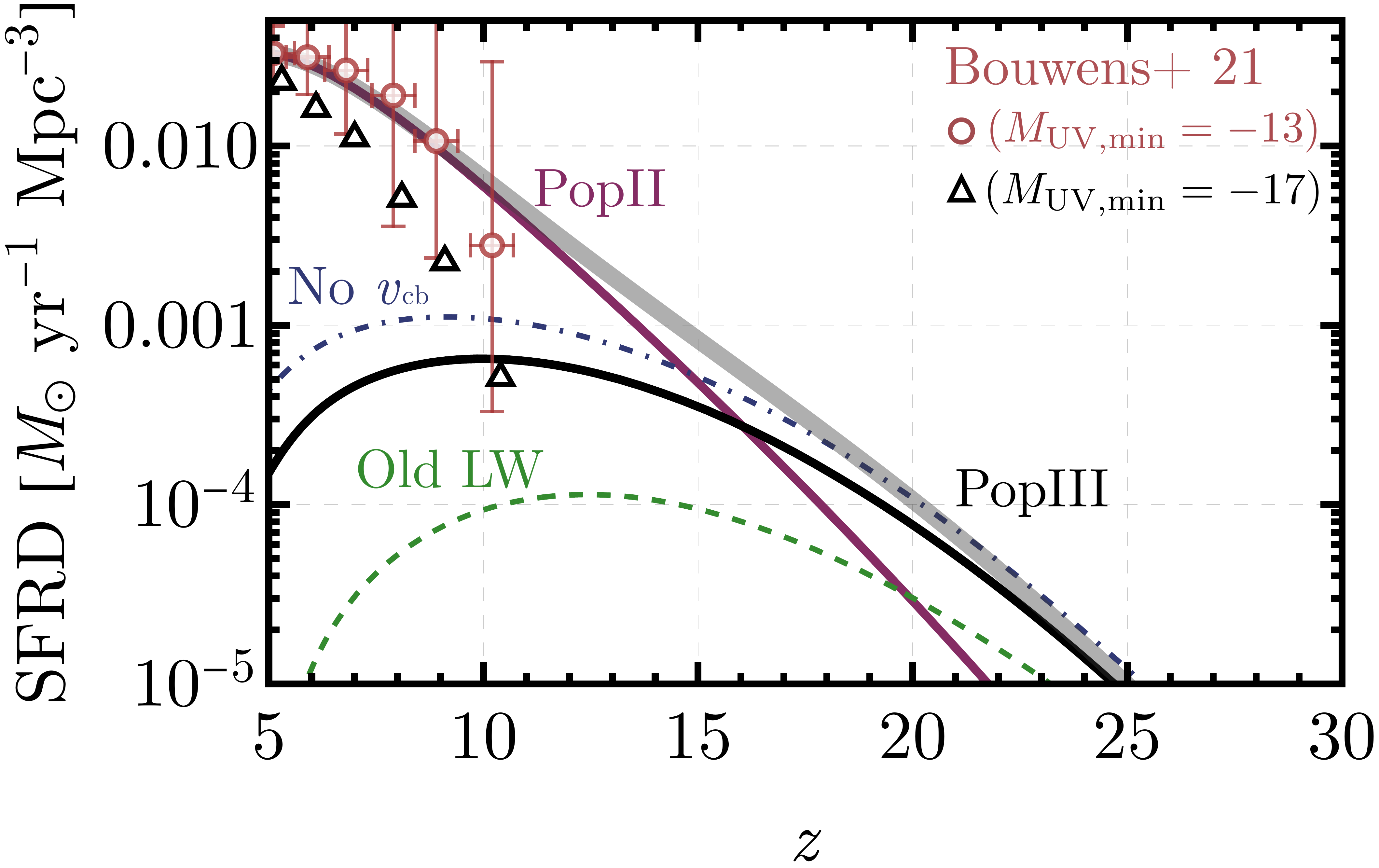}
	\caption{Star-formation rate density (SFRD) as a function of $z$ in an average-density region of the universe.
	The purple and black lines show our prediction for PopII and PopIII stars hosted in ACGs and MCGs, respectively, and the gray thick line is their sum.
	The green-dashed line shows the result for PopIII stars if the \citet{Machacek:2000us} LW feedback prescription was utilized (given the same LW flux), whereas the blue dash-dotted line represents the PopIII result in the absence of relative velocities.
	The red data-points are the result of extrapolating the \citet{Bouwens2021} UVLFs to $M_{\rm UV,min}=-13$, corresponding to $M_h\approx 10 \, M_{\rm atom}$ (or $M_h=10^{9-10} \, \Msun$) in our model. 
	We also show, as black triangles, the result when only integrating down to $M_{\rm UV,min}=-17$, where we have displaced each triangle by $\Delta z=+0.3$ for better visualization (as well as the lowest-$z$ circle for the same reason).
	}	
	\label{fig:SFRD}
\end{figure}

Given our fiducial choices, we can compute the SFRD for both PopII and PopIII on a representative (i.e., average-density) patch of the universe~\citep{Madau:1996yh}.
We do so in Fig.~\ref{fig:SFRD}, where we show the individual PopII and PopIII contributions, as well as their sum.
As is clear from this figure, PopIII star formation in MCGs dominates over ACGs at higher redshifts, and for our fiducial parameters this transition takes place at $z\approx 15$.
We also show two alternative scenarios of PopIII star formation to illustrate the impact of feedback.
In one, we turn off the relative-velocity ($\vcb$) effect,
and as a result we would overestimate the SFRD of PopIII stars by $50\%$.
In the other, we use the previous feedback formula from \citet[][corresponding to $A_{\rm LW}=22$ in Eq.~\eqref{eq:fLW}]{Machacek:2000us}, which did not include self-shielding.
In that case the SFRD is reduced by roughly an order of magnitude, with the discrepancy increasing at lower $z$, where $J_{21}$ is larger (though we note that we use the same $J_{21}$ flux as in the fiducial simulation).

Additionally, we compare our predicted SFRDs with data, obtained by extrapolating the recent measurements from \citet{Bouwens2021}.
We convert our SFRD to UV luminosities using a constant $\kappa=1.15\times 10^{-28} \, \Msun$ s yr$^{-1}$ erg$^{-1}$~\citep{SunFurlanetto_2016,Oesch_2018},
and extrapolate the Schechter fit from \citet[][with errors inherited from the uncertainty in the Schechter parameters]{Bouwens2021} to a minimum UV magnitude $M_{\rm UV,min}=-13$~\citep{Park:2018ljd}, which corresponds to haloes with $M_h=10^{9-10}\,\Msun$ for the redshifts of interest.
As expected, ACGs fit all the current  ($z\lesssim10$) data well on their own, as observations cannot reach the fainter MCGs (see for instance~\citealt{Sun:2021drn} for a SPHEREx forecast).

\subsection{Cosmic radiation fields}

Using the (inhomogeous) ACG and MCG SFRDs from Eqs.~(\ref{eq:SFRDII},\ref{eq:SFRDIII}), we compute the corresponding  cosmic radiation fields that are relevant for the thermal and ionization evolution of the IGM: soft UV (LW and Lyman series), ionizing and X-ray.  {\tt 21cmFAST} calculates the radiation field incident on each simulation cell through a combination of excursion-set photon-counting (for ionizing photons) and lightcone integration (for soft UV and X-rays), accounting also for IGM attenuation/absorption.  These procedures are described in detail in, e.g., \citet{Mesinger:2010ne, Mesinger:2012ys, Sobacchi:2014rua}, and we encourage the interested reader to consult these works for further details.  Here we only summarize the free parameters that are the most relevant for our analysis.

We assume the emissivity for all of the above radiation fields scales with the SFR (Eqs.~\ref{eq:SFRDII} and \ref{eq:SFRDIII}).  To calculate UV emission, we take PopII/PopIII SEDs from \citet{Barkana2005ApJ...626....1B}, normalized to have $N_{\gamma/b}^{(i)}=5000$ and $44000$ ionizing photons per stellar baryon for PopII and III, respectively.  We assume only a fraction $f_{\rm esc}$ of ionizing photons that are produced manage to escape the host galaxy and ionize the IGM.  We take a power-law relation for the typical $f_{\rm esc}$ as a function of halo mass (\citealt{Park:2018ljd}):
\be
\fesc^{(i)} = f_{ {\rm esc},m_i}^{(i)} \left(\dfrac{M_h}{M_i}\right)^{\alpha_{\rm esc}^{(i)}}
\ee
for both stellar populations ($i=$\{II, III\}), with $M_i=\{10^{10}, 10^7\}\Msun$ and $m_i = \log_{10}(M_i)$ as before.\footnote{While there is no consensus on how the ionizing escape fraction depends on galaxy properties, simulations suggest that a generic power law captures the mass behavior of the population-averaged $\fesc$ (e.g., \citealt{PKD15, Kimm17, Lewis20}).}
We choose fiducial parameters in broad accordance with the best fits from the latest observations in \citet{Qin:2021gkn}, setting
\begin{align}
   \log_{10} f_{\rm esc,10}^{\rm (II)} &= \log_{10} f_{\rm esc,7}^{\rm (III)} = -1.35, \nonumber
\end{align}
so the escape fractions from MCGs and ACGs are comparable at their pivot points.
We have lowered the ACG escape fraction normalization ($f_{\rm esc,10}^{\rm (II)}$) by $-0.15$ with respect to \citet{Qin:2021gkn}, in order to allow for an increased (though overall small) MCG contribution to reionization.
We assume the same scaling with mass
\be
\alpha_{\rm esc}^{(\rm II)}=\alpha_{\rm esc}^{(\rm III)}=-0.3 \nonumber
\ee
for both  populations, as these agree with Lyman-$\alpha$ forest + CMB data~\citep{Qin:2021gkn}.  Under these assumptions, we find that ACG-hosted PopII stars dominate the ionizing photon budget at $z\leq 15$.
This is to be expected, as MCGs are subdominant at lower $z$, and we agnostically set a (relatively) low value of $\fesc^{(\rm III)}$ for PopIII stars.

To calculate X-ray emission we assume a power-law SED with a spectral energy index, $\alpha_X$, and a low-energy cutoff $E_0$.  X-ray photons with energies below $E_0$ are absorbed within the host galaxies and do not contribute to ionizing and heating the IGM.  This was shown to be an excellent characterization of the X-ray SED from either the hot interstellar medium (ISM) or high-mass X-ray binaries (HMXBs), when emerging from simulated, metal-poor, high-redshift galaxies \citep{Fragos:2013bfa, Pacucci:2014wwa, Das:2017fys}.  Both $\alpha_X$ and $E_0$ control the hardness of the emerging X-ray spectrum, and thus the patchiness of IGM heating.  Here we fix $\alpha_X=1$, and vary $E_0$.  For our fiducial value, we choose $E_0=0.5$ keV, based on the ISM simulations of \citet{Das:2017fys}, though we also explore a more optimistic model with a softer SED.
The normalization of the X-ray SED is determined by the soft-band (with energies less than 2 keV)\footnote{Only the soft-band X-ray emission efficiently heats the IGM, as X-rays with $E\gtrsim$ 1.5--2 keV have mean free paths longer than the size of the universe at the redshifts of interest.} X-ray luminosity to SFR parameter, $L_{X,\rm <2keV}/{\rm SFR}$.  Consistent with simulations of HMXBs in metal-poor environments \citep{Fragos:2013bfa}, here we take  $\log_{10}( L_{X,\rm <2keV}/{\rm SFR} )\approx 40.5$ erg s$^{-1}$ per unit SFR ($\Msun\,$ yr$^{-1}$) for both ACGs and MCGs.  Such high values are further supported by recent 21-cm power spectrum upper limits at $z=8$ from HERA~\cite{HERA:2021noe}.

\section{Observables during the CD and EoR}
\label{sec:CDEoR}

We now use the models for ACGs and MCGs outlined above to predict the evolution of the thermal and ionization state of the IGM at high redshifts.
We focus on the contribution of PopIII star-forming MCGs, as their corresponding feedback is the main improvement of this work.

We use two sets of galaxy parameters, {\it Fiducial (EOS)} and {\it Optimistic (OPT)}, summarized in Table~\ref{tab:Fids}.  The fiducial (EOS) uses the same ACG parameters as in  \citet{Qin:2021gkn}, where such a model was shown to reproduce EoR observables (at redshifts $z\lesssim10$ when the MCG contribution is negligible).  However, we do lower the escape fraction of ACGs slightly ($\Delta \log f_{\rm esc,10}^{\rm (II)} = -0.15$), so as to allow for a modest contribution of MCGs to the high-redshift tail of the EoR (as could be slightly preferred by the CMB EE PS at $l\sim20$--30; e.g., \citealt{Qin:2020xrg,Wu:2021kch,Ahn:2020btj}).

Our Optimistic (OPT) parameter set was chosen in order to enhance the relative contribution of MCGs to the SFRD.  Although the actual parameter values are fairly arbitrary (given the lack of MCG constraints), we tuned the OPT model to reproduce the timing of the putative EDGES global 21-cm detection at $z\sim17$~\citep{Bowman:2018yin}.  This is mainly done by increasing the SFRD of minihalos (through $f_{\ast, 7}^{\rm (III)}$) as well as the softness of the X-ray SED emerging from galaxies (through $E_0$).

In this section we introduce the main high-redshift observables using our Fiducial (EOS) model.  
We will show the impact of PopIII stars and present the 2021 installment of the Evolution Of 21-cm Structure (EOS) project, whose goal is to show the state of knowledge of the astrophysics of cosmic dawn and reionization.
Later in Secs.~\ref{sec:astroparams} and \ref{sec:VAOs}, when we study parameter variations and VAOs, we will mainly focus on the optimistic (OPT) model.

\begin{table}
\centering
\begin{tabular}{l|cc}
Parameter                                                                                       & Fiducial (EOS2021) & Optimistic (OPT) \\ \hline
$\log_{10} f_{*,10}^{\rm (II)}$                                                                & $-1.25$            & $-1.50$         \\
$\log_{10} f_{*,7}^{\rm (III)}$                                                                & $-2.5$         &     $-1.75 $       \\
$\alpha_*^{\rm (II)}$                                                                          & 0.5             & 0.5           \\
$\alpha_*^{\rm (III)}$                                                                         & 0               & 0             \\ \hline
$\log_{10} f_{\rm esc,10}^{\rm (II)}$                                                          & $-1.35 $          &  $-1.20 $        \\
$\log_{10} f_{\rm esc,7}^{\rm (III)}$                                                          & $-1.35 $          &   $-2.25 $      \\
$\alpha_{\rm esc}^{\rm (II)}=\alpha_{\rm esc}^{\rm (III)}$ & $-0.3$            & $-0.3 $        \\ \hline
$L_X^{(\rm II)}=L_X^{(\rm III)}$                                                    & 40.5            & 40.5          \\
$E_0$ [keV]                                                                                & 0.5             & 0.2          
\end{tabular}
\caption{Summary of our choices for some of the main free parameters related to the SFRD (top part), ionizations (middle), and X-ray emission (bottom).
The EOS 2021 (EOS) fiducial model is used throughout the text, except in the last two sections (\ref{sec:astroparams} and \ref{sec:VAOs}) where we use OPT.
Other parameters are fixed to the values motivated in the text.
}
\label{tab:Fids}
\end{table}

\subsection{UV Luminosity Functions}

The first observable we show are UVLFs.
Although limited to comparably brighter galaxies during the EoR/CD, UVLFs detected with the HST provided invaluable insights into galaxy formation and evolution at $z \leq 10$.

We follow~\citet{Park:2018ljd}, where the 1500 \AA\, luminosity is obtained from the SFR with a conversion factor of $\kappa=1.15\times 10^{-28} \, \Msun$ s yr$^{-1}$ erg$^{-1}$~\citep{SunFurlanetto_2016,Oesch_2018}, and for simplicity we take $\kappa$ to be the same for both PopII and PopIII populations.  Detailed population-synthesis models suggest that there could be a factor of $\sim$ 2 variation in this conversion, based on the IMF, metalicity and star formation history (e.g., \citealt{Wilkins19}).

\begin{figure}
	\includegraphics[width=0.48\textwidth]{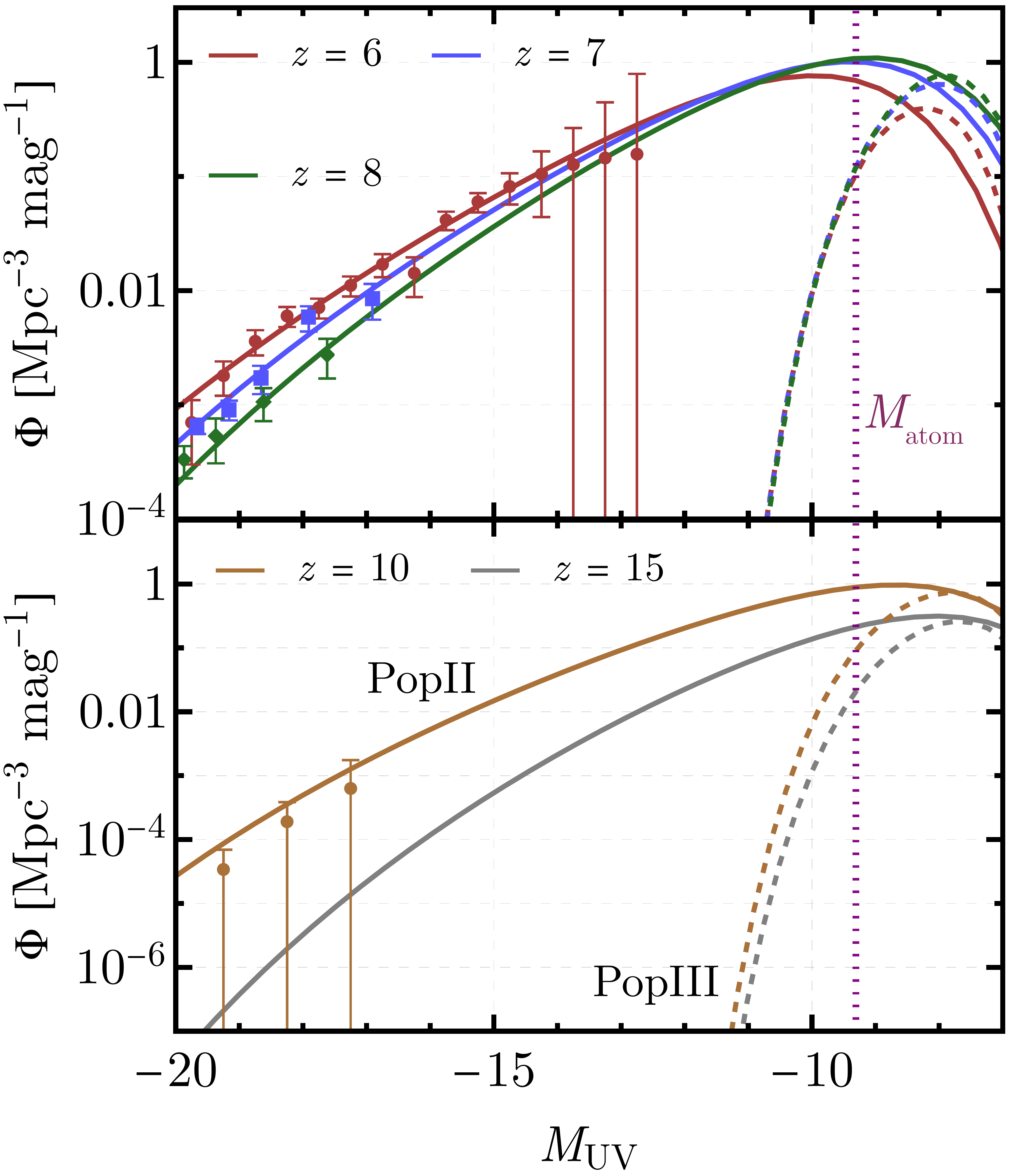}
	\caption{UVLFs at five redshifts for our EOS fiducial parameters, where the solid and dashed lines represent the ACG and MCG contribution, respectively.
	The data points are from \citet{Bouwens:2014fua} \citet{Bouwens2016} and \citet{Oesch_2018}, where only galaxies with $M_{\rm UV}\geq-20$ are shown as these dominate the cosmic radiation backgrounds at the high redshifts of interest. 
	The top panel shows $z=6-8$, where the data covers a broader magnitude range, whereas the bottom panel focuses on earlier epochs, where only some data are available at $z=10$.
	The vertical dotted line shows the magnitude expected of a halo with $M_h=M_{\rm atom}$ at $z=6$, illustrating that in our model MCGs are far too faint to be directly detectable through the UVLF.
	Nevertheless, these data allow us to constrain the SHMR of ACGs, and thus isolate the contribution of MCGs to cosmic radiation fields using 21-cm observations.
	}	
	\label{fig:UVLF}
\end{figure}

We show the predicted UVLFs for our fiducial (EOS2021) parameters in Fig.~\ref{fig:UVLF}, which matches very well the observational data from \citet{Bouwens:2014fua} \citet{Bouwens2016} and \citet{Oesch_2018}.
This is by construction, as our parameters are motivated by the maximum a posteriori (MAP) model  from \citet{Qin:2021gkn}, which included UVLFs (in addition to other EoR observables) in the likelihood.

We note that the atomic-cooling threshold $M_{\rm atom}=2\times 10^8\,\Msun$ at $z=6$ corresponds to $M_{\rm UV}\approx -9$ in our model, making it difficult to directly observe the even fainter MCGs.
This is clear in our Fig.~\ref{fig:UVLF}, where MCGs only dominate at fainter magnitudes, beyond the reach of even {\it JWST}. 
We note that previous work (e.g., \citealt{Oshea15, Xu16galaxies, Qin:2020pdx}), found a turnover for MCGs at even fainter magnitudes ($M_{\rm UV}\gtrsim-7$ -- $-6$).  This difference is due to our feedback prescriptions, including relative velocities that dominate at early times (c.f.~Fig.~\ref{fig:Mturns}). As a consequence, the UVLF for MCGs is always below (or comparable to) that of ACGs in Fig.~\ref{fig:UVLF}.
We stress that even though we are unlikely to observe such ultra faint magnitudes, UVLFs provide an invaluable dataset by allowing us to anchor the SHMR scaling relations at the brighter end that is well probed by observations ($-20 \lesssim M_{\rm UV} \lesssim -15$).

Through the rest of this section we will study how the inclusion of PopIII-hosting MCGs---and the $\vcb$ feedback on them---affects reionization and the 21-cm signal.
We will do so by comparing our best-guess EOS fiducial simulation to one without MCGs, as well as one with both ACGs and MCGs but no relative velocities (similar to \citet{Qin:2020xyh} though with an updated LW feedback prescription).

\subsection{EoR history}

\begin{figure}
	\includegraphics[width=0.48\textwidth]{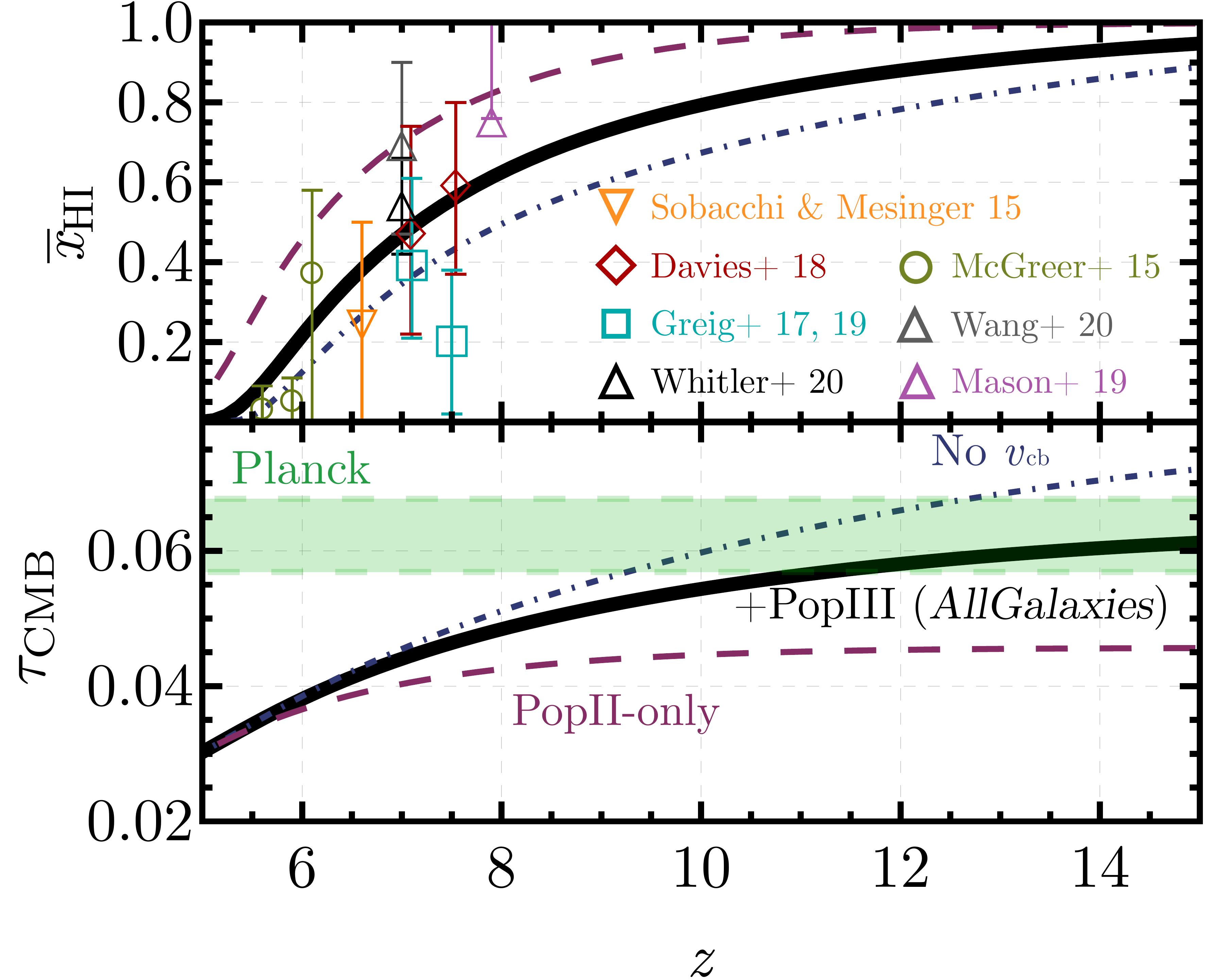}
	\caption{Neutral-hydrogen fraction ({\bf top}) and optical depth to reionization ({\bf bottom}) as a function of redshift $z$.
	The black thick line shows our fiducial (EOS2021) model, including both PopII and PopIII contributions, whereas the purple line is the result from a PopII-only simulation.
    The blue dot-dashed curve shows the result of a simulation with no $\vcb$ feedback, which produces  larger stellar formation at earlier times, and would be above the 1$\sigma$ measurement of $\tau_{\rm CMB}$ with {\it Planck} data~\citep{deBelsunce:2021mec}, shown as the green band.
	}	
	\label{fig:xHItau}
\end{figure}

We show the evolution of the EoR from our Fiducial model (EOS2021) in Fig.~\ref{fig:xHItau}.  We plot both $\bar{x}_{\rm HI}$ and the optical depth $\tau_{\rm CMB}$ of the CMB due to reionization, as a function of $z$.
Given our fiducial parameters, MCGs only make a small contribution to cosmic reionization, and chiefly at high $z$. 
The overall evolution of $\xHI$ agrees broadly with current measurements from \citet{McGreer:2014qwa,Greig:2016vpu,Greig:2018rts,Mason:2019ixe,Whitler:2019nul,Wang:2020zae} and {\it Planck} 2018 temperature and polarization (reanalyzed in~\citealt{deBelsunce:2021mec}).
This is mostly by design, as our ACG (PopII) parameters were chosen to be consistent with \citet{Qin:2021gkn}, who used various EoR observables to constrain the EoR history.  In particular, the final overlap stages of reionization are constrained by observations of the Lyman-$\alpha$ opacity fluctuations.  \citet{Qin:2021gkn} found that the latest forest spectra require a late end to reionization at $z\sim 5.5$~(see also e.g., \citealt{Kulkarni:2018erh,Nasir:2019iop,Keating:2019qda, Choudhury:2020vzu}).

The contribution of MCGs would however be largest during the earliest stages of the EoR.  In our fiducial model PopIII-hosting MCGs drive a modest increase of the CMB optical depth: $\Delta \tau_{\rm CMB}\approx 0.015$ (though the precise value is sensitive to our fairly arbitrarily chosen MCG parameters).  As seen from the figure, this contribution is mostly sourced at $z\sim$ 8 -- 12, with the very high-$z$ tail only contributing $\tau(z=15-30)=2\times 10^{-3}$, well below the limit of 0.02 from {\it Planck} 2018~\citep{Millea:2018bko,Heinrich:2021ufa}.
While further data from CMB experiments will more precisely pin-point $\tau_{\rm CMB}$~\citep{Abazajian:2016yjj,SimonsObservatory:2018koc}, it will be difficult to isolate the high-$z$ contribution that could be caused by MCGs~\citep{Wu:2021kch}.
We note that ignoring $\vcb$ feedback roughly doubles the contribution of MCGs to reionization, for our fiducial parameter choices.

\begin{figure}
	\includegraphics[width=0.46\textwidth]{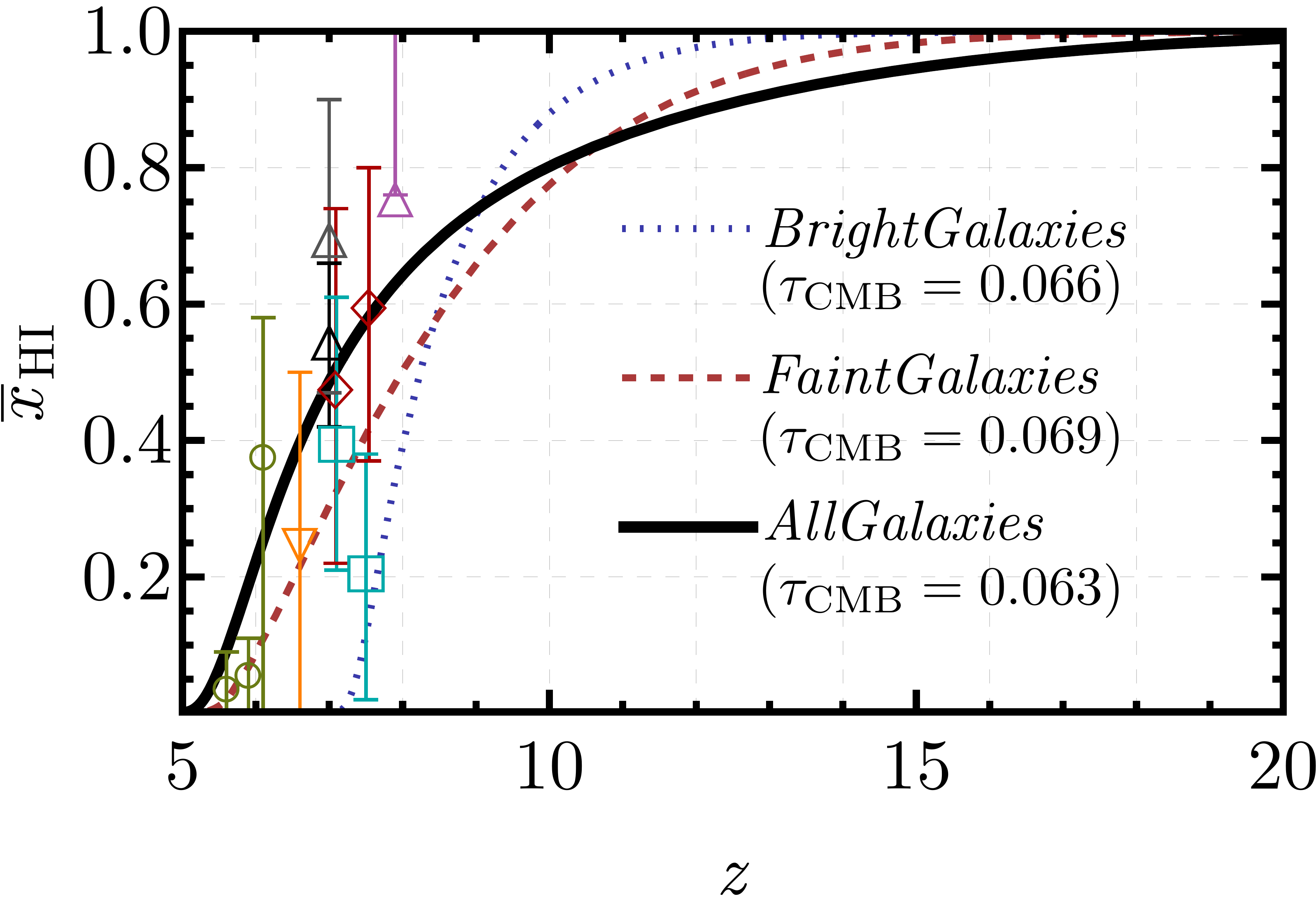}
	\caption{
	Global neutral-hydrogen fraction for our new 2021 EOS ({\it AllGalaxies}) simulation in black, compared against the 2016 EOS simulations of  \citet[][{\it Bright} and {\it FaintGalaxies} in blue and red, respectively]{Mesinger:2016ddl}.
	The $\bar{x}_{\rm HI}$ datapoints are the same as in Fig.~\ref{fig:xHItau}, and we write the integrated optical depth $\tau_{\rm CMB}$ for each simulation.
	}	
	\label{fig:EOSxH}
\end{figure}

In Fig.~\ref{fig:EOSxH} we compare our fiducial, EOS2021 EoR history to that of the EOS2016 release \citep{Mesinger:2016ddl}.  The 2016 release was comprised of two models, {\it FaintGalaxies} and {\it BrightGalaxies}, both of which only included PopII-hosting ACGs but assumed different turnover mass scales for SNe feedback.  In this work we assume SNe feedback does not induce a turnover, and include also PopIII-hosting MCGs with the associated LW and $v_{\rm bc}$ feedback followed self-consistently.  To make this distinction explicit, we denote our EOS2021 model as {\it AllGalaxies} in the figure.

All EOS releases are ``tuned" to reproduce the current state of knowledge.  In 2016, our estimate of $\tau_e$ and forest data suggested an earlier middle/end of reionization, as is reflected in this comparison plot.  The shapes of the EoR histories are also notably different.  By including MCGs, {\it AllGalaxies} (EOS2021) produces a more extended tail to higher redshifts, with percent level ionization up to $z\sim20$.  Despite this earlier start, the ACG-driven mid and late stages of reionization occur {\it more rapidly} in {\it AllGalaxies} than in {\it FaintGalaxies}.  This is because both EOS2016 models assumed a constant mass-to-light ratio ($\alpha_*^{\rm (II)}=0$).  This assumption, albeit common, is inconsistent with the latest UVLF observations and overestimates star formation in small galaxies, thus resulting in a slower evolution of the SFRD and associated cosmic epochs (see also \citealt{Mirocha2016_UVLF_GS,Park:2018ljd}).  This figure highlights the importance of using the latest observations to guide our models of the early universe.

\subsection{The 21-cm line during the Cosmic Dawn}

The biggest impact from MCGs will be to the cosmic-dawn epoch, which we mainly observe through the 21-cm line of neutral hydrogen.
We now explore this observable.

We first show our definitions, and refer the reader to \citet{Furlanetto:2006jb} and \citet{Pritchard:2011xb} for detailed reviews of the physics of the 21-cm line.
We use the full expression for the 21-cm brightness temperature~\citep{Barkana:2000fd}
\be
T_{21} = \dfrac{T_S-\Tcmb}{1+z}\left(1-e^{-\tau_{21}}\right), 
\label{eq:T21def}
\ee
where $\Tcmb$ is the temperature of the CMB (which acts as the radio back light), $T_S$ is the spin temperature of the IGM, and
\be
\tau_{21} = (1+\delta) x_{\rm HI} \dfrac{T_0}{T_S}
\dfrac{H(z)}{\partial_r v_r} (1+z),
\ee
where $\partial_r v_r$ is the line-of-sight gradient of the velocity.
We have defined a normalization factor
\be
T_0 = 34 \, {\rm mK} \left(\dfrac{1+z}{16}\right)^{1/2} \times \left(\dfrac{\Omega_b\,h^2}{0.022} \right)
\left(\dfrac{\Omega_m\,h^2}{0.14} \right)^{-1/2}
\ee 
anchored at our {\it Planck} 2018 cosmology.

The spin temperature of hydrogen, which determines whether 21-cm photons are absorbed from the CMB (if $T_S < \Tcmb$) or emitted (for $T_S > \Tcmb$), is set by competing couplings to the CMB and to the gas kinetic temperature $T_K$, and can be found through
\be
T_S^{-1} = \dfrac{T_{\rm CMB}^{-1}+x_\alpha T_\alpha^{-1} + x_c T_K^{-1}}{1+x_\alpha + x_c},
\ee
where $T_\alpha$ is the color temperature (which is closely related to $T_K$,~\citealt{Hirata:2005mz}), and
$x_i$ are the couplings to $T_K$ due to Lyman-$\alpha$ photons ($x_\alpha$) through the Wouthuysen-Field effect~\citep{Wout, Field} and through collisions ($x_c$, which are only relevant in the IGM at $z\gtrsim 30$,~\citealt{Loeb:2003ya}).

There are two main avenues for measuring the 21-cm signal.
The first is by through its monopole against the CMB, usually termed the {\it global signal} (GS hereafter).
The second is through the fluctuations of the signal, commonly simplified into the Fourier-space two-point function or {\it power spectrum} (PS hereafter).
We now explore each in turn.

\subsubsection{Global Signal}

The 21-cm GS (denoted by $\overline T_{21}$)
is currently targeted by experiments such as EDGES~\citep{Bowman:2018yin}, LEDA~\citep{LEDA}, SARAS~\citep{Singh:2017syr}, Sci-Hi~\citep{Voytek:2013nua}, and Prizm~\citep{PRIZM}.
We note however that the lack of angular information makes the GS especially difficult to disentangle from foregrounds, which can be several orders of magnitude stronger than the cosmic signal.

\begin{figure}
	\includegraphics[width=0.46\textwidth]{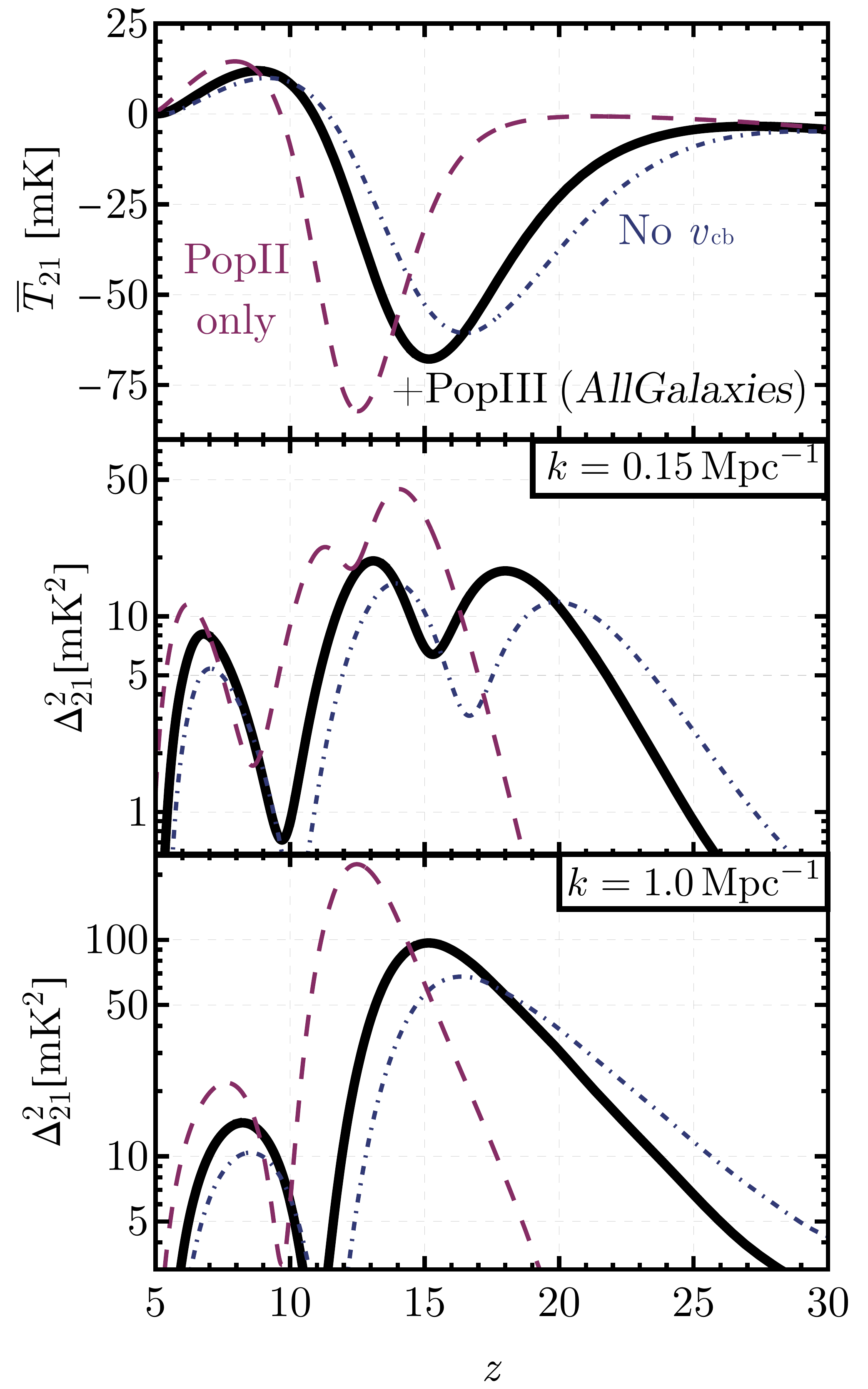}
	\caption{
	Evolution of the 21-cm signal, as a function of redshift $z$, across the entire CD and the EoR, for different models.
	The {\bf top} panel shows the 21-cm global signal (GS), the whereas the {\bf middle} and {\bf bottom} panels show the 21-cm power spectrum (PS) at two different wavenumbers: $k=0.15\,$ and $1.0\,\Mpcinv$, respectively.
	As before, the black line shows our fiducial EOS model with PopII and PopIII stars with all feedback considered, whereas the blue dash-dotted line does not consider feedback due to the streaming velocities $\vcb$, and the purple dashed line only has PopII stars.
	The PopII-only case shows a markedly delayed 21-cm evolution, as the GS in the top panel does not turn into absorption until $z\sim 15$, whereas the case with PopIII stars begins its descent around $z\sim 22$, reaching a minimum at $z\approx 15$.
	The no-$\vcb$ case has much faster evolution, especially at higher $z$, due to the absence of $\vcb$-induced feedback on the first galaxies.
	The 21-cm PS at large scales ($k=0.15\,\Mpcinv$) and small scales ($k=1.0\,\Mpcinv$) follow a similar pattern, though the large-scale one close to vanishes at the minima and maxima of the GS~\citep{Munoz:2020itp}.
	}	
	\label{fig:T21P21_fid}
\end{figure}

We show our predictions for the 21-cm GS in Fig.~\ref{fig:T21P21_fid}.
The 21-cm signal is characterized three well-known eras at high $z$.
First, there is the epoch of coupling (EoC, $z\sim 15-25$ for our EOS parameters), where the GS $\overline T_{21}$ becomes more negative due to the Wouthuysen-Field (WF) coupling sourced by the Lyman-series photons from the first galaxies~\citep{Wout,Field}.
Then, there is the epoch of X-ray heating (EoH), where X-rays emitted by galaxies heat up the IGM, slowly increasing $\overline T_{21}$ until it is above zero.
Finally, during the EoR $\overline T_{21}$ is driven towards zero following $\overline {x}_{\rm HI}$.

The difference between models with PopII only and with PopIII stars is dramatic, as shown in the top panel of Fig.~\ref{fig:T21P21_fid}.
This is to be expected, as the high-$z$ SFRD is dominated by the smaller (and thus more abundant) MCGs.
Ignoring PopIII-dominated MCGs delays the minimum in the GS from $z\approx 15$ to $z \approx 12.5$, and results in a more rapid evolution of all epochs (EoC, EoH, and EoR).
The top panel of Fig.~\ref{fig:T21P21_fid} also shows that neglecting $\vcb$ feedback shifts the EoC and EoH earlier by $\Delta z\approx +2$, which highlights the importance of this effect during the cosmic dawn.

\begin{figure}
	\includegraphics[width=0.46\textwidth]{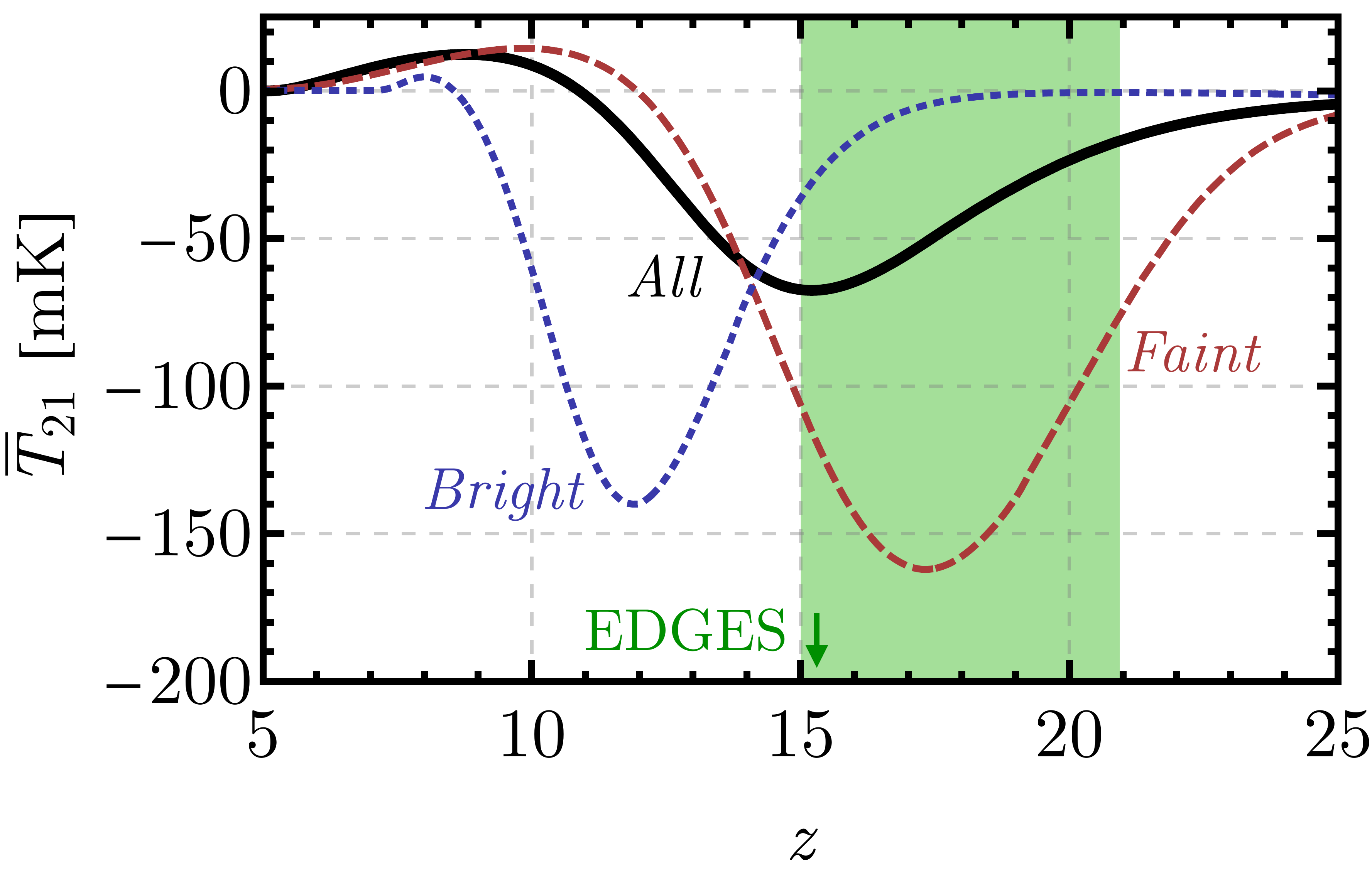}
\caption{
	Global 21-cm brightness temperature for the 2016 EOS simulations of {\it Bright} and {\it FaintGalaxies} (from \citealt{Mesinger:2016ddl}), in blue and red, respectively, as well as the new {\it AllGalaxies} simulation presented here, in black.
 	The green region shows the range of the claimed EDGES detection~\citet[][though its depth of $\overline T_{21}\approx -500$ mK is below the range of this plot]{Bowman:2018yin}.
 	}	
	\label{fig:EOSglobalT21}
\end{figure}

We also compare the GS of the fiducial EOS2021 simulation against the previous two models of EOS2016.
The most striking distinction between these three simulations is that the  depth for the new EOS2021 ({\it AllGalaxies}) model is a factor of two shallower than for both EOS 2016 simulations, only reaching values of $\overline T_{21}\approx -70$ mK.
As already discussed, the emissivity in the 2016 simulations did not follow the SHMR implied by UVLFs but was instead proportional to the collapsed-fraction (equivalent to a constant $f_*$ in Fig.~\ref{fig:fstar}).
As a consequence, the SFRD in those models evolved more rapidly, producing more distinct EoC and EoH epochs, whereas in the realistic {\it AllGalaxies} model these overlap (see also~\citealt{Mirocha2016_UVLF_GS, Park:2018ljd,Qin:2020xyh}).

Such a shallower absorption trough 
also has implications for exotic cosmic explanations of the recent EDGES detection~\citep{Bowman:2018yin}; though we caution that a cosmological interpretation of the detected signal remains very controversial \citep{Hills:2018vyr,Sims:2019kro}.
The lowest point of our trough ($\overline T_{21}\approx -70$ mK) is roughly a factor of 7 shallower than claimed by EDGES ($\overline T_{21}\approx -500$ mK), requiring either stronger dark-matter electric charges (e.g.,~\citealt{Munoz:2018pzp,Barkana:2018lgd}), or a brighter extra radio background (e.g.,~\citealt{Ewall-Wice:2018bzf,Pospelov:2018kdh}) than previously assumed.
However in terms of timing, our fiducial EOS model (with PopIII stars) peaks at $z\sim 15$, only slightly later than the timing of the first claimed EDGES detection (at $z\approx 17$).
We will study a different set (OPT) of MCG parameters in Secs.~\ref{sec:astroparams} and \ref{sec:VAOs}, which give rise to an absorption trough at an earlier $z\approx 17$.

\subsubsection{Power Spectrum}

We now study the 21-cm fluctuations.
For simplicity, we will focus on the spherically-averaged PS summary statistic, defined through
\begin{equation}
    \VEV{\delta T_{21} (\mathbf k)\delta T_{21} (\mathbf k')} = (2\pi)^3 \delta_D(\mathbf k + \mathbf k') P_{21}(\mathbf k),
\end{equation}
although in practice we will employ the reduced power spectrum $\Delta^2_{21} = k^3 P_{21}(k)/(2\pi^2)$, with units of mK$^2$, for convenience.
Interferometers can measure many 21-cm modes at each $z$, and thus the PS (and other spatially-dependent statistics) can provide more detailed insights into the early universe, compared to the GS~\citep{Pritchard:2006sq,Munoz:2019hjh,Fialkov:2014wka,Parsons:2012qh,Pober:2013ig,Cohen:2017xpx,Jones:2021mrs}.
Experiments such as HERA~\citep{DeBoer:2016tnn}, LOFAR~\citep{vanHaarlem:2013dsa}, MWA~\citep{Tingay:2012ps}, LWA~\citep{Eastwood:2019rwh}, and the SKA~\citep{Koopmans:2015sua} are aiming to measure the 21-cm PS.

We show the evolution of the 21-cm PS at two different Fourier-space wavenumbers $k$ in Fig.~\ref{fig:T21P21_fid} (along with the GS for visual aid).
The large-scale ($k=0.15\,\Mpcinv$) power has three bumps,
corresponding to the three eras outlined above, when fluctuations in the Lyman-$\alpha$ background, IGM temperature, and ionization fractions dominate the 21-cm PS, respectively.  Between these, the negative contribution of the cross power between these fields gives rise to relative troughs in the PS~\citep{Lidz:2007az,Pritchard:2006sq,Mesinger:2012ys} (and as a result $\Delta^2_{21} \propto d T_{21}/dz$ on large scales ~\citealt{Munoz:2020itp}).
At smaller scales ($k=1.0\,\Mpcinv$), however, this cancellation does not take place, and the power is larger overall.
The power is larger for the PopII-only model at both large and small scales, as the smaller-mass MCGs that host PopIII stars are less biased, producing smaller 21-cm fluctuations.
The absence of $\vcb$ feedback shifts all curves towards earlier times.
Moreover, as we will see in Sec.~\ref{sec:VAOs}, the $\vcb$ fluctuations become imprinted onto the 21-cm PS~\citep{Munoz:2019rhi,Visbal:2012aw,Dalal:2010yt}, giving rise to sizable wiggles on the 21-cm power spectrum.

\begin{figure}
	\includegraphics[width=0.46\textwidth]{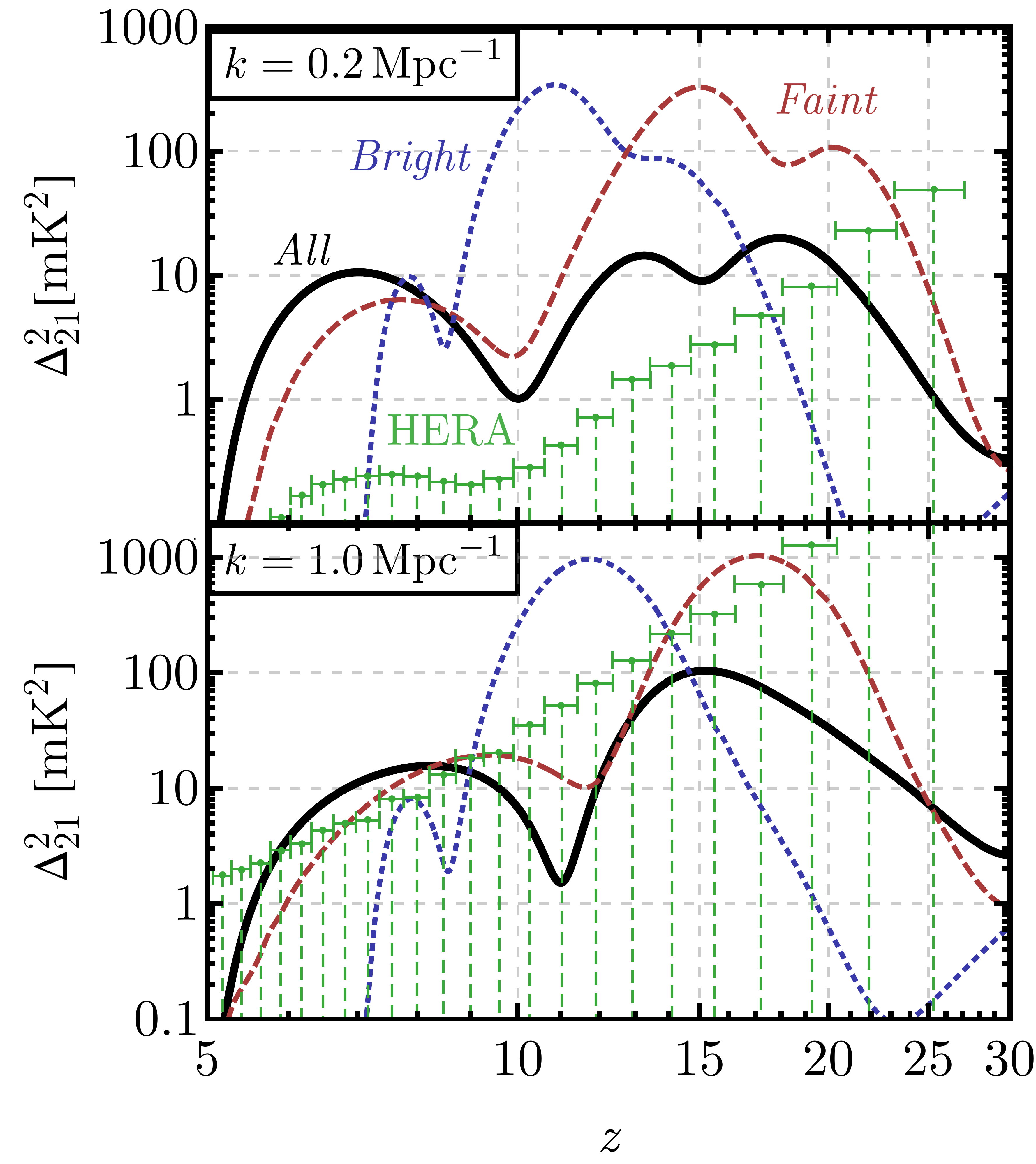}
	\caption{
	The 21-cm power spectrum at two wavenumbers $k=0.2\,$ and 1.0 Mpc$^{-1}$ for all EOS simulations, with the same color coding as Fig.~\ref{fig:EOSglobalT21}.
	We additionally show the expected noise from HERA in different $z$ bands (all corresponding to a bandwidth of $B=8$ MHz), where we assume 180 days of observation, moderate foregrounds, and $\Delta k =0.1\,\rm Mpc^{-1}$.
	We have taken {\it AllGalaxies} as the fiducial for the cosmic-variance noise, and find a total SNR $=183$ added in quadrature over $k$ and $z$.
	}	
	\label{fig:EOSP21}
\end{figure}

In Fig.~\ref{fig:EOSP21} we compare the PS from our fiducial EOS2021 model ({\it AllGalaxies}), to the previous EOS2016 models ({\it BrightGalaxies} and {\it FaintGalaxies}).
The newer {\it AllGalaxies} model shows significantly smaller power during the cosmic dawn than both EOS 2016 predictions.
That is because of both the shallower absorption---and slower evolution---of the 21-cm global signal.
This reduction reaches an order of magnitude at $z\geq 10$, and will hinder the observation of the 21-cm power spectrum with interferometers.

Nevertheless, our {\it AllGalaxies} 21-cm power spectrum is still significantly above the thermal noise level forecasted for upcoming interferometers like HERA and SKA.
We show in Fig.~\ref{fig:EOSP21} the expected noise after 1 year (1080 hours) of integration with HERA, calculated with {\tt 21cmSense}\footnote{\url{https://github.com/jpober/21cmSense}}~\citep{Pober2013,Pober:2013jna} under the moderate-foreground assumption with a buffer $a=0.1\, h \Mpcinv$ above the horizon.
We assume a fixed bandwidth of 8 MHz (corresponding to $\Delta z=0.7$ at $z=10$), spherical bins of $\Delta k = 0.1\,\rm \Mpcinv$, and a system temperature~\citep{DeBoer:2016tnn}
\begin{equation}
    T_{\rm sys} (\nu)= 100\,{\rm K} + 120 \,{\rm K} \times \left( \dfrac{\nu}{150\,\rm MHz}\right)^{-2.55}.
    \label{eq:TsysHERA}
\end{equation}
This results in a noise that is below the signal up to $z\approx 20$ at large scales, and comparable to the signal at small scales.
At low $k$, the noise is dominated by cosmic variance, which tracks the amplitude of the PS, whereas for at high $k$ it is largely thermal. 

We further quantify the detectability of our EOS 2021 model ({\it AllGalaxies}), by computing the signal-to-noise ratio (SNR).
We calculate this quantity for each wavenumber $k$ and redshift bin $z$ considered, and add them in quadrature.
We consider two values for $T_{\rm sys}$, that of Eq.~\eqref{eq:TsysHERA} and a more pessimistic one of
\begin{equation}
    T_{\rm sys}^{\rm pess.} (\nu)= 100\,{\rm K} + 400 \,{\rm K} \times \left( \dfrac{\nu}{150\,\rm MHz}\right)^{-2.55},
    \label{eq:Tsyspess}
\end{equation}
following~\cite{Dewdney:SKA}, which results in a noise larger by a factor of $\sim 3$ at the redshifts of interest.
We also calculate the SNR for the fiducial SKA-LOW 1 design \cite{Dewdney:SKA} using a tracked observing strategy. Specifically, we assume a 6 hour per-night tracked scan for a total of 1000 hours.
Table~\ref{tab:SNR_EOS} shows the SNRs for the different setups, where using Eq.~\eqref{eq:TsysHERA} we find SNR $=186$ for HERA and SNR $=164$ for the SKA, both of which would provide detections at high significance.
These SNRs would be reduced by a factor of $\sim 2-3$ for the pessimistic $T_{\rm sys}$ from Eq.~\eqref{eq:Tsyspess}.
Divided into epochs, the SNR is significantly dominated by the EoR, with the EoH contributing a factor of $\sim 5$ less, and the EoC only showing SNR $\sim 10$.
Interestingly, for our fiducial ``narrow and deep" SKA survey, the SKA can reach larger SNR at high $z$ where thermal noise dominates, while HERA performs better at lower $z$ where cosmic variance can be dominate the noise.  Assuming different SKA observing strategies can shift the balance between cosmic-variance and thermal-noise errors by considering either larger observing volumes or deeper integration times \citep[see e.g][]{Greig:2020ska}.
We will explore in~\citet{Mason_21cmFisher} the range of constraints that such a detection would provide for astrophysical and cosmological parameters.

\begin{table}
\centering
\begin{tabular}{l c c c c}
SNR for EOS2021 & Total & EoR  & EoH  & EoC \\
\hline
HERA          & 186   & 183         & 34             & 9           \\
HERA (pess.)  & 87    & 86          & 8              & 1           \\
\hline
SKA           & 164   & 157         & 41             & 22          \\
SKA (pess.)   & 85    & 84          & 12             & 4          
\end{tabular}
\caption{Signal-to-noise ratios (SNR) for the two interferometers we consider, under a regular system-noise assumption, given by Eqs.~\eqref{eq:TsysHERA}, and a pessimistic one, from Eq.~\eqref{eq:Tsyspess}.
The three epochs cover the ranges of $z\leq10$ for the EoR, $10<z<15$ for the EoH, and $z\geq15$ for the EoC.
}
\label{tab:SNR_EOS}
\end{table}

\begin{figure}
	\includegraphics[width=0.46\textwidth]{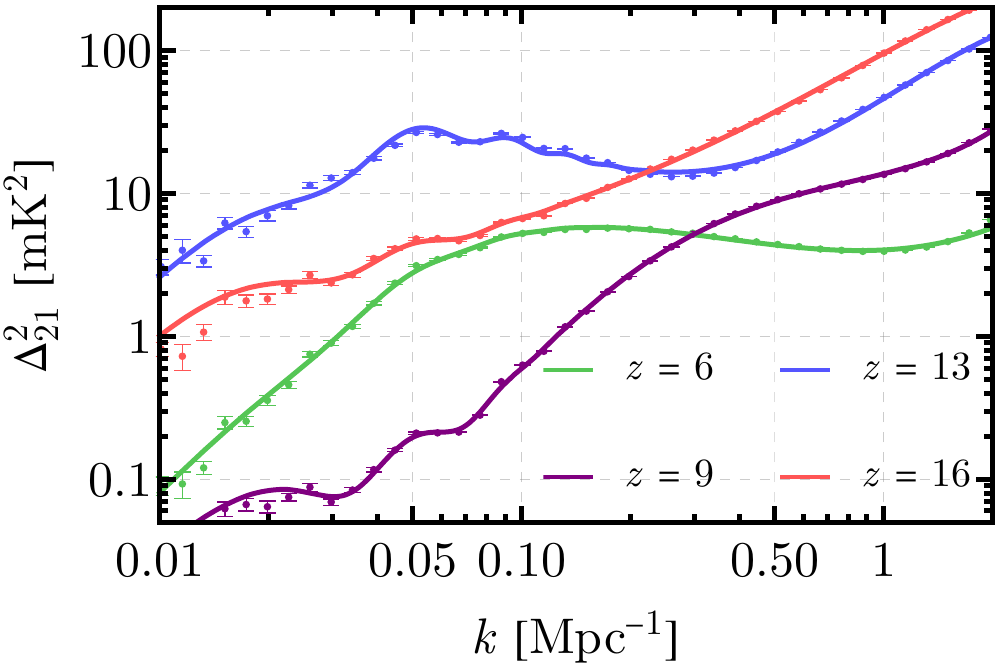}
	\caption{Power spectrum of 21-cm fluctuations as a function of wavenumber $k$ for our {\it AllGalaxies} (EOS2021) simulation.
	The four redshifts are chosen to be during the EoC $(z=16)$, halfway through the EoH $(z=13)$, in the transition between the EoH and the EoR $(z=9)$, and finally during the EoR $(z=6)$.
	The small error bars come from the Poisson noise in our box, which is 1.5 Gpc comoving in size.
	The lines represent a fit (using $k=0.02-0.5\,\Mpcinv$) to a smooth polynomial added to the wiggles from the VAOs (sourced by the streaming velocities $\vcb$), as we will explain in Sec.~\ref{sec:VAOs}.
	}	
	\label{fig:P21_k_EOS}
\end{figure}

Finally, we show the scale dependence of the 21-cm PS at four redshifts in Fig.~\ref{fig:P21_k_EOS}.
These are chosen to illustrate the power spectrum during each of the three epochs of interest (EoR, EoH, and EoC), as well as in the transition between the EoR and EoH.
The power is relatively flat with $k$ except in the transition case ($z=9$), where the large-scale power drops dramatically due to the negative contribution of the cross-terms~\citep{Lidz:2007az,Pritchard:2006sq,Mesinger:2012ys,Munoz:2020itp}.
Interestingly, at $z=13$ the power peaks at $k=0.1\,\Mpcinv$, where there are wiggles in the 21-cm power spectrum.
These are due to the streaming velocities $\vcb$, which have acoustic oscillations that become imprinted onto the SFRD (through the feedback described in Sec.~\ref{sec:FirstGalaxies}), and thus on the 21-cm signal.
We will describe these velocity-induced acoustic oscillations (VAOs) in detail in Sec.~\ref{sec:VAOs}.

\begin{figure*}
	\includegraphics[width=0.98\textwidth]{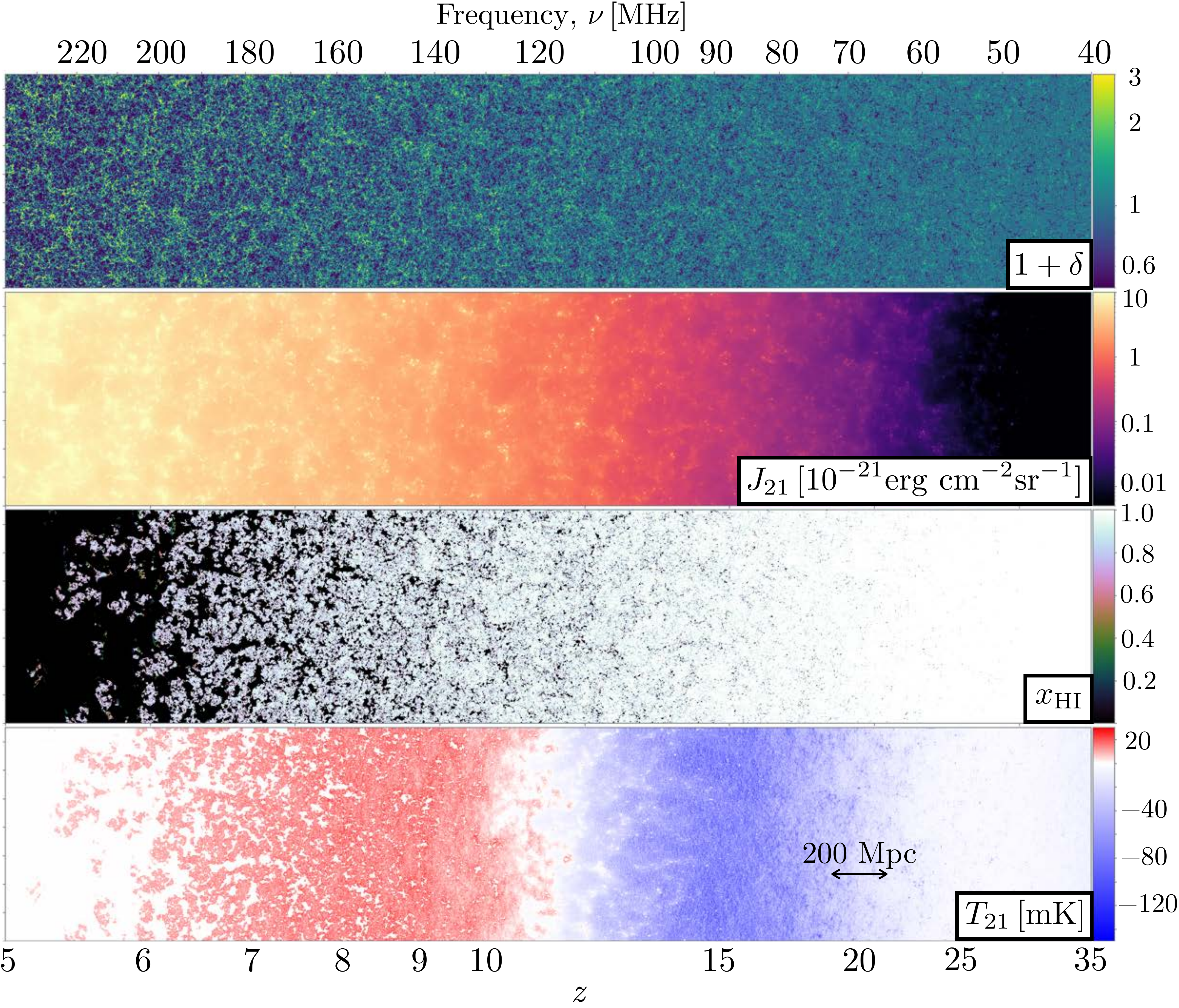}
	\caption{
	Lightcone from a simulation with the EOS 2021 parameters from Table~\ref{tab:Fids}.
	The slices are 750 Mpc in height (half of the full simulation, for better visualization of small scales), and 1.5 Mpc in depth.
	The top panel shows the density field, in log-normal scale around unity.
	The second panel is the LW flux $J_{21}$, in the customary units of $10^{-21}$ erg s$^{-1}$ Hz$^{-1}$ cm$^{-2}$ sr$^{-1}$, which dissociates molecular hydrogen and halts PopIII stellar formation on MCGs.
	The third panel is the neutral-hydrogen fraction $x_{\rm HI}$, which shows the overall evolution of reionization, as well as the large-scale neutral patches at $z\sim6$ that can explain opacity fluctuations seen in the Lyman-$\alpha$ forest.
	Finally, the bottom panel is the 21-cm signal $T_{21}$ in mK, where the absorption (blue) trough occurs at $z\sim 15$, the transition to emission (red) at $z\sim 10$, and the signal disappears (white) due to reionization by the end of the simulation.
	}	
	\label{fig:LCa0}
\end{figure*}

\subsection{Visualizations}

We end this section with some visualizations of our EOS 2021 simulation.  These consist of slices through various lightcones and simulation cubes.

We begin with Fig.~\ref{fig:LCa0}, which shows 2D slices through cosmic lightcones from our fiducial EOS 2021 simulations.  The horizontal axis shows evolution with cosmic time,  while the vertical axis corresponds to a fixed comoving length (here taken to be half of the full EOS size in order to more easily identify small scale features).
While we only track a few variables in that Figure, we note that {\tt 21cmFAST} can output other relevant quantities such as the local recombination rate, the intensity of the UV, X-ray, Lyman-$\alpha$ backgrounds, the velocity fields, and kinetic and spin temperatures.
We describe each of the panels in turn.

$\bullet$ The top panel shows the matter over/under-densities, which grow due to gravity as the universe evolves, forming the cosmic web that we see today.

$\bullet$ The second panel of Fig.~\ref{fig:LCa0} shows the Lyman-Werner flux, which dissociates H$_2$ molecules and thus impedes star formation in MCGs. 
The overall LW flux $J_{21}$ grows rapidly over time (roughly following the SFRD), with notable spatial fluctuations.  The LW flux is largest in regions corresponding to the largest matter overdensities, which host the first generations of (highly biased) galaxies.  Therefore these anisotropies result in a stronger suppression of PopIII star formation, than would be expected assuming a homogeneous background, and imprint structure in the 21-cm signal.

$\bullet$ The third and fourth panels of Fig.~\ref{fig:LCa0} show the evolution of the two quantities most directly observable: the neutral hydrogen fraction $x_{\rm HI}$ and the 21-cm brightness temperature $T_{21}$. The neutral fraction $x_{\rm HI}$ is homogeneously close to unity until $z\sim 10$, as chiefly MCGs form at those early times, which are not very efficient at reionizing hydrogen.
The EoR takes place during $z=5-10$, accelerating at later times.
This panel reveals large-scale neutral patches at $z\sim 5.5-6$, as required by recent Lyman-$\alpha$ data~\citep{Becker:2014oga,Bosman:2018xxh}.
These last neutral regions trace under-dense environments.

$\bullet$ The 21-cm signal $T_{21}$, in the last panel, shows a dramatic evolution during cosmic dawn, beginning with absorption at $z\sim 15-20$ due to the Wouthuysen-Field effect, followed by a transition to emission at $z\sim12$ as the IGM is heated by X-rays from the first galaxies, and finishing with a slow decay towards zero as reionization takes place.
The late-time ($z\lesssim 10$) behavior of the 21-cm signal is dominated by the reionization bubbles, as the $x_{\rm HI}$ and $T_{21}$ fields are clearly correlated.
The early-time ($z\gtrsim 10$) behavior, however, is related to the UV and X-ray flux from the first galaxies, which depends on the densities $\delta$ in a non-linear and non-local way.

\begin{figure*}
	\includegraphics[width=1.0\textwidth]{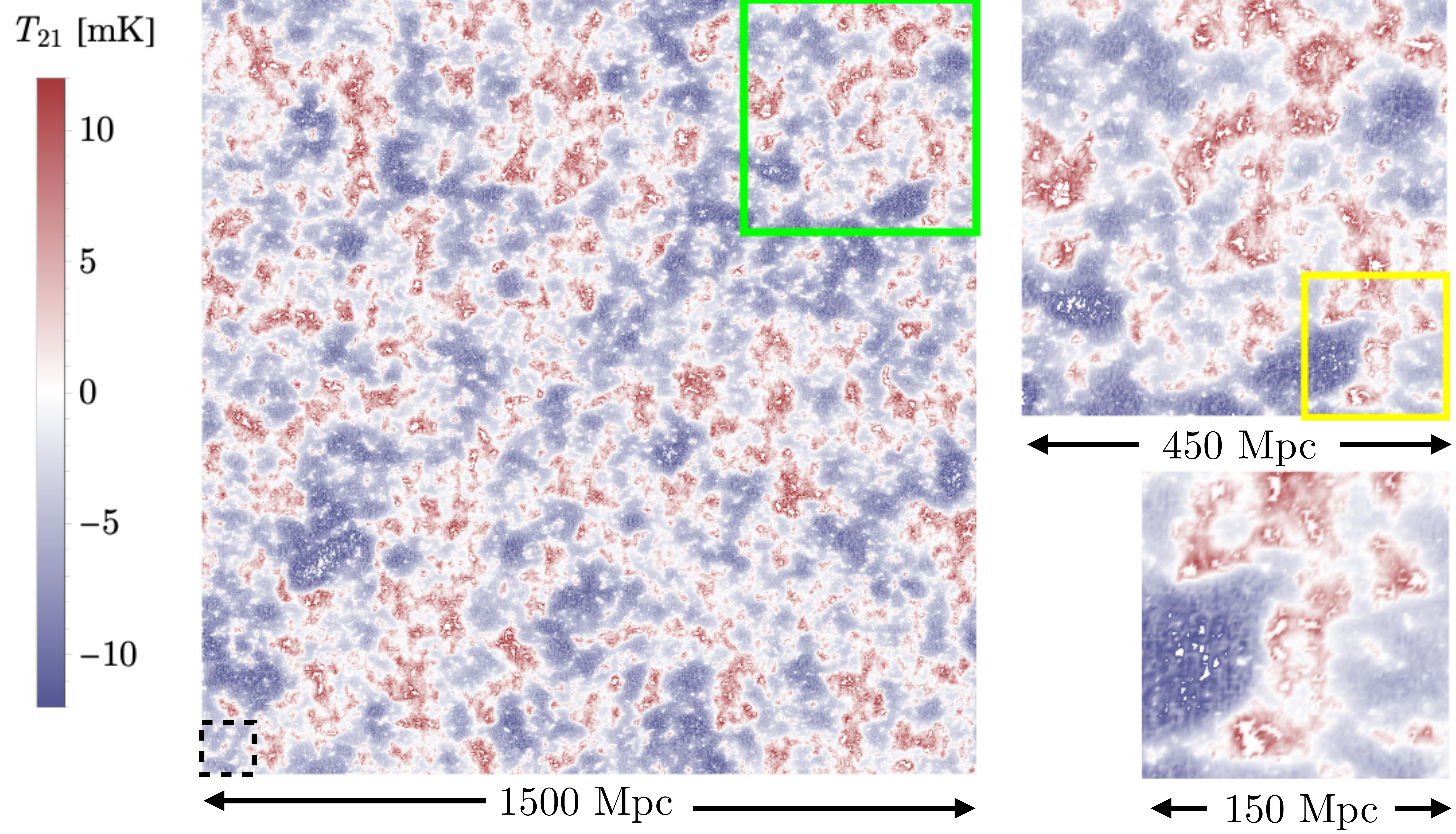}
	\caption{
	Slice through the 21-cm temperature in our {\it AllGalaxies} simulation (with the EOS 2021 parameters of Tab.~\ref{tab:Fids}) at $z=11$.
	The left panel shows the entire simulation (1.5 Gpc comoving in side), whereas the two right panels represent successive zoom-ins of each of the highlighted regions. 
	For reference, state-of-the-art radiative-transfer hydrodynamical simulations of reionization that resolve ACGs (e.g., \citealt{Kannan21,Garaldi:2021gfo,Smith:2021hqg}) can typically cover volumes $\lesssim (100\,\rm Mpc)^3$, shown as the black dashed square on the bottom left.
	Red and blue correspond to 21-cm emission and absorption, as in Fig.~\ref{fig:LCa0}, and all slices are 1.5 Mpc in thickness.
	At this redshift our model is transitioning from the EoH to the EoR. We see that these two eras overlap, causing the centers of heated regions ($T_{21}>0$, red), which  have the highest densities of galaxies, to show no signal ($T_{21}=0$, white).
	}	
	\label{fig:zoominEOS}
\end{figure*}

In Fig.~\ref{fig:zoominEOS} we show a zoom-in slice through the 21-cm map at $z=11$.
This redshift corresponds roughly to the transition between the EoH and the EoR.
It is clear that there is a large overlap between these two eras, as the overdense heated regions (with $T_{21}>0$, shown in red), are beginning the process of reionization from the inside
(producing $T_{21}\to0$, in white).  We note that the underdense regions, exposed to a smaller X-ray flux, are remain colder than the CMB at this redshift ($T_{21}<0$, in blue).

Fig.~\ref{fig:zoominEOS} also shows the dynamic range of our simulations, which can resolve structure as large as the simulation box (1.5 Gpc) and down to the cell size (1.5 Mpc).
A zoom-in animation of the entire cosmic dawn and EoR evolution is provided at \href{https://scholar.harvard.edu/julianbmunoz/eos-21}{this url}.

\section{Learning about the first galaxies}
\label{sec:astroparams}

In the previous sections we have demonstrated that MCGs drive the 21-cm signal from the early cosmic dawn, given our fiducial set of parameters.
As a consequence, 21-cm studies are a promising avenue to learn about the properties of the first PopIII stars and their host galaxies.
Here we perform a brief exploratory study of how the 21-cm signal varies as a function of the different MCG stellar and feedback parameters in our model.

Throughout this section, we will vary parameters around a more optimistic set of galaxy properties, labeled {\it OPT} in Table~\ref{tab:Fids}.
This set allows for a more significant contribution from MCGs compared to EOS, making it is easier to learn about these first galaxies.
In particular, we increase the stellar efficiency of MCGs by $\Delta \log_{10}f_{*,10}^{(\rm II)}=-0.75$ (roughly a factor of $\approx 5$), and decrease it for ACGs by $\Delta \log_{10}f_{*,10}^{(\rm II)}=-0.25$, in both cases compensating their $\fesc$ to keep a similar EoR evolution.
This higher MCG contribution pushes the trough of the 21-cm GS to an earlier $z\sim 17$ (in line with the central redshift claimed by EDGES in~\citealt{Bowman:2018yin}), rather than the $z\sim 15$ of our EOS parameters.  We also find that in OPT, PopIII stars dominate the SFRD for $z\gtrsim10$, thus dictating the evolution of the entirety of cosmic dawn.

We begin by studying the slope of the SHMR, here parametrized with the $\alpha_*^{\rm (III)}$ parameter.  The SHMR slope likely holds clues about SNe and other galactic feedback mechanisms.  For example, assuming star formation is regulated by SNe-driven outflows with a constant energy coupling efficiency, can result in $\alpha_* \sim$ 2/3 (e.g., \citealt{Wyithe2013MNRAS.428.2741W}).  This is remarkably close to the empirically determined value from $z\gtrsim6$ UVLFs of $\alpha_*^{\rm (II)} \sim 0.5$ (for ACGs hosting PopII stars).

As discussed previously, ACGs are too faint to allow us to directly measure $\alpha_*^{\rm (III)}$ from UVLFs.  Although it is tempting to assume the same SHMR slope for ACGs and MCGs, this could be incorrect due to, e.g., their different IMFs and associated SNe energies.
For example, a fixed mass of stars forming in each MCG (e.g., \citealt{Kulkarni:2018erh}) would result in $\alpha_*^{\rm (III)} < 0$, which would be very different from the $\alpha_*^{\rm (II)} \sim 0.5$ that is empirically determined from $z\gtrsim6$ UVLFs.

\begin{figure}
	\includegraphics[width=0.46\textwidth]{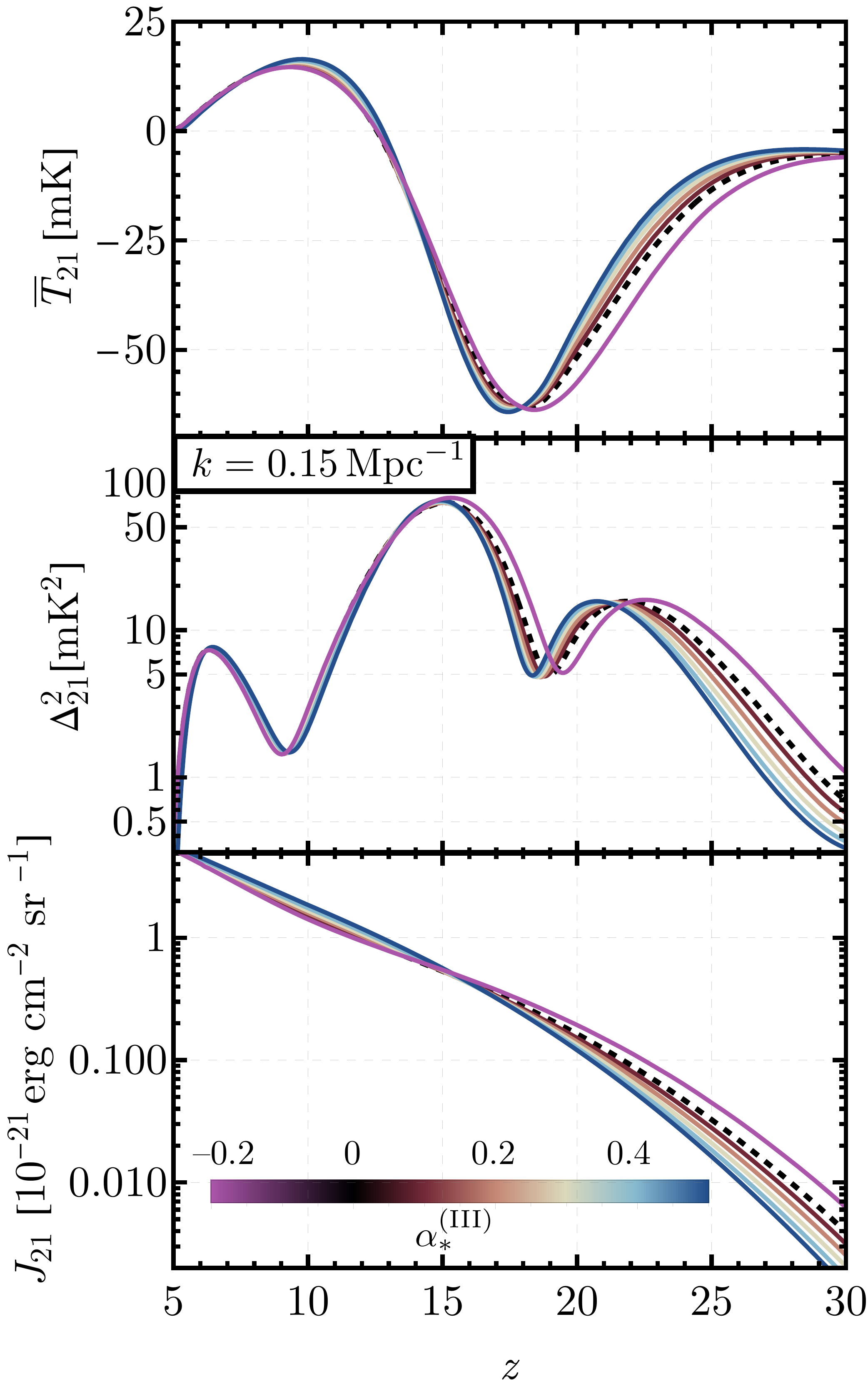}
	\caption{
	Predictions for the evolution of the 21-cm GS ({\bf top}), PS ({\bf middle}), and the LW flux ({\bf bottom}) across CD and the EoR, when varying the slope of the SHMR of MCGs.
	We consider a range of power-law indices $\alpha_*^{\rm (III)}$ of the SHMR, as defined in Eq.~\eqref{eq:SFRDIII}, 
	from $\alpha_*^{\rm (III)}=0$ (our fiducial, in black) to 0.5 (red to blue lines), as well as $\alpha_*^{\rm (III)}=-0.2$ (in pink, corresponding to a reversed SHMR).
	Steeper $\alpha_*^{\rm (III)}$ indices produce less star formation at early times,
	delaying the onset of cosmic dawn (top two panels).
	However, this also produces less LW radiation, as shown in the bottom panel, dissociating H$_2$ more weakly.
	}	
	\label{fig:T21alphas}
\end{figure}

In Fig.~\ref{fig:T21alphas} we show the impact of varying the power-law index $\alpha_*^{\rm (III)}$ on the 21-cm CD signal (global and power spectrum evolution).
We consider $\alpha_*^{\rm (III)}$ in the $-0.2$ -- 0.5 range.
It is clear that
the 21-cm signal does not vary dramatically over this range of SHMR slopes.
Steeper indices (larger $\alpha_*^{\rm (III)}$) effectively result in steeper SFRD evolutions at very high redshifts. Although initially there are fewer PopIII stars in those models, delaying the 21-cm signal, subsequently the cosmic evolution accelerates.

Interestingly, all models in Fig.~\ref{fig:T21alphas} agree at $z\approx 15$, when the contribution from ACGs starts being relevant.  However, we do see that the $J_{21}$ evolution of some models crosses-over around this redshift (c.f.~the green curve).  This is due to LW feedback:  models that initially have a larger $J_{21}$ are subsequently able to more strongly quench MCG star formation, thus resulting in a weaker $J_{21}$ at $z<15$.

\begin{figure*}
	\includegraphics[width=0.98\textwidth]{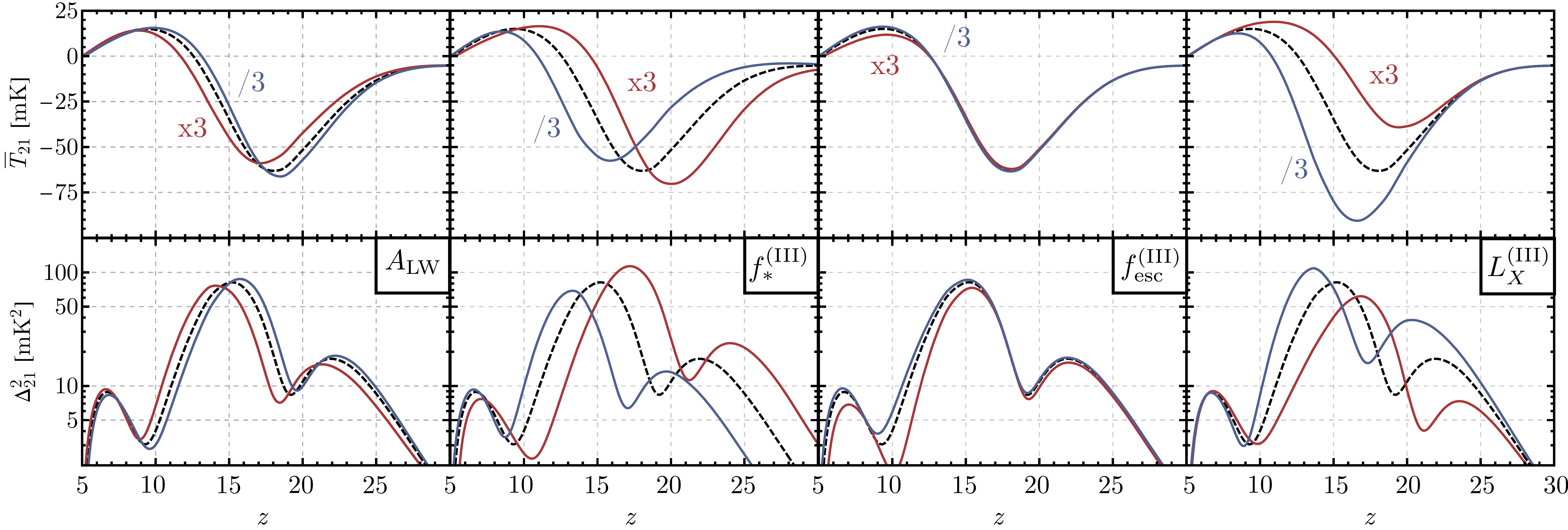}
	\caption{The effect of varying four PopIII parameters on the 21-cm global signal ({\bf top}) and power spectrum  ({\bf bottom}, at $k=0.23\,\Mpcinv$). 
	These parameters encode the strength of LW feedback ($A_{\rm LW}$), the stellar efficiency ($f_*^{(\rm III)}$) and escape fraction ($f_{\rm esc}^{(\rm III)}$) of MCGs, as well as  the X-ray luminosity $L_X^{(\rm III)}$ per unit SFR of PopIII stars.
	Black dashed lines show our OPT fiducial model, and in each panel we indicate the parameter we vary, increasing (decreasing) it threefold in red (blue).
	}	
	\label{fig:paramvary}
\end{figure*}

We now study the impact of the other parameters regulating the UV and X-ray emissivities of MCGs.  Specifically, in Fig.~\ref{fig:paramvary}
we show how the 21-cm GS and PS (at $k=0.15\,\Mpcinv$) vary with: (i) $A_{\rm LW}$, the amplitude of LW feedback; (ii) $f_{*,7}^{(\rm III)}$, the normalization of the SHMR; (iii) $f_{\rm esc,7}^{(\rm III)}$, the ionizing escape fraction; and (iv) $L_{X,\rm <2keV}^{\rm (III)}/{\rm SFR}$, the X-ray luminosity to SFR relation.
Unlike for $\alpha_*^{\rm (III)}$ discussed above, the uncertainty on these parameters is better sampled in log space.  Therefore in Fig.~\ref{fig:paramvary} we show results when increasing or decreasing each parameter by a factor of three around the {\it OPT} values from Table~\ref{tab:Fids}.

The first panel of Fig.~\ref{fig:paramvary} shows that stronger LW feedback (larger $A_{\rm LW}$) translates into a delayed 21-cm GS and PS, especially at high $z$.
The larger impact of other feedback sources, chiefly the relative velocities, makes the signal depend only weakly on $A_{\rm LW}$, though this parameter can still delay the cosmic-dawn milestones by $\Delta z\approx 1$ within the range of values we study.

Changing $f_{\rm esc,7}^{(\rm III)}$ also has a modest impact, as MCGs are generally negligible contributors to the EoR in our models.  However, the largest values of the escape fraction shown here do result in an earlier start to the EoR (driven by MCGs), but with a similar end (driven by ACGs).

On the other hand, varying the stellar fraction $f_{*,7}^{(\rm III)}$ (second panel) or $L_{X,\rm <2keV}^{\rm (III)}/{\rm SFR}$ (fourth panel) notably changes the signal during the two CD epochs driven by MCGs: the EoH and EoC.  Changing the stellar fraction impacts both epochs, as star formation drives all of the cosmic radiation fields in our models.  Higher stellar fractions shifts the CD to earlier times, resulting in a higher effective bias of the sources driving each epoch, and thus a higher 21-cm PS on large scales.

Changing $L_{X,\rm <2keV}^{\rm (III)}/{\rm SFR}$ only impacts the relative timing of the EoH.  Increasing the X-ray luminosity of the first galaxies results in a larger overlap of the EoH and EoC, as the coupling is not completed before the IGM is heated.  
Consequently, the GS absorption trough is shallower, and the large-scale power decreases from the increased negative contribution of the cross-power in these two fields (e.g., \citealt{Pritchard:2006sq, Mesinger:2012ys, Schneider:2020xmf}).

Each of the parameters impacts the signal differently as a function of redshift and scale, which may allow us to distinguish between them.
However, in order to forecast parameter uncertainties,  one has to capture the correlations between them, for instance through an MCMC~\citep{Greig:2015qca} or Fisher matrix~\citep{Mason_21cmFisher}.  We leave this question for future work.
We note that the expected SNRs for the OPT model are similar to the EOS ones reported in Sec.~\ref{sec:CDEoR}. 
That is because the OPT model shows slightly larger fluctuations, though at lower frequencies where noise is larger.
Table~\ref{tab:SNR_OPT} contains our forecasted SNRs for the OPT parameters under each of the assumptions considered.

\begin{table}
\centering
\begin{tabular}{l c c c c}
SNR for OPT  & Total & EoR & EoH & EoC \\
\hline
HERA         & 206   & 193 & 66  & 31  \\
HERA (pess.) & 94    & 93  & 16  & 6   \\
\hline
SKA          & 195   & 178 & 69  & 37  \\
SKA (pess.)  & 99    & 95  & 25  & 11 
\end{tabular}
\caption{Same as Tab.~\ref{tab:SNR_EOS} but for the OPT parameters.
}
\label{tab:SNR_OPT}
\end{table}

\section{Velocity-induced Acoustic Oscillations}
\label{sec:VAOs}

The last study we perform is on the unique signature of the DM-baryon relative velocities on the 21-cm fluctuations.
We quantify to what extent the streaming velocities produce velocity-induced acoustic oscillations (VAOs) on the 21-cm signal in our simulations, for the first time jointly including inhomogeneous LW feedback with self shielding.
Throughout this section we will assume astrophysical parameters from the Optimistic (OPT) set, unless otherwise specified.

\begin{figure*}
	\includegraphics[width=0.98\textwidth]{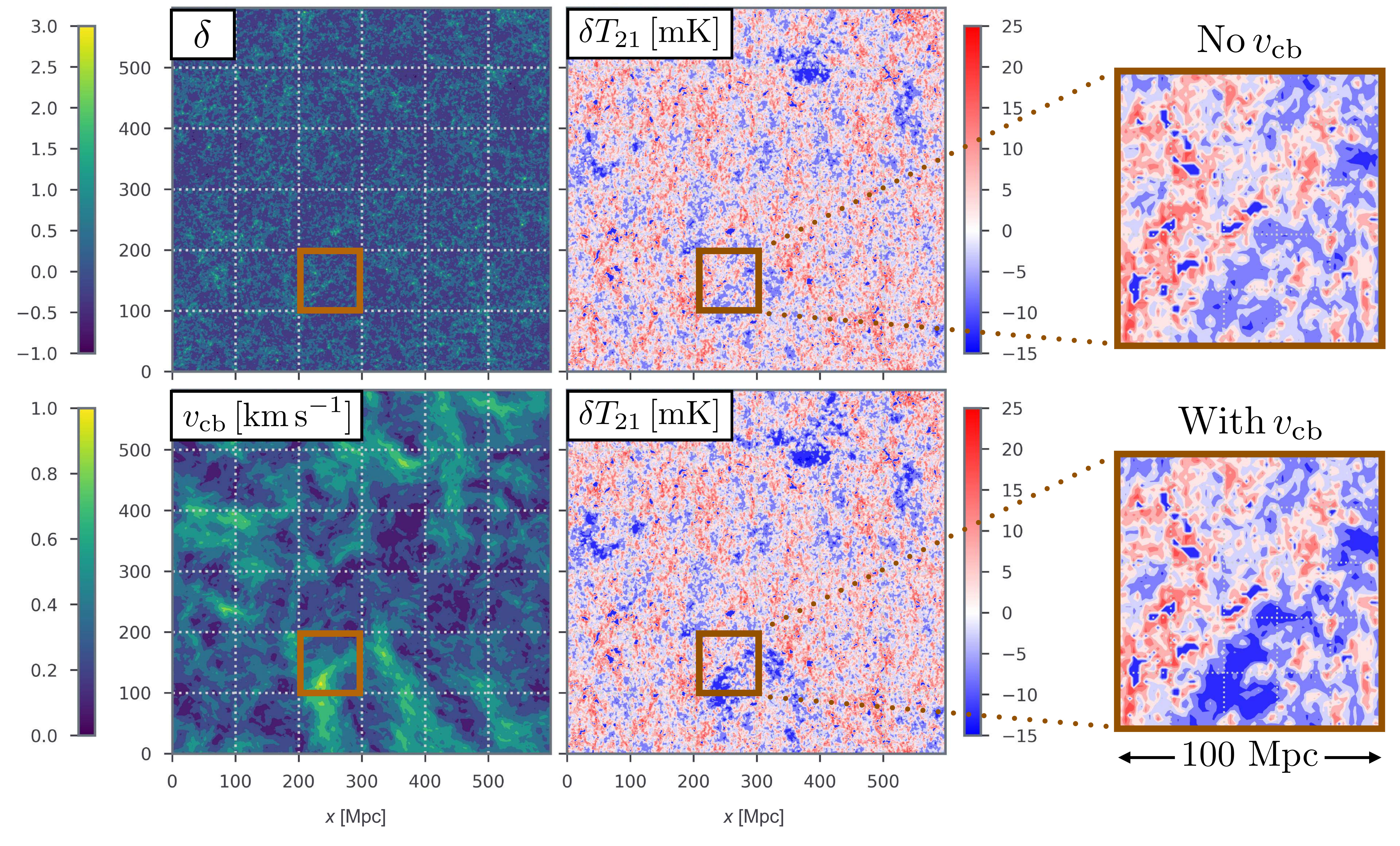}
	\caption{
	Effect of relative velocities on the 21-cm signal with our OPT parameters and the most up-to-date feedback prescriptions.
	We show a slice through our simulations, 1.5-Mpc deep and 600 Mpc on a side, at $z=11$, where different panels show different quantities.
	Top left shows the matter density, and bottom left the DM-baryon relative velocity. 
	The two central panels plot the 21-cm brightness temperature, on the top without fluctuating relative velocities (i.e., $\vcb=v_{\rm avg}$), and in the bottom with the full $\vcb$ effect.
	We also show a zoom-in (100 by 100 Mpc) region of large relative velocity on the right, where the suppressive effect of $\vcb$ on the first galaxies---and thus on $T_{21}$---is readily apparent.
	}	
	\label{fig:relvel_4panel}
\end{figure*}

\subsection{VAOs across Cosmic Dawn}

The interactions between baryons and photons give rise to the well-known baryon acoustic oscillations (BAOs), which at low $z$ are observed in the matter distribution as an excess correlation at the baryon acoustic scale (here used interchangeably with the baryon drag scale $r_{\rm drag}$; ~\citealt{Eisenstein:1997ik}).
The dark matter, however, does not partake in these BAOs, which gives it different initial conditions than baryons at recombination.
This produces a relative (or streaming) velocity between dark matter and baryons, which fluctuates spatially with the same $r_{\rm drag}$ scale, due to their acoustic origin~\citep{Tseliakhovich:2010bj}.
In Fourier space, the power spectrum of $\vcb$ presents large wiggles, which are inherited by the radiation fields, as regions of large relative velocity suppress the formation of the first stars (chiefly PopIII, see Sec.~\ref{subsec:MCGs}).
Consequently, the 21-cm signal becomes modulated by these streaming velocities during cosmic dawn, giving rise to velocity-induced acoustic oscillations~\citep[VAOs,][]{Munoz:2019rhi,Dalal:2010yt,Visbal:2012aw,Fialkov:2012su}, with the same acoustic origin as the BAOs, though sourced by velocity---rather than density---fluctuations.
\cite{Munoz:2019fkt} showed that these VAOs provide us with a standard ruler to measure physical cosmology during cosmic dawn, independently of galaxy astrophysics.

Until recently it was not known how the feedback from $\vcb$ interacted with the LW dissociation of molecular hydrogen.
As we showed in Sec.~\ref{sec:FirstGalaxies}, however, recent hydrodynamical simulations from~\citet{Kulkarni:2020ovu} and \citet{Schauer:2020gvx} indicate that for the regime of interest $(J_{21}\leq 1)$ these two processes act coherently.
Furthermore, self shielding in the first galaxies produces weaker LW feedback, and thus a larger impact of the relative velocities.
Together, these two effects give rise to sizable VAOs, as we now show.

\subsubsection*{Slices}

We begin by showing the impact of $\vcb$ directly on the 21-cm maps.
For that we compare a standard simulation (with OPT parameters) against one with no fluctuating $\vcb$ (achieved by setting {\tt FIX\_VAVG $=$ True} in {\tt 21cmFAST}).
The latter simulation just uses a homogeneous value of $\vcb=v_{\rm avg}\approx 26$ km s$^{-1}$, corresponding to the mean of its distribution.
The reason for this choice, rather than setting $\vcb=0$, is that the background evolution in the latter case would be significantly different (see e.g., Fig.~\ref{fig:T21P21_fid}), making it difficult to compare results at a fixed redshift.

We plot slices (1.5 Mpc thick) through our simulations at $z=11$ (during the EoH) in Fig.~\ref{fig:relvel_4panel}.
The slice through the relative-velocity field clearly shows large-scale acoustic structure, with islands of large $\vcb$ separated by roughly $r_{\rm drag}\approx 150$ Mpc.
In contrast, the matter field ($\delta$) has power on all visible scales, down to our cell size.
We also show the 21-cm map resulting from our two simulations with and without fluctuating $\vcb$ (but with otherwise identical OPT parameters). 
Regions of large $\vcb$ have a colder IGM, as they form fewer stars, and thus emit fewer X-rays.
In order to illustrate this effect, we zoom into a patch 100 Mpc in size near a region of large $\vcb$, where the full-physics simulation clearly presents deeper absorption correlated with the velocity map.

\begin{figure*}
	\includegraphics[width=0.98\textwidth]{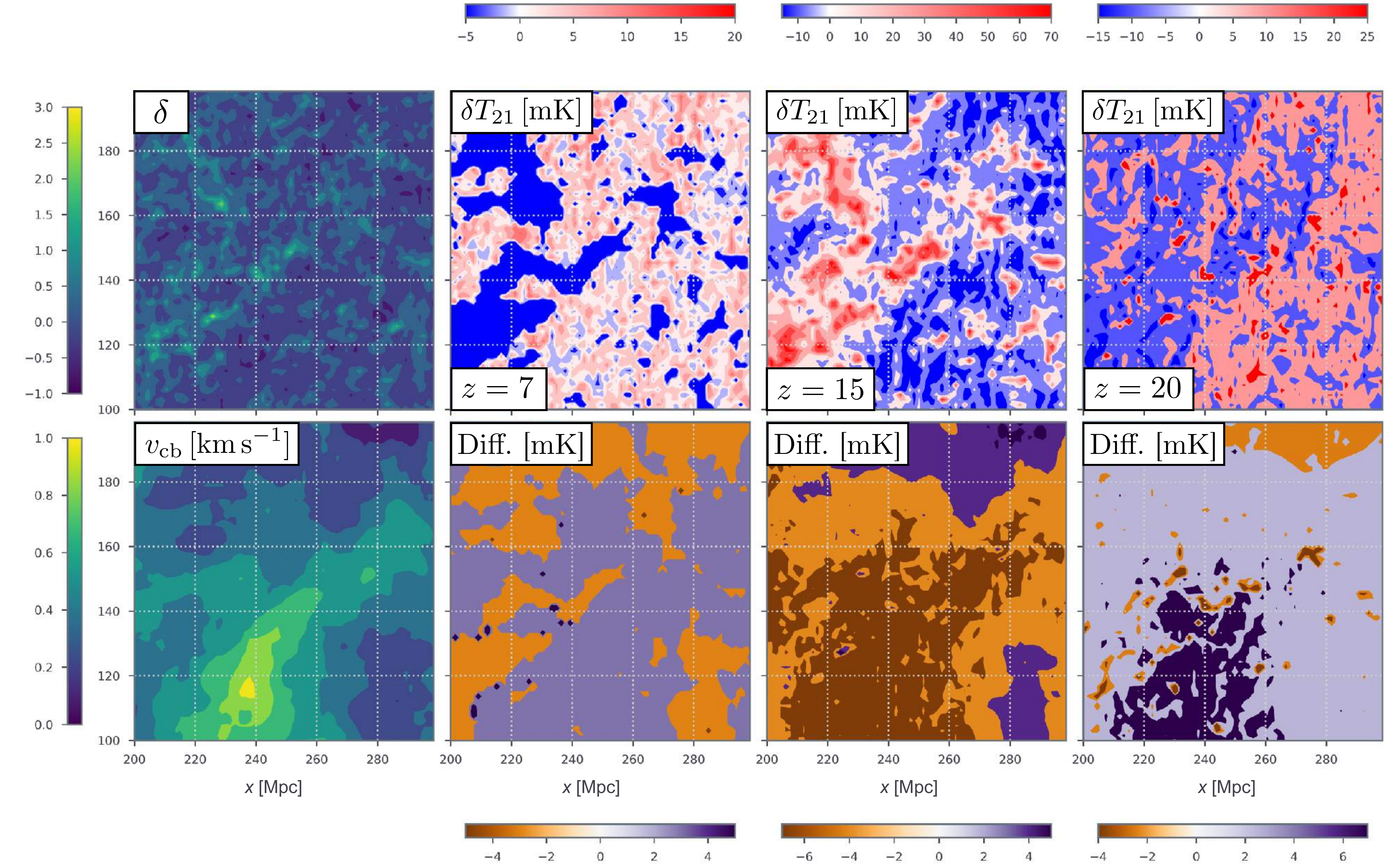}
	\caption{
	Zoom-in slice, 100\,Mpc on a  side and 1.5 Mpc in depth, from the same region as Fig.~\ref{fig:relvel_4panel}.
	We show the zoomed-in density $\delta$ and relative velocity $\vcb$ at $z=11$ in the left two panels.
	The rest of top panels show the 21-cm fluctuations $\delta T_{21}$ for no VAOs ($\vcb=v_{\rm avg}$), whereas the bottom panels show its difference with the full (i.e., with VAO) case, defined as Diff $=\delta T_{21}^{\rm full}-\delta T_{21}^{\rm no\,VAO}$.
	The effect of VAOs is subtle during the EoR ($z=7$, second column), as differences are at the $\lesssim 2$ mK level,
	but more noticeable during the EoH ($z=15$, third column) and the EoC ($z=20$, fourth column).
	}	
	\label{fig:relvel_zoomonly}
\end{figure*}

We now move on to study the effect of $\vcb$ during other cosmic eras.
In Fig.~\ref{fig:relvel_zoomonly} we show a 100 Mpc zoom-in of our simulations, at the same location as in Fig.~\ref{fig:relvel_4panel}.
Rather than showing maps of $\delta T_{21}$ with and without VAOs, we plot the difference (Diff  $\equiv \delta T_{21}^{\rm full}-\delta T_{21}^{\rm no\,VAO}$) between the two cases, which allows for a closer comparison, at three redshift snapshots.
The first is at $z=7$, during the EoR, where the effect of the relative velocities is rather small, affecting the signal at the $\sim 1-2$ mK level.
The second is at $z=15$, at the peak of heating, and shows a large impact of $\vcb$, causing Diff $\approx \pm5$ mK, with the largest differences taking place in the lowest-$\vcb$ regions, which heated more slowly.
The last snapshot is at $z=20$, which is during the EoC and where $\vcb$ impacts the signal moderately, but in the opposite direction (as fewer photons produces less coupling, and thus more positive $\delta T_{21}$).
As is clear from Fig.~\ref{fig:relvel_4panel}, the profile of the relative velocity (left panel) is smeared when observed in the 21-cm signal, due to photon propagation.
This is especially true in the $z=20$ panel, where the difference has a homogeneous value of $\sim +2$ mK in the entire (100-Mpc) zoom-in region.  This is because the photons just redward of Lyman-$\beta$ that drive WF coupling have mean free paths comparable to the 100-Mpc scale of these zoom-ins.
Below we provide a simple analytic expression for the 21-cm PS including VAOs and accounting for such photon diffusion via window functions.

\begin{figure}
	\includegraphics[width=0.48\textwidth]{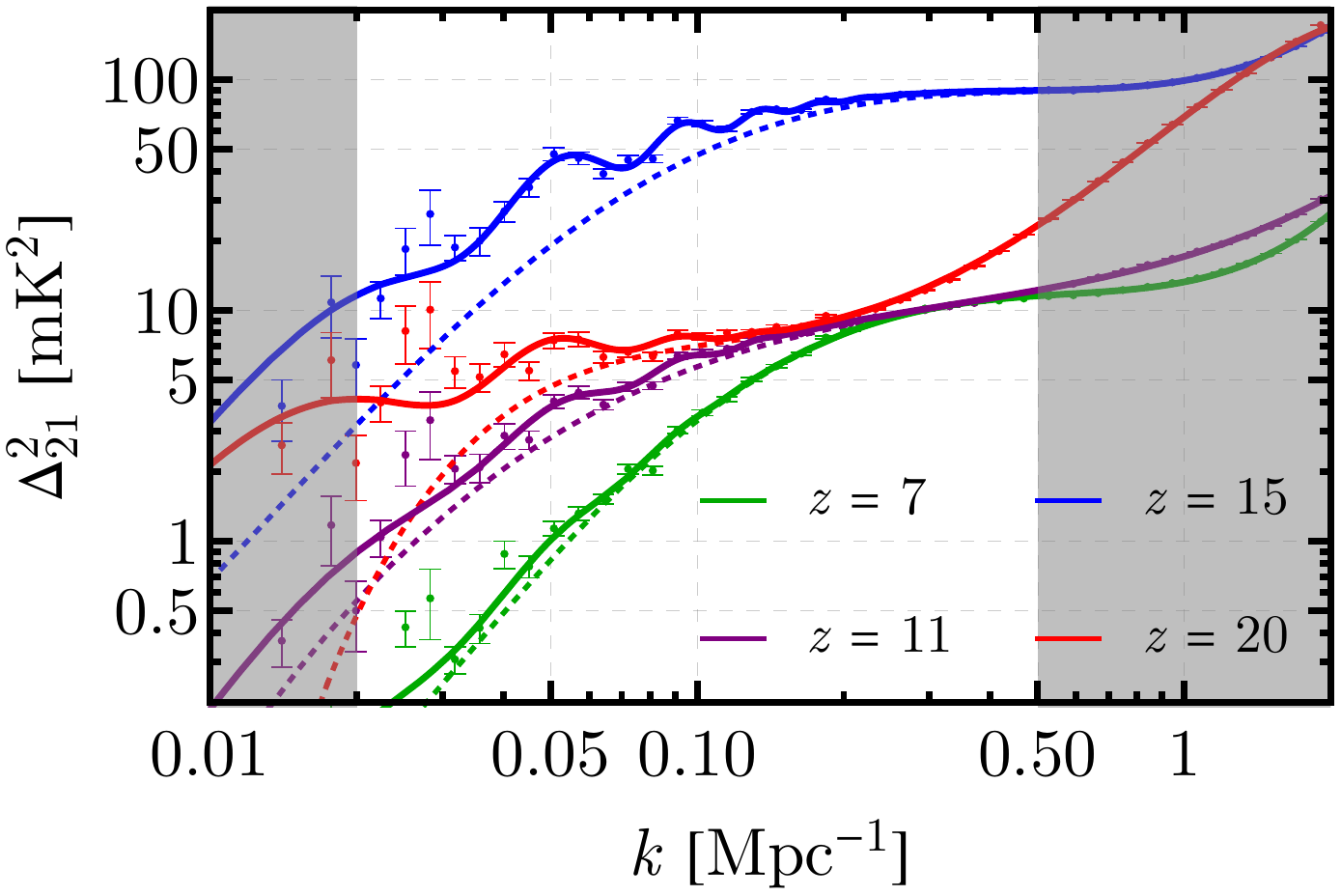}
	\caption{The 21-cm power spectrum as a function of wavenumber $k$ at four redshifts $z$.
	The data-points show our simulation results (and Poisson noise) for the OPT parameters, and the lines present fits obtained using Eq.~\eqref{eq:VAOfit}.
	The solid line contains VAOs, whereas the dashed line does not (i.e., it has $b_{\vcb}=0$), which has lower power at large scales and no wiggles.
	Simulation data-points in the gray shaded regions have not been included in the fit.
	The VAOs are most obvious during the EoH ($z\approx 10-15$), are somewhat visible during the EoC ($z\gtrsim 15$, though suppressed at high $k$), and very small during the EoR $z\lesssim 10$.
	}	
	\label{fig:21cmPS_VAOs_a0}
\end{figure}

\subsubsection*{Power Spectrum}

The simulation slices studied above give us an idea of the impact of the relative velocities on the 21-cm signal at different epochs.
We  now calculate the 21-cm PS as a function of $k$, quantifying the observable signature of the VAOs.
We plot this observable from our simulations in Fig.~\ref{fig:21cmPS_VAOs_a0} at the same redshifts as were shown in Figs.~\ref{fig:relvel_4panel} and \ref{fig:relvel_zoomonly}.
The amplitude of those power spectra trace the overall redshift evolution that we studied in Sec.~\ref{sec:CDEoR}.
However, more interestingly from the point of view of VAOs is the shape of the PS with $k$.
The simulation data points show marked acoustic oscillations (i.e., wiggles) at $k=0.05-0.5\,\Mpcinv$, inherited from the $\vcb$ fluctuations.
These VAOs are most pronounced during the X-ray heating era, increasing the power spectrum by an $\mathcal O(1)$ factor both at $z=11$ and $z=15$.
They also appear during the EoC, at $z=20$, and to a much lesser degree in the EoR at $z=7$ (though we do not consider the effect of $\vcb$ on ionizing sinks, as described in~\citep{Park:2020ydt,Cain:2020npm}, which may enhance the late-time VAOs).

The relative velocity $\mathbf v_{\rm cb}$ is a vector field, so due to isotropy it can only affect observables through $v_{\rm cb}^2$ to first order.
We define the VAO shape $\Delta^2_{\vcb}$ to be the power spectrum of
\begin{equation}
    \delta_v = \sqrt{\dfrac{3}{2}} \left(\dfrac{v_{\rm cb}^2}{v_{\rm rms}^2} - 1\right).
    \label{eq:deltavcb}
\end{equation}
This quantity has unit variance, zero mean, and is redshift independent.
In \citet{Munoz:2019rhi} we showed that the shape of the VAOs is unaltered by the complex astrophysics of cosmic dawn\footnote{This is not true for all scenarios, as for instance the sharp cutoffs in the primordial-black-hole accretion model of \citet{Jensen:2021mik} do not always follow the VAO shape.}, although its amplitude is damped if the X-ray or UV photons that affect the 21-cm signal travel significant distances (comparable to $1/k$).
Thus, not only does the amplitude of the VAOs change between eras, but also the $k$ range where they are visible.
This is clear from Fig.~\ref{fig:21cmPS_VAOs_a0}, as the $z=15$ power spectrum has $\sim 3-4$ visible wiggles, whereas at $z=20$ only 2 can be distinguished.
That is because during the EoC the mean-free path of the relevant Lyman band photons is rather large ($\sim 100$ Mpc; \citealt{Dalal:2010yt}), whereas during the X-ray heating era it is much shorter for realistic SEDs \citep{Pacucci:2014wwa, Das:2017fys}.
This is especially true for our OPT set of parameters, which has an optimistic value of $E_0=0.2$ keV and thus X-rays travel shorter distances.
For the EOS parameter set on the other hand, we set $E_0=0.5$ keV (see \citealt{Das:2017fys}), resulting in longer X-ray mean free path and thus more damped VAOs (c.f.~Fig.~\ref{fig:P21_k_EOS}, where only 2-3 wiggles are apparent).

In order to analytically model the VAOs we follow the approach of \citet[][see also~\citealt{Hotinli:2021xln}]{Munoz:2019fkt,Munoz:2019rhi}, and use the fact that density- and $\vcb$-sourced fluctuations are uncorrelated to first order to write
\be
\Delta^2_{21} = \mathcal P_{\rm nw}(k) + b_{\vcb}^2 W_i^2(k) \Delta^2_{\vcb}(k),
\label{eq:VAOfit}
\ee
where $\Delta^2_{\vcb}$ is the power spectrum of $\delta_v$ in Eq.~\eqref{eq:deltavcb} and contains the VAO shape, $b_{\vcb}$ is a bias factor that determines its amplitude, and $W_i$ is a window function that accounts for photon propagation.
The $\mathcal P_{\rm nw}(k)$ ``no-wiggle" term, instead, accounts for the usual density-sourced 21-cm fluctuations, and we model it as a simple 4-th order polynomial,
\begin{equation}
    \log \mathcal P_{\rm nw}(k) = \sum_{j=0}^4 c_j \left[\log(k)\right]^j,
\end{equation}
which suffices to capture its behavior in the region of interest ($k=0.02-0.5\,\Mpcinv$).

For the ``wiggle" VAO part we know $\Delta^2_{\vcb}$, but need to find both the window function $W_i$ that accounts for damping of VAOs, and the bias $b_{\vcb}$ that parametrizes their amplitudes. Let us begin with the window function.

\begin{figure}
	\includegraphics[width=0.48\textwidth]{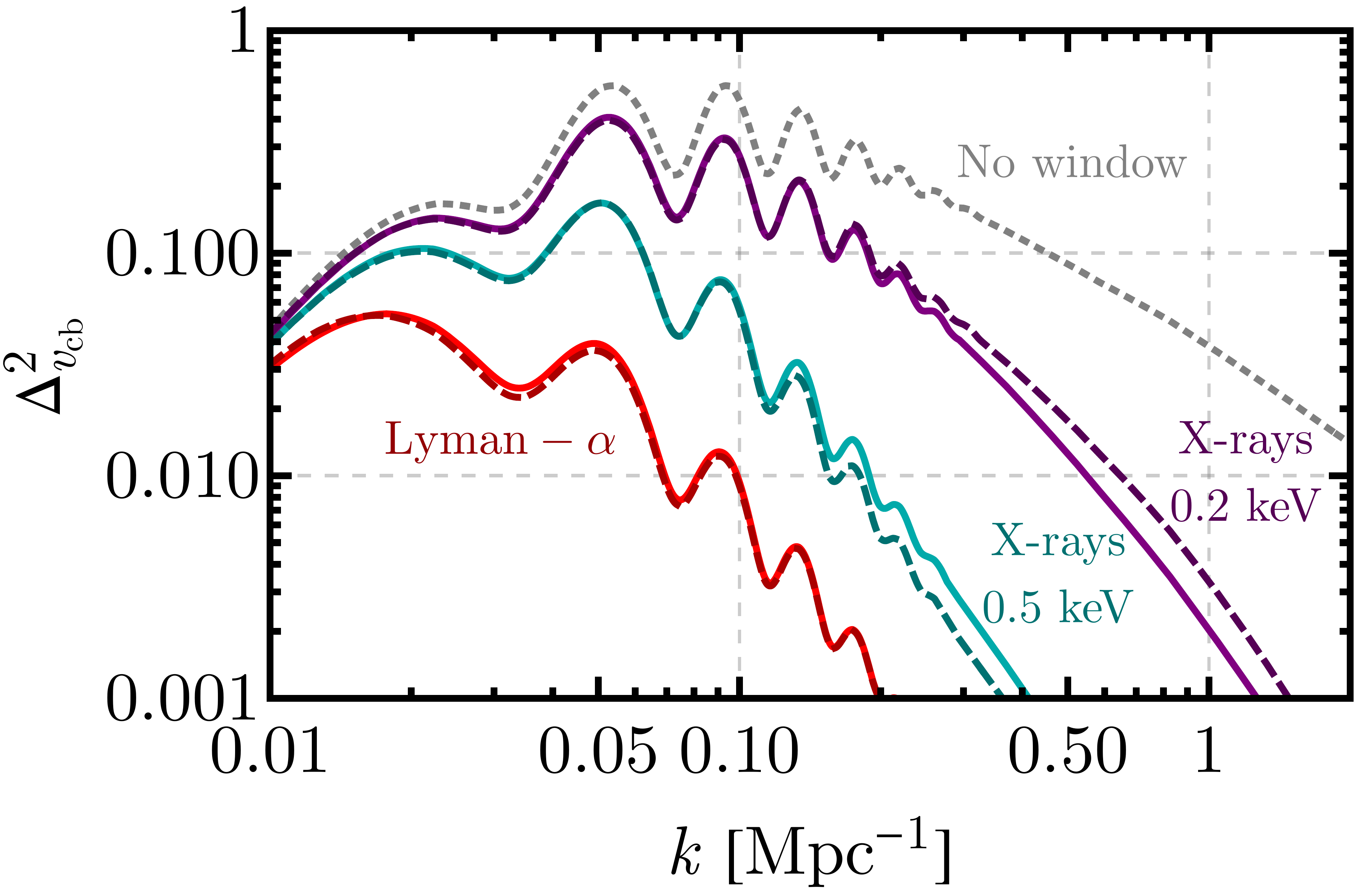}
	\caption{Relative-velocity power spectrum, from Eq.~\eqref{eq:deltavcb},
	in the case that it is not modulated by any window function (gray dotted), as well as damped by the mean free path of X-rays with a cutoff at 0.2 keV (purple), 0.5 keV (teal), and of the UV photons that produce WF coupling upon re-entering the Lyman-$\alpha$ transition (red).
	For these three cases, dashed lines show the numeric result from \citet{Munoz:2019rhi}, whereas the solid lines follow the simpler fit presented in Eq.~\ref{eq:Win_VAO_fit}. 
	}	
	\label{fig:Pv_window}
\end{figure}

We follow the approach of \citet{Munoz:2019rhi} and define separate window functions $W_i(k)$ for the EoC ($i=\alpha$) and the EoH ($i=X$).
Rather than computing them numerically, as is done in \citet{Dalal:2010yt,Munoz:2019rhi}, here we use a parametrized form that provides a good fit across the entire range of interest.
We write
\be
W_i(k) = \left[\dfrac{1}{1+(k/k_{{\rm cut},i})^{\beta_i}
} \right]^{1/\beta_i},
\label{eq:Win_VAO_fit}
\ee
which has two free parameters ($\beta_i$ and $k_{{\rm cut},i}$) for each era, fitted to the results of \citet{Munoz:2019rhi} but kept independent of $z$ otherwise.
We find that for Lyman-$\alpha$ photons $\beta_a = 2$ and $k_{{\rm cut},a}=0.015\,\Mpcinv$ provide an excellent fit.
For X-rays the damping depends on the assumed energy cutoff scale (i.e., the minimum X-ray energy $E_0$ escaping the ISM of host galaxies).
First, for $E_0 = 0.2$ keV (as set in our OPT parameters), a good fit is provided by $\beta_X = 1$ and $k_{{\rm cut},X}=0.3\,\Mpcinv$.
Second, when setting $E_0 = 0.5$ keV (for the EOS fiducial) the X-rays  have longer mean free paths, and we find
$\beta_X = 1.5$ and $k_{{\rm cut},X}=0.04\,\Mpcinv$.
We show the VAO power spectrum in Fig.~\ref{fig:Pv_window} multiplied by each of these window functions, together with the numerical result from \citet{Munoz:2019rhi}, finding good agreement.
This figure also shows that, as expected, longer travel distances yield more suppression of VAO amplitudes.  The Lyman-$\alpha$ VAOs are more damped than X-rays with $E_0=0.5$ keV, which in turn are more damped than X-rays with a lower-energy cutoff at $E_0=0.2$ keV.
Numerically, at $z=15$ the distance a Lyman-$\beta$ photon travels until entering the Lyman-$\alpha$ resonance is roughly 300 Mpc comoving, whereas the mean-free path of X-rays is a significantly shorter 30 Mpc for $E_0=0.5$ keV,  or 3 Mpc  for 0.2 keV~\citep{McQuinn:2012bq}.
As a consequence, these
latter cases have higher $k_{{\rm cut},i}$, and a shallower suppression index $\beta_i$.
We emphasize that the parameters of this fit would technically vary with $z$, and have not been fit to high precision, but instead to round numbers, as that suffices for our purposes of studying the detectability of VAOs and their extraction from simulated power-spectrum data.

The amplitude of the VAOs, parametrized in our analytic PS expression through the bias $b_{\vcb}$, depends on the astrophysics driving the 21-cm signal at any given epoch.
Star formation feedback from streaming velocities preferentially impacts smaller scales, and thus MCGs are more affected than ACGs  (though see Sec.~\ref{sec:FirstGalaxies} for their impact on ACGs).
As a consequence, $b_{\vcb}$ is larger (i.e., with more prominent VAOs) for models in which the SFRD is driven by the smallest halos.
In Fig.~\ref{fig:21cmPS_VAOs_a0} we assume the optimistic (OPT) parameters for the MCG SFR from Table~\ref{tab:Fids}, and thus VAOs are evident during both the EoH and EoC.

\begin{figure}
	\includegraphics[width=0.48\textwidth]{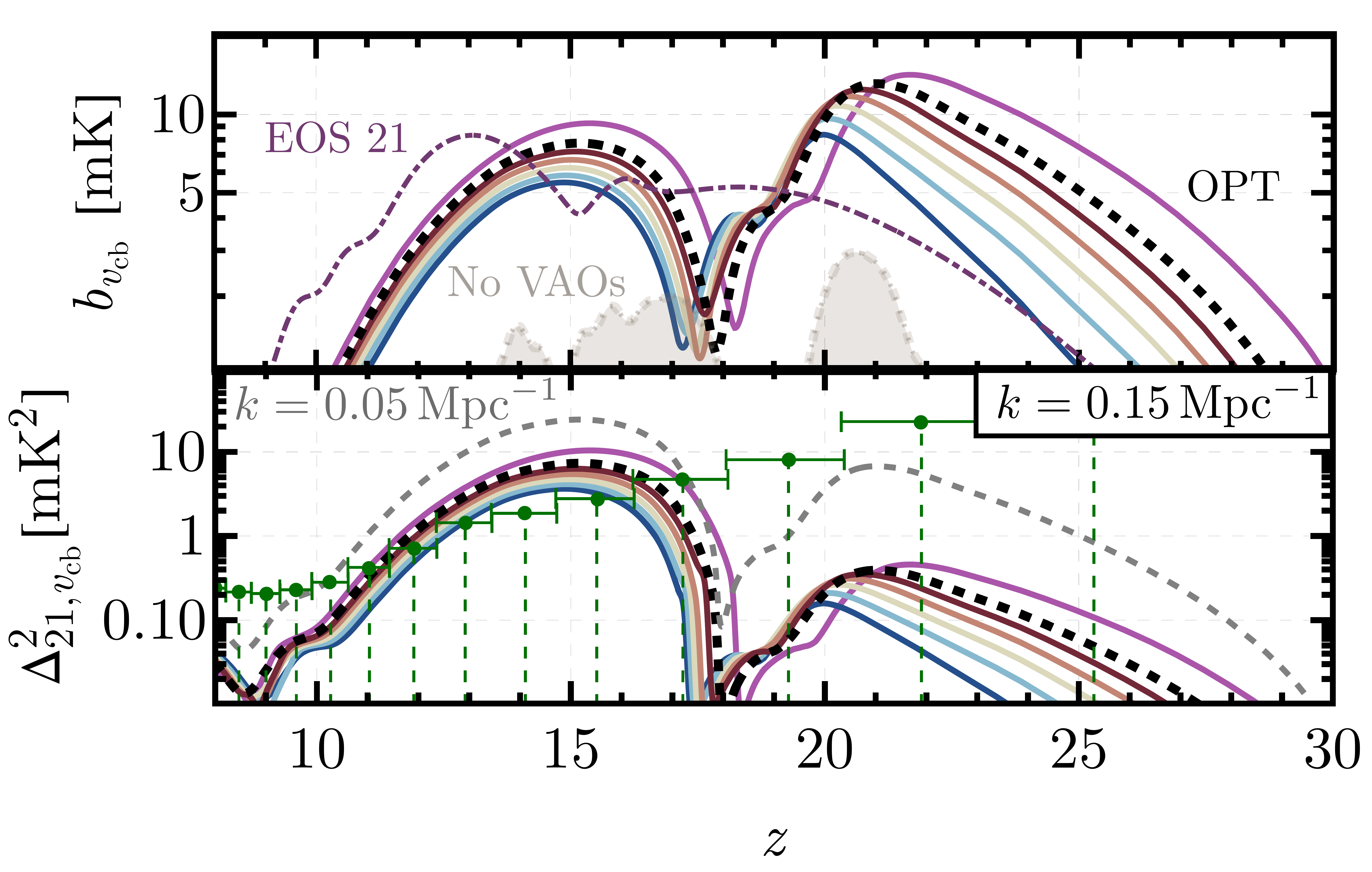}
	\caption{Size of the VAOs under different assumptions.
	As before, the black dashed line shows our fiducial SHMR for MCGs, with an index $\alpha_*^{\rm (III)}=0$, whereas the red-blue and pink lines represent different values of that index, following the same convention as Fig.~\ref{fig:T21alphas}.
	The {\bf top} panel shows the velocity bias $b_\vcb$, defined in Eq.~\eqref{eq:VAOfit}, where the shaded brown region is the result of fitting a simulation with no VAOs (fixing $\vcb=v_{\rm avg}$), and can be taken as an error estimate of the fitting procedure.
	The purple dash-dotted line shows the results for the EOS parameters, rather than OPT.
	The {\bf bottom} panel shows the VAO-only power spectrum at $k=0.15\,\Mpcinv$, in the range observable by HERA (whose noise is shown in green for $k=0.05-0.15\,\Mpcinv$).
	We additionally show the power in our fiducial ($\alpha_*^{\rm (III)}=0$) case at $k=0.05\,\Mpcinv$, as a dashed gray line, which grows significantly for the EoC ($z>17$) as VAOs are less damped at smaller $k$ during that epoch.
	}	
	\label{fig:bias_VAOs}
\end{figure}

We fit for the bias parameter $b_{\vcb}$ for each redshift independently  over the $k$-ranges shown in white in Fig. \ref{fig:21cmPS_VAOs_a0}.  The results are shown in Fig.~\ref{fig:bias_VAOs}.
The amplitude of VAOs has two peaks, corresponding to the EoC (at $z\sim 20$ for our OPT parameters) and the EoH ($z\sim 15$), and all but vanishes in the transition between the two, as the effect of $\vcb$ goes in opposite directions between these two eras, producing less coupling at earlier times (and thus higher $T_{21}$), and less heating at late times (lower $T_{21}$). 
We also show $b_{\vcb}$ for our EOS fiducial ({\it AllGalaxies}) simulation in that Figure, which shows somewhat smaller VAOs, delayed to later times; as well as a null test $b_{\vcb}$ for a simulation with no VAOs as an error-bar estimation of our fitting procedure.
For comparison, the HERA~\citet{HERA:2021noe} found that $b_\vcb<\{50,180\}$ mK at 95\% CL, using their phase-1 limits at $z=\{8,10\}$.
These data only cover lower redshifts, where VAOs are not expected to be important (c.f.~Fig.~\ref{fig:bias_VAOs}), but they highlight the need for further sensitivity to reach the level of VAOs ($b_\vcb\approx 10$ mK) predicted in our models.

In the bottom panel of Fig.~\ref{fig:bias_VAOs} we plot the $\vcb$-only component of the 21-cm power spectrum, defined as $\Delta^2_{21, \vcb} = b_{\vcb}^2 W_i^2(k) \Delta^2_{\vcb}(k)$,
at a scale $k=0.15\,\Mpcinv$ (roughly corresponding to a ``sweet spot" in terms of foreground contamination and thermal noise for interferometers; e.g., \citealt{Greig:2020ska, HERA:2021noe,vanHaarlem:2013dsa,Tingay:2012ps}).
This VAO power is relatively high during the EoH, reaching $\Delta^2_{21,\vcb}\approx 10$ mK$^2$.
During the EoC, however, it only has values $\Delta^2_{21,\vcb}\approx 0.5$ mK$^2$, as the photon propagation in the latter strongly suppresses large-$k$ fluctuations.
To illustrate this point, we also plot the VAO-only power for $k=0.05\,\Mpcinv$ (where the deepest LOFAR limits lie~\citealt{Mertens:2020llj}) in that Figure, which grows by nearly two orders of magnitude during the EoC (and one during the EoH), showing that reaching lower $k$ by careful foreground cleaning is ideal for detecting acoustic wiggles in the high-$z$ 21-cm signal.
We predict a smaller VAO power across all of cosmic dawn than our previous work.
In \citet{Munoz:2019rhi} we had found $\Delta^2_{21,\vcb}\approx 50$ mK$^2$ during the EoH, a factor of a few larger.
Part of the reason is the inclusion of inhomogeneous LW feedback, which tends to suppress VAOs~\citep{Fialkov:2012su}.
The largest factor, however, is the new parametrization of the SFRD (see Eqs.~\ref{eq:SFRDII},\ref{eq:SFRDIII}).
In \citet{Munoz:2019rhi} we considered a mass-independent SHMR shared for PopII and PopIII stars (i.e.,~$\alpha_*^{(i)}=0$ for $i=$ II and III) , which produces a much faster evolution of CD and larger 21-cm fluctuations (see discussion in Sec.~\ref{sec:CDEoR}).
With the more-realistic SHMR considered here both the overall 21-cm PS and the VAOs are smaller, so VAOs are still an $\mathcal O(1)$ component of the large-scale 21-cm power spectrum in Fig.~\ref{fig:21cmPS_VAOs_a0}.

\subsection{VAOs and the SHMR of PopIII hosts}

So far we have shown VAOs in simulations with either our OPT and EOS parameter sets.
Given our lack of knowledge about cosmic dawn, however, the first galaxies could have much different parameters than we expected.
We now perform a brief exploratory study of how the amplitude of the VAOs can be used to learn about the astrophysics of cosmic dawn.

We focus on the slope of the SHMR for MCGs, parametrized with $\alpha_*^{\rm (III)}$, which we showed in the previous section has a very modest impact on the redshift dependence of the PS (at fixed $k$) and the GS.  Here we study its impact on the VAO component of the PS, shown in Fig.~\ref{fig:bias_VAOs} for the same values of $\alpha_*^{\rm (III)}$ as in Fig.~\ref{fig:T21alphas}.
We see that steeper SHMRs (larger $\alpha_*^{\rm (III)}$) suppress the VAO amplitude, especially at high $z$.  This is understandable since steeper SHMRs decrease the relative contribution of the smallest halos to the SFRD, and these smallest halos are the most sensitive to the streaming velocities.
Interestingly, the impact of $\alpha_*^{\rm (III)}$ on the VAOs component of the power appears more noticeable than in the overall 21-cm power spectrum or global signal (c.f.~Fig.~\ref{fig:T21alphas}).  Therefore, the amplitude of VAOs can provide a cleaner view of the halo - galaxy connection of MCGs.

We also study variations of the amplitude $A_{\rm LW}$ of the LW feedback.
We find that increasing or decreasing $A_{\rm LW}$ by a factor of up to 3 does not change the amplitude of the VAOs, only its $z$ dependence.
This is because the LW and $\vcb$ feedback effects multiply coherently, so they are rather independent. 
Given that the 21-cm GS and the complete PS amplitude are also insensitive to $A_{\rm LW}$ (see Fig.~\ref{fig:paramvary}), the best avenue for studying this parameter may be further hydrodynamical simulations, instead of inferring it from 21-cm data.

\subsection{Detectability}

The VAOs that we study here have been shown to be a robust standard ruler during cosmic dawn, allowing 21-cm interferometers to measure the cosmic expansion rate at $z\sim 10-20$~\citep{Munoz:2019fkt}.
However, we would first need to detect them.

As opposed to Sec.~\ref{sec:CDEoR}, where we considered the entire 21-cm power spectrum, here we forecast SNRs for the $\Delta^2_{21,\vcb}$ component.
For that we use the same noise as before, which includes the cosmic variance from the full $\Delta^2_{21}$ signal.
This noise is shown in Fig.~\ref{fig:bias_VAOs} along with the VAO-only power spectrum.
We find a SNR $=5$ for the EOS parameters assuming 1080 hours of HERA data at moderate foregrounds.
For the OPT parameters, instead, we find more optimistic estimates, with SNR = $9$.
In both cases the SNR is only above unity over the range $z\approx10-15$, showing that the EoH is the most promising epoch to detect VAOs, and thus to measure $H(z)$~\citep{Munoz:2019fkt}.

\subsection{VAOs in the lightcone}

All work on VAOs thus far has been on co-eval boxes (i.e., at fixed $z$).
In reality, however, 21-cm fluctuations are measured in the lightcone, as the fluctuations along the line-of-sight (LoS) direction evolve with $z$.
This is particularly important for using VAOs as a standard ruler, as mainly LoS modes are observed by interferometers, which are then used to measure $H(z)$.

\begin{figure}
	\includegraphics[width=0.48\textwidth]{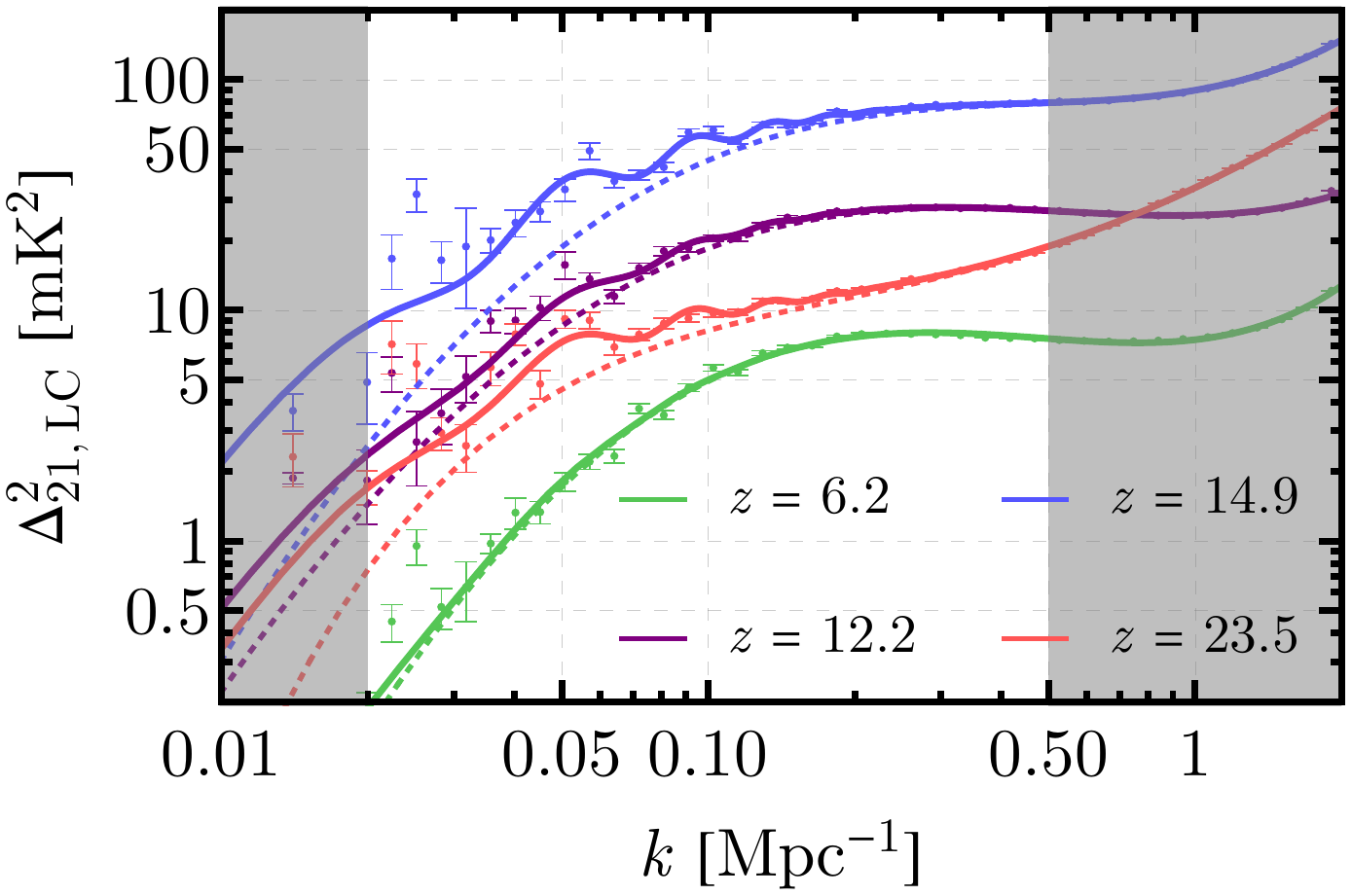}
	\caption{Power spectra in the lightcone (i.e., including LoS effects) at four bins marked by the central redshift, where each bin has a LoS comoving width of 383 Mpc, as well as their fits using Eq.~\eqref{eq:VAOfit}.
	This figure shows that VAOs are still present in the lightcone (LC) power spectra.
	}	
	\label{fig:LC_VAOs}
\end{figure}

It is expected that LoS effects slightly change the large-scale 21-cm power~\citep{Datta:2011hv,LaPlante:2013jhi,Ghara:2015yfa,Greig:2018hja}.
In order to include these effects we divide our lightcone, which is 600 Mpc on a side and $3830$ Mpc in length, into 10 blocks (each $383$ Mpc along the line of sight), and compute the power spectrum in each of them.
We show the resulting full-lightcone 21-cm power spectra in Fig.~\ref{fig:LC_VAOs}, where the VAOs are still clearly apparent (with larger Poisson noise from the simulations as each $z$ box corresponds to a smaller comoving volume).
While a full comparison of lightcone and co-eval boxes is beyond the scope of this work, we find that during the EoH the bias $b_{\vcb}$ is reduced by 20\%, showing a small but not negligible LoS effect on the amplitude of VAOs, though not on their shape.
This is an important cross-check for using VAOs as a standard ruler.
For a brief study of how to measure VAOs without Poisson variance from the simulations we encourage the reader to visit Appendix~\ref{app:VAODiff}.

\section{Discussion and Conclusions}
\label{sec:Conclusions}

The first generation of PopIII stars 
heralded the transition from the dark ages to the cosmic dawn.
In this work we have improved the treatment of these stars in the public code {\tt 21cmFAST}, including the combined impact on star formation from baryon-DM streaming velocities ($v_{\rm cb}$) and the inhomogeneous LW background.
In our model, PopIII stars are hosted in molecular-cooling galaxies (MCGs, with halo masses $M_h <M_{\rm atom}\sim 10^{7-8}\,\Msun$), whereas PopII stars form in atomic-cooling galaxies (ACGs, with masses above $M_{\rm atom}$).
Thus, PopIII stars dominate the photon budget in the early universe, and are expected to set the timing of the cosmic-dawn 21-cm signal at $z\sim 10-20$.
Later on, however, feedback and the natural appearence of heavier haloes will make MCGs subdominant, and PopII stars are expected to drive cosmic reionization, at $z\sim5-10$.

Many questions remain about the formation of the first galaxies.
As such, here we implement flexible models that account for two distinct (PopII and III) stellar populations with different SHMRs, building upon~\citet{Park:2018ljd,Munoz:2019rhi,Qin:2020xyh}.
We provide generic fitting formulae for the impact of LW and $\vcb$ feedback (in Sec.~\ref{sec:FirstGalaxies}), and calibrate them with results from state-of-the-art hydrodynamical simulations.
This allows us to generate a new EOS 2021 model covering the evolution of the 21-cm signal across cosmic dawn and reionization, which encapsulates our current knowledge of these epochs.
We have dubbed this model {\it AllGalaxies}, and it enhances the previously available {\it Bright} and {\it FaintGalaxies} (from EOS 2016) by including PopIII-hosting MCGs.
The parameters of this {\it AllGalaxies} model are chosen to give rise to late reionization (finishing at $z\approx 5.5$, as expected from recent Lyman-$\alpha$ forest data), and its  CMB optical depth is $\tau_{\rm CMB}=0.063$, in line with recent determinations from {\it Planck}.
The 21-cm signal in our {\it AllGalaxies} simulation is shallower (only reaching $\overline T_{21}\approx -75$ mK during cosmic dawn) than in the previous EOS models.
This is due to both the inclusion of MCGs and 
a steeper SHMR for ACGs, the later being required to match UVLF observations.
As a consequence, both the expected 21-cm global signal and its fluctuations will be more difficult to detect (see also \citealt{Mirocha2016_UVLF_GS,Park:2018ljd}).
Nevertheless, we expect both the HERA and SKA interferometers to 
reach a 21-cm power-spectrum detection of this model at a  SNR $\approx 200$ with 1000h of integration.
We make public the detailed lightcones of this model, as well as associated visualizations.

We also performed an exploratory study of how PopIII stars affect the cosmological 21-cm signal.
We have found that for the redshifts of interest for MCGs ($z\gtrsim 12$), $\vcb$ feedback likely dominates over LW feedback (cf.~Fig.\ref{fig:Mturns}).
As a consequence, the amplitude $A_{\rm LW}$ of LW feedback only has a modest impact on the 21-cm signal.
Similarly, the signal is not sensitive to the escape fraction $f_{\rm esc}^{\rm (III)}$ of ionizing photons from MCGs, as MCGs do not significantly contribute to reionization in our model.
On the other hand, the star formation efficiency and the X-ray luminosities of MCGs do impact the 21-cm signal significantly.

The streaming velocities $\vcb$
fluctuate spatially with an acoustic signature inherited from recombination.
As a consequence, the distribution of the first galaxies (and thus the 21-cm signal) shows velocity-induced acoustic oscillations (VAOs): large wiggles in their power spectrum at $k\lesssim 0.1\Mpcinv$.
We showed that VAOs are present and detectable even when including inhomogeneous LW feedback, and considering lightcone effects.
This will allow us to use the 21-cm  power spectrum as a standard ruler during cosmic dawn.
Moreover, the amplitude of the 21-cm VAO oscillations can be used to study the SHMR of MCGs.  The slope of the SHMR can provide insight about the  stellar content of MCGs and the associated feedback mechanisms; without VAOs, this important quantity would be very difficult to detect.

Our results and public simulations can be used to guide 21-cm observing strategies and data pipelines.  
With our current state of knowledge we expect the 21-cm power spectrum to be detected at high significance by upcoming interferometric observations.
Such a detection will provide us with a new window on the stellar and energy content of our cosmos at unprecedented early times.

\begin{table*}
\begin{tabular}{l|llll}
Name & PopII parameters & PopIII parameters               & Size {[}Mpc{]} &  \\
\hline
\tt EOS2021$^*$ (AllGalaxies)                                                 & EOS 2021 (EOS)               & EOS                              & 1500            &  \\
\tt 600\_EOS  & EOS  & EOS  & 600             &  \\
\tt 600\_novel  & EOS  & EOS ($A_{v_{\rm vcb}}=0$) & 600            &  \\
\tt 600\_noMINI  & EOS  & None & 600            &  \\ 
\tt 600ptX  &  Optimistic (OPT)       & OPT ($\alpha_*^{(\rm III)}=$ X, for $-0.2$-0.5) & 600            &  \\ 
\tt 600alwY                                             & OPT               & OPT ($A_{\rm LW}=Y$)             & 600            &  \\
 $\tt 400\_i\_hi^*$                                          & OPT               & OPT ($\theta_i\times3$)         & 400            &  \\
$\tt 400\_i\_lo^*$                                         & OPT               & OPT ($\theta_i/3$)              & 400           &  
\end{tabular}
\caption{Table with the simulations that we make publicly available. The EOS fiducial has a more conservative SHMR for MCGs compared to ACGs, as opposed to our Optimistic (OPT) fiducial (see Tab.~\ref{tab:Fids} for parameter values).
All simulations have a resolution of 1.5 Mpc per cell, and an asterisk denotes that we make full lightcones available, in addition to global quantities and power spectra.
\label{tab:simnames} }
\end{table*}

\subsection*{Data Availability}

All the simulations presented in this work are freely available for download (see footnote~\ref{footnote:linkEOS}).
We make lightcones from the {\it AllGalaxies} simulation (initial conditions, perturbed densities, relative velocities, spin and kinetic temperatures, ionization fractions, and 21-cm brightness temperatures) publicly available.
In addition to the {\it AllGalaxies} simulation (1.5 Gpc in size), we share the output of the simulations used in Sec.~\ref{sec:CDEoR} (which are either 600 or 400 Mpc in size, at the same resolution).
We compile the names of the simulations, and their parameters, in Table~\ref{tab:simnames}.

\section*{Acknowledgements}

We are grateful to Greg Bryan, Daniel Eisenstein, and Jordan Mirocha for enlightening discussions.
JBM is supported by a Clay fellowship at the Smithsonian Astrophysical Observatory. A.M. acknowledges funding from the European Research
Council (ERC) under the European Union’s Horizon 2020
research and innovation programme (grant agreement No
638809 – AIDA). The results presented here
reflect the authors’ views; the ERC is not responsible for
their use. Parts of this research were supported by the Australian Research Council Centre of Excellence for All Sky Astrophysics in 3 Dimensions (ASTRO 3D), through project number CE170100013.  We gratefully acknowledge computational resources of the Center for High Performance Computing (CHPC) at Scuola Normale Superiore (SNS).
This work was performed in part at the Aspen Center for Physics, which is supported by National Science Foundation grant PHY-1607611.
This work was partially supported by a grant from the Simons Foundation.
C.A.M. acknowledges support by the VILLUM FONDEN under grant 37459 and from NASA Headquarters through the NASA Hubble Fellowship grant HST-HF2-51413.001-A awarded by the Space Telescope Science Institute, which is operated by the Association of Universities for Research in Astronomy, Inc., for NASA, under contract NAS5-26555.

\bibliographystyle{mnras}
\bibliography{21cmVAO}

\appendix

\section{Streaming-velocity suppression on the matter power spectrum}
\label{app:Pmattervcb}

In this appendix we show a simple but accurate fit to the suppression on the small-scale matter power spectrum induced by the relative velocities.

\begin{figure}
	\includegraphics[width=0.48\textwidth]{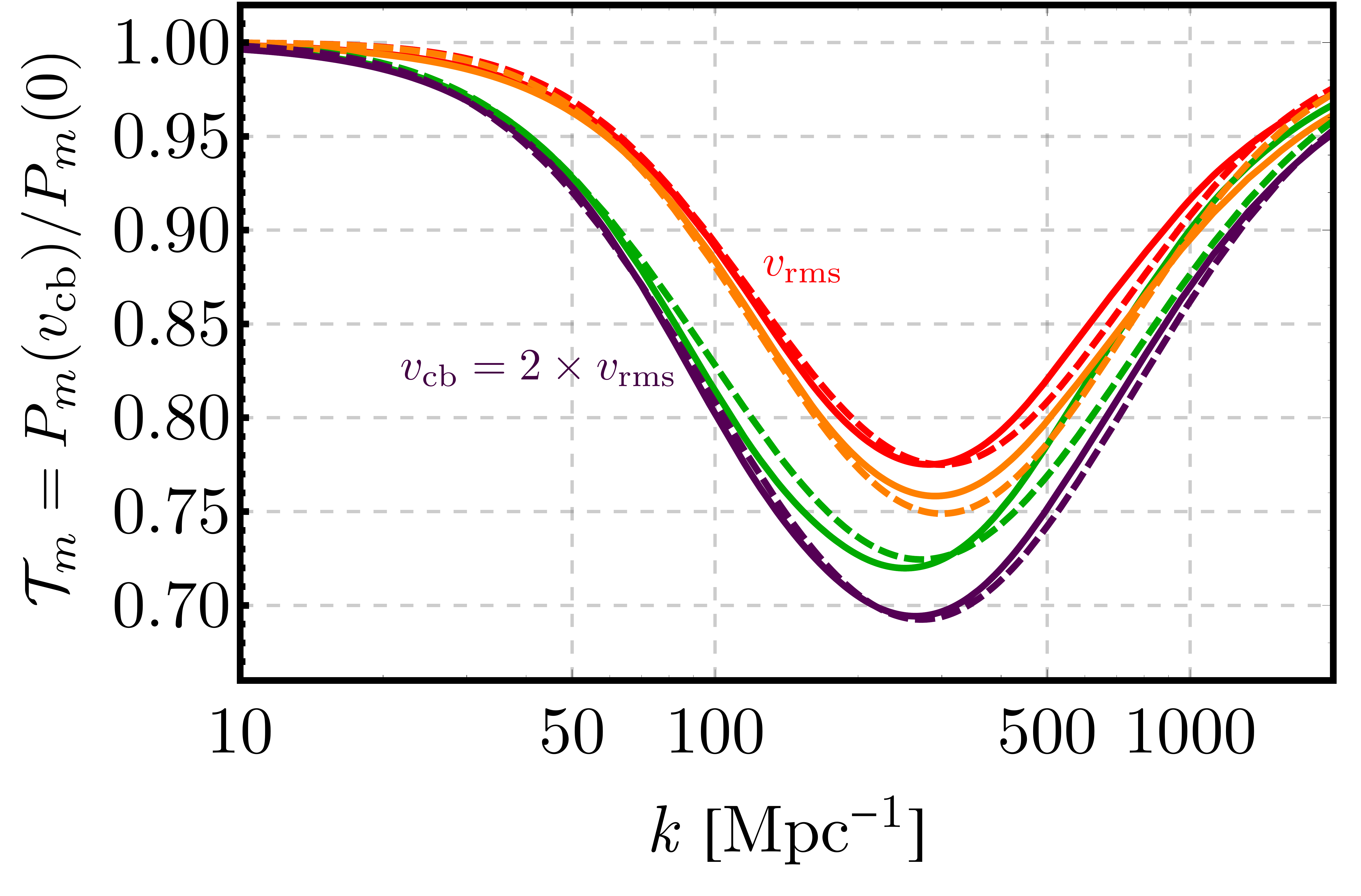}
	\caption{
	Transfer ratio of matter fluctuations at small scales due to the effect of the DM-baryon relative velocities $\vcb$, defined in Eq.~\eqref{eq:AppTmatter}.
	The solid lines show the exact result from solving the ODEs (as in \citealt{Munoz:2019rhi}), whereas the dashed lines show our Gaussian approximation from Eq.~\eqref{eq:Pmatterfit_z_app}.
	The red and green lines are evaluated at $z=30$, whereas the orange and purple are at $z=15$.
    Upper (red and orange) lines have	
	$\vcb=\vrms$, whereas lower(purple and green) ones have $\vcb=2\times\vrms$.
	In the main text we use the $z=20$ result, described in Eq.~\eqref{eq:Pmatterfit}.
	}	
	\label{fig:matterfit}
\end{figure}

Following~\citet{Tseliakhovich:2010bj,Munoz:2019rhi}, we solve for the evolution of the DM and baryon densities for different initial values of $v_{\rm cb}$ and its inner-product cosine $\mu$ with each wavenumber $k$, in order to compute the density fluctuations of baryons and DM at each $z$ and $k$.
From there, we can calculate the matter power spectrum as a function of $\vcb$ by averaging over $\mu$, to find $P_m^2(k,z,\vcb)$~\citep{Tseliakhovich:2010bj}.
We show in Fig.~\ref{fig:matterfit} the ratio of this quantity to its no-velocity counterpart
\be
\mathcal T_m = \dfrac{P_m(k,z;\vcb)}{P_m(k,z;0)}
\label{eq:AppTmatter}
\ee
as well as the fit defined in Eq.~\eqref{eq:Pmatterfit}.
The main feature of these curves is that larger values of $\vcb$ produce a bigger drop in the power spectrum, as the larger relative velocity puts DM and baryons out of phase (or equivalently, it does not allow baryons to collapse to DM overdensities, slowing growth at small scales).
However, at much smaller scales the baryonic fluctuations are damped due to the Jeans pressure, driving $\delta_b$ down for $k>k_J\approx 500\Mpcinv$, and recovering the DM-only fluctuations regardless of $\vcb$.
Put together, these two effect produce a dip at $k\approx 300\Mpcinv$ in Fig.~\ref{fig:matterfit}, which turns back to unity at larger $k$.

We have used Eq.~\eqref{eq:Pmatterfit} to fit this dip as a Gaussian with three parameters (as described in Sec~\ref{sec:FirstGalaxies} of the main text), where
\be
\mathcal T_m(\vcb) =  1 - A_p  \exp \left[ - \dfrac{(\log[k/k_p])^2}{2 \sigma_p^2}, \right],
\ee
and we find the following fits
\ba
A_p(\vcb,z) &= 0.24 \times [(1+z)/21]^{-1/6} \left(\vcb/\vrms\right)^{\alpha_p}, \nonumber\\
k_p(\vcb) &= 300 \,\Mpcinv \exp\left[- 0.2 \,\left(\vcb/\vrms - 1 \right)\right], \nonumber\\
\sigma_p(\vcb) &= 0.8 (\vcb/\vrms)^{1/3},
\label{eq:Pmatterfit_z_app}
\end{align}
for the {\it Planck} 2018 cosmology,
where $\alpha_p=1$ for $\vcb\leq \vrms$ and $\alpha_p=0.5$ otherwise.
This heuristic fit was calibrated at $z=20$, and while neither $k_p$ nor $\sigma_p$ depend on $z$ we see it provides a good fit during the relevant $z$ at cosmic dawn.
To illustrate this point, we show the real transfer function $\mathcal T_m(\vcb)$ in Fig.~\ref{fig:matterfit} at $z=15$ and 30, as well as our fit, where the two are in good agreement.

\begin{figure}
	\includegraphics[width=0.47\textwidth]{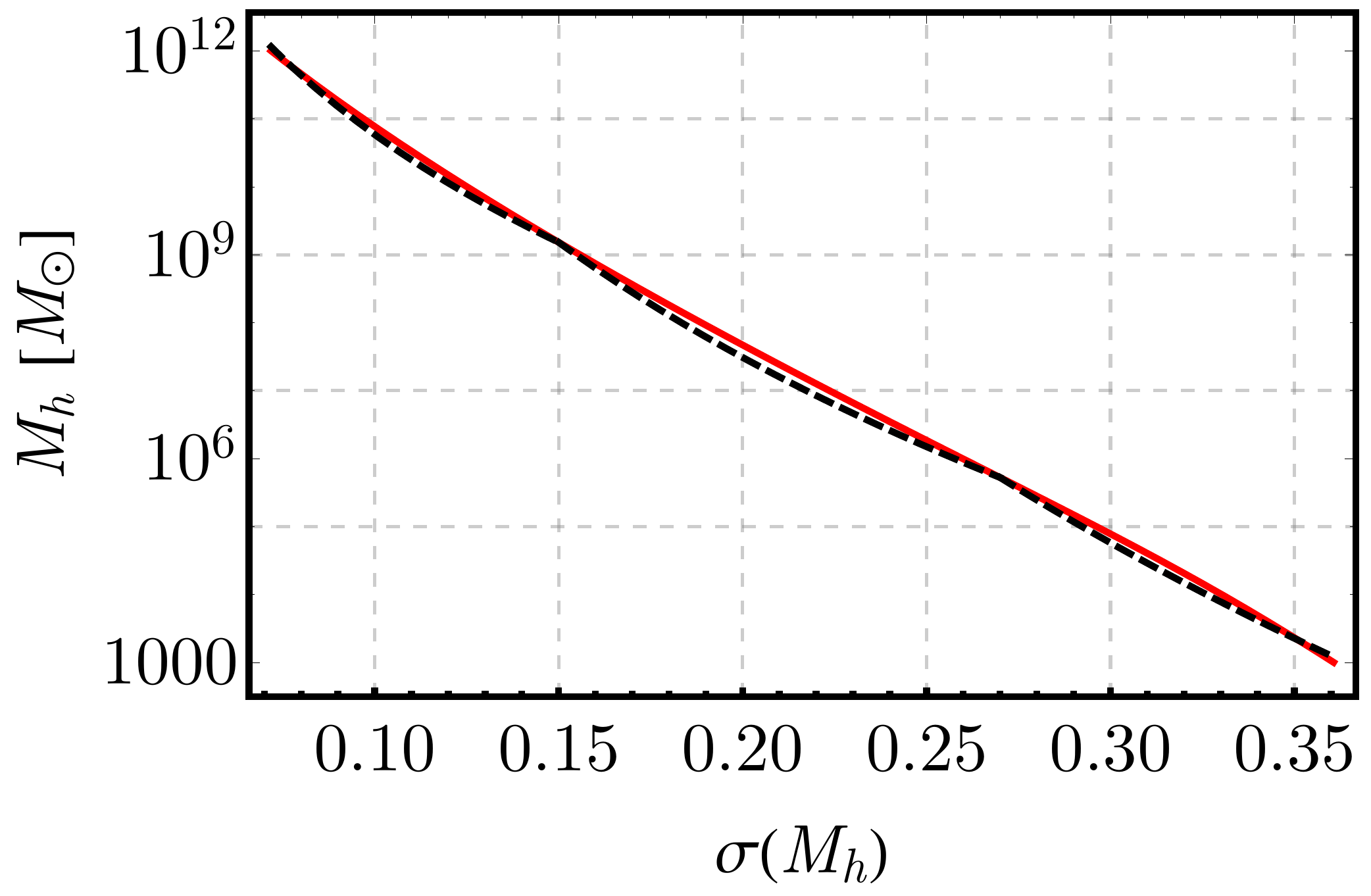}
	\includegraphics[width=0.47\textwidth]{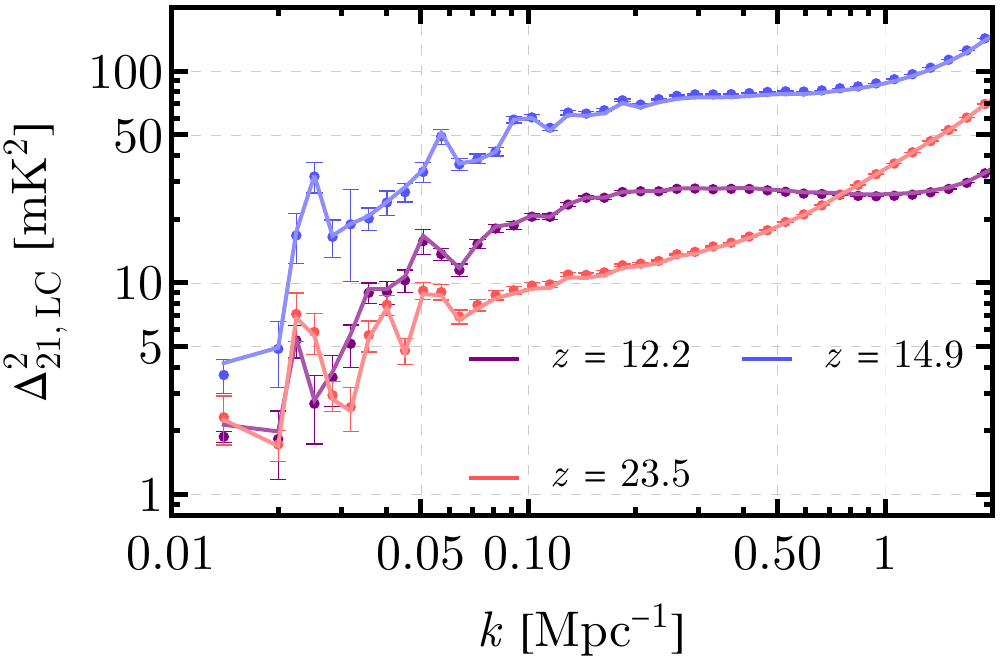}
	\caption{
{\bf Top:}	Relation between the halo mass $M_h$ and the rms fluctuation $\sigma(M_h)$ at their scales. 
	The red line is the result using the full power spectrum, whereas the black dashed line is the broken power-law from Eq.~\eqref{eq:sigmaMvsM}.
	Note that we need to go to low $M$ values (large $\sigma$) since in some cells $\delta_R\approx \delta_{\rm crit}$, which corresponds to $\nu\to0$ (or $\sigma \gg 1$). All $\sigma$ are linearly extrapolated to $z=0$.
	{\bf Bottom:} Lightcone power spectra with the {\tt FAST\_FCOLL\_TABLES} turned on (points with errors) and off (solid line) at three redshifts, for our OPT parameters.
	The different between the two is negligible at high redshifts, and below 10\% at low redshifts.
	The global signal is  identical to the percent level between the two cases.
	}
	\label{fig:MsigmaM}
\end{figure}

\section{Fast SFRD Tables}
\label{app:fasttabs}

The excursion-set algorithm in {\tt 21cmFAST} takes advantage of the extended Press-Schechter approach~\citep{Bond:1990iw} to find the SFRD (and its derived quantities such as the ionizating flux) at a given scale and overdensity.
This requires tabulating the SFRD for many values of the overdensity $\delta_R$ and variance $\sigma_R$ for different radii $R$, so as to not compute it in every cell at every $z$.
This is not extremely computationally expensive for runs with only atomic-cooling galaxies (ACGs, taking approximately $\sim2$ minutes to generate the tables down to $z=6$), though it is for runs with minihaloes hosting molecular-cooling galaxies (MCGs, taking $\sim4$ hours for the same settings), as the tables ought to be generated for many different values of $M_{\rm turn}^{\rm(III)}$.

Here we show an analytic approximation that allows us to compute those tables much faster, cutting the generation time of tables by a factor of $\approx 30$.

In {\tt 21cmFAST} the SFRD (and derived quantities such as UV and X-ray emissivities) is modulated using extended PS theory, 
where we have
\be
{\rm SFRD}^{\rm EPS}(R,\mathbf x) \propto \int dM_h \dfrac{dn}{dM_h}(\mathbf x) \dot M_*(M_h) f_{\rm duty}(M_h),
\label{eq:fcollgen}
\ee
for each cell at $\mathbf x$ and radius $R$ around it (which will enter an integral over previous times).
The HMF depends on position through the overdensity $\delta_R(\mathbf x)$ and the variance $\sigma_R$, and the pre-factor is position-independent so it cancels out when dividing by the average over all cells.
Our assumption is that feedback produces a power-law SHMR, Eqs.~(\ref{eq:SFRDII},\ref{eq:SFRDIII}), so
\be
\dot M_*(M_h) \propto M_h \times (M_h/10^{10} M_\odot)^\alpha
\ee
for a power-law index $\alpha$ (which we can identify with $\alpha_*$ for X-ray and non-ionzing UV fluxes, but for ionizations is $\alpha = \alpha_*+\alpha_{\rm esc}$).
In these calculations the HMF $dn/dM$ is given by the standard old PS formula, as that is the only one for which the EPS formalism has been properly tested.

We will speed up the computation by using an approximate analytic solution.
We know how to compute the integral in Eq.~\eqref{eq:fcollgen} analytically for the case of $\alpha=0$ (so $\dot M_*(M_h)\propto M_h$) with a sharp cutoff (i.e., a Heaviside rather than exponential functional form for $f_{\rm duty}(M_h)$). 
In that case we can use the variable $\nu = \delta_{\rm crit}^2/\sigma^2$ to re-write the integral as
\be
{\rm SFRD}^{\rm EPS}(\mathbf x,z) \!\! \propto \!\!\! \int_{\rm \nu_{\rm min}}^\infty\!\!\! \dfrac{d\nu}{\sqrt{\nu}} e^{-\nu/2}  = 
\sqrt{2\pi}\rm{erfc}\left(\sqrt{\dfrac{\nu_{\rm min}}{2}}\right)
\label{eq:Fcollalpha0}
\ee
where erfc is the complementary error function, and we have defined
\be
\nu_{\rm min} = \delta_{\rm crit}^2/\sigma(M_{\rm turn})^2.
\ee

The EPS formalism~\citep{Bond:1990iw} shows that for regions of radius $R$ that are overdense by $\delta_R$, given a variance of matter fluctuations of $\sigma_R^2$, then all the $\nu$ are corrected to be $\tilde \nu$, defined as 
\be
\tilde \nu = \tilde \delta_{\rm crit}^2/\tilde \sigma^2,
\label{eq:nutilde}
\ee
with $\tilde \delta_{\rm crit} =  \delta_{\rm crit}  - \delta_R$, and $\tilde \sigma^2 = \sigma^2 - \sigma^2_R$.
This allows us to compute the SFRD at an arbitrary point through a simple erfc, as in Eq.~\eqref{eq:Fcollalpha0}.
This expression is exact, and extremely fast to evaluate (compared to performing many numerical integrals).
So, while it cannot be used directly in our tables (as we do not have a Heaviside  $f_{\rm duty}$ or $\alpha=0$ always),
it gives us a good scaffolding upon which to build our analytic solution.

First, we will use a Heaviside  $f_{\rm duty}$ in our EPS, though we keep the usual (exponential) duty factor in the average result that normalizes at each $z$.

Second, we can use the following analytic integral, 
\be
\int_{\rm \nu_{\rm min}}^\infty \dfrac{d\nu}{\sqrt{\nu}} e^{-\nu/2}  \nu^\beta  = 
2^{\beta + 1/2} \Gamma\left(\beta+1/2,\nu_{\rm min}/2\right)
\label{eq:nuint}
\ee
which is a generalization of Eq.~\eqref{eq:Fcollalpha0} adding a power-law $\nu^\beta$.
Then, the trick is to approximate the function
\be
\sigma(M_h)/\sigma_p = (M_h/M_p)^{1/\gamma},
\label{eq:sigmaMvsM}
\ee
for some power-law index $\gamma$ and pivot scale $M_p$ (for which $\sigma(M_p) \equiv \sigma_p$).
In practice, this can only be done over a small range of variances $\sigma$ (or halo masses $M_h$), as otherwise the functional form does not work.
For LCDM models, and the mass range of interest, a good fit is a broken power-law, with $\gamma=9$ for masses below a first pivot ($M_h<M_{p1}=5.3\times 10^5 M_\odot$), $\gamma=21$ for masses above the second pivot ($M_h>M_{p2}=1.5\times 10^9 M_\odot$), and where the index between the two pivots is set to $\gamma=13.6$ by continuity.
This approximates $\sigma(M_h)$ as a broken power-law of $M_h$ to $\sim10$\% precision, as shown in Fig.~\ref{fig:MsigmaM} which suffices for our purposes.
For each of the power-law indices, we take $\beta = \alpha \times \gamma/2$ in Eq.~\eqref{eq:nuint}.

\begin{figure}
	\includegraphics[width=0.48\textwidth]{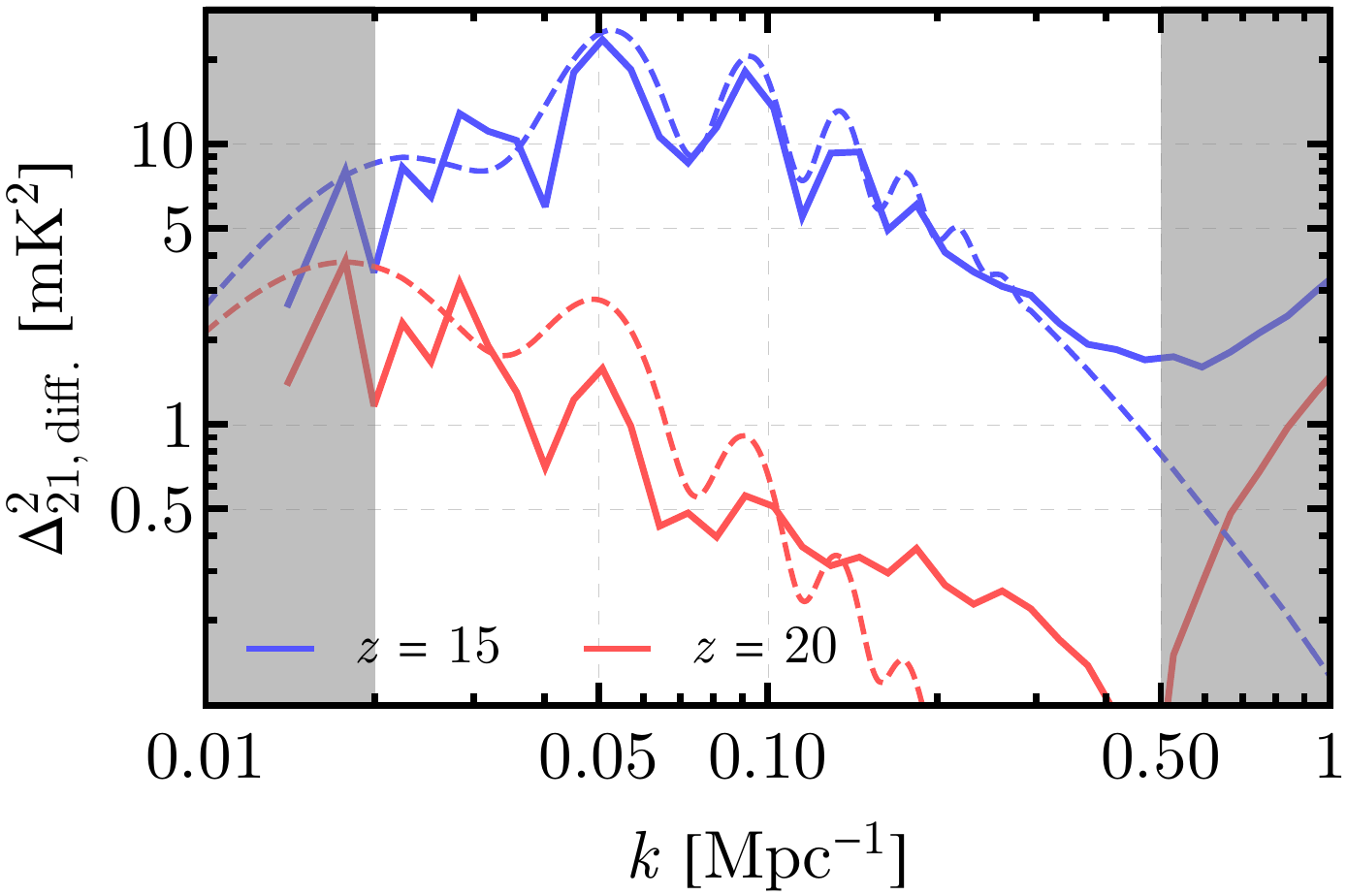}
	\caption{Difference between the full relative-velocity case and one with fixed $\vcb=v_{\rm avg}$ ({\tt FIX\_VAVG} in the code), in solid blue at $z=15$ and red at $z=20$.
	These two cases have a similar background, but the VAOs are absent on the average case, as the velocities do not fluctuate.
	The dashed lines correspond to the fit from Eq.~\eqref{eq:VAOfit}, with the biases and window functions from that section.
	We do not show error-bars on the simulation results here as it is the difference of two runs with the same initial conditions.
	}	
	\label{fig:vavgP21}
\end{figure}

A subtletly is that for regions of radius $R$ we will not just have a power-law, but instead, following Eq.~\eqref{eq:nutilde} we will have
\be
{\rm SFRD}^{\rm EPS} \propto \int_{\tilde \rm \nu_{\rm min}}^\infty \dfrac{d\tilde  \nu}{\sqrt{\tilde  \nu}} e^{-\tilde \nu/2} \left(\nu/\nu_p\right)^\beta \left(\dfrac{1}{1+\nu/\nu_p}\right)^\beta,
\label{eq:nuinttilde}
\ee
with $\nu_p = \tilde \delta^2/\sigma^2$ (notice the lack of tilde on the $\sigma^2$ here).
A further approximation is to break $(1+\nu/\nu_p)^{-\beta}$ into the two regimes of 1 and $(\nu/\nu_p)^{-\beta}$ for $\nu<\nu_p$ and $\nu>\nu_p$, respectively.
Therefore, for $\nu \gg \nu_p$ the power laws cancel (recovering an erfc), whereas below it follows Eq.~\eqref{eq:nuint}.
This allows us to write the collapsed fraction through this broken power-law as a sum of $\Gamma$ functions.

The code allows the user to turn on this functionality with the flag {\tt FAST\_FCOLL\_TABLES}.
By default they are not used for ACGs, as the speed boost is only $\sim \times 2$, but for MCGs it can reach $\times 30$, so it is recommended.
The calculation for the ionization tables proceeds identically, with $\alpha = \alpha_* + \alpha_{\rm esc}$.
We show an example of how the SFRD compares to the exact calculation in Fig.~\ref{fig:MsigmaM}, which for the MCG regime of interest agrees to $\approx 10\%$, and $\approx$ few \% for $\alpha=0$, which is the MCG-assumed value.
The agreement is even closer in practice, since we only use these tables for the EPS part of the calculation, so an overall offset is cancelled out, and only the $\delta$ behavior remains.
We note that we are not explicitly enforcing $f_*$ or $f_{\rm esc}$ to be less than unity in these approximations.
Although models that violate that condition are disfavored by observations, we caution the reader to set off the {\tt FAST\_FCOLL\_TABLES} when exploring a wide parameter space of extreme models.

\section{VAOs without Poisson noise}
\label{app:VAODiff}

Here we show a simple check that the VAOs can be recovered from our simulations with small Poisson variance.
Our goal is to show that one need not run very large 21-cm boxes to obtain the large-scale VAOs, if two simulations are compared: one with full relative velocities and one with a fixed $\vcb=v_{\rm avg}$.
These two cases share a similar background evolution but have different large-scale power spectrum, as the $\vcb$ anisotropies become imprinted onto the 21-cm fluctuations.
We show the difference in the power spectra during the EoH and EoC in Fig.~\ref{fig:vavgP21}, as well as the fit we found in Sec.~\ref{sec:VAOs}, which provides a reasonable approximation to this difference.
Notice that the high-$k$ part of the power spectrum still deviates in the two cases, as the background matching is not perfect for the average-velocity case.
Subtracting these two cases severely reduces the simulation Poisson error (as the two simulations share initial conditions), though it does not fully cancel it, as clear from the data variation in Fig.~\ref{fig:vavgP21}.
This helps extracting VAOs from more modestly sized simulations.

\bsp	
\label{lastpage}
\end{document}